\documentclass[twocolumn,table,tighten,twocolappendix]{aastex63}
\usepackage{xcolor}

\mathchardef\mhyphen="2D
\usepackage{ mathrsfs }

\received{November 11, 2020}
\revised{December 22, 2020}
\accepted{December 24, 2020}
\submitjournal{the Astrophysical Journal Supplement Series}

\usepackage{tipa}
\usepackage{url, soul, bm}
\usepackage{amsmath,comment}      
\usepackage[T1]{fontenc}
 \usepackage{longtable}
\usepackage{calligra}
\usepackage{CJK}
\usepackage{rotating}

\usepackage{textgreek}
\shorttitle{Spitzer/IRAC Candidate YSOs}

\begin{document} 

\title{SPICY: The Spitzer/IRAC Candidate YSO Catalog for the Inner Galactic Midplane}

\correspondingauthor{Michael A. Kuhn and Rafael S. de Souza}
\email{mkuhn@astro.caltech.edu, drsouza@shao.ac.cn
}  

\author[0000-0002-0631-7514]{Michael A. Kuhn}
\affiliation{Department of Astronomy, California Institute of Technology, Pasadena, CA 91125, USA}

\author[0000-0001-7207-4584]{Rafael S. de Souza}
\affiliation{Key Laboratory for Research in Galaxies and Cosmology, Shanghai Astronomical Observatory,\\
Chinese Academy of Sciences, 80 Nandan Rd., Shanghai 200030, China}

\author[0000-0002-2308-6623]{Alberto Krone-Martins}
\affiliation{Donald Bren School of Information and Computer Sciences, University of California, Irvine, CA 92697, USA}
\affiliation{CENTRA/SIM, Faculdade de Ci\^{e}ncias, Universidade de Lisboa, Ed. C8, Campo Grande, 1749-016, Lisboa, Portugal}

\author[0000-0002-9419-3725]{Alfred Castro-Ginard}
\affiliation{Institut de Ci\`{e}ncies del Cosmos, Universitat de Barcelona (IEEC-UB), Mart\'{i} i Franqu\`{e}s 1, 08028 Barcelona, Spain}

\author[0000-0002-0406-076X]{Emille E. O. Ishida}
\affiliation{Universit\'{e} Clermont Auvergne, CNRS/IN2P3, LPC, F-63000 Clermont-Ferrand, France}

\author[0000-0001-9062-3583]{Matthew S. Povich}
\affiliation{Department of Physics and Astronomy, California State Polytechnic University Pomona, 3801 West Temple Avenue, Pomona, CA 91768, USA}
\affiliation{Department of Astronomy, California Institute of Technology, Pasadena, CA 91125, USA}

\author{Lynne A. Hillenbrand}
\affiliation{Department of Astronomy, California Institute of Technology, Pasadena, CA 91125, USA}

\collaboration{8}{for the COIN Collaboration}

\begin{abstract} 
We present $\sim$120,000 Spitzer/IRAC candidate young stellar objects (YSOs) based on surveys of the Galactic midplane between $\ell\sim255^\circ$ and $110^\circ$, including the GLIMPSE~I, II, and 3D, Vela-Carina, Cygnus~X, and SMOG surveys (613~square degrees), augmented by near-infrared catalogs. We employed a classification scheme that uses the flexibility of a tailored statistical learning method and curated YSO datasets to take full advantage of IRAC's spatial resolution and sensitivity in the mid-infrared $\sim$3--9~$\mu$m range. Multi-wavelength color/magnitude distributions provide intuition about how the classifier separates YSOs from other red IRAC sources and validate that the sample is consistent with expectations for disk/envelope-bearing pre--main-sequence stars. We also identify areas of IRAC color space associated with objects with strong silicate absorption or polycyclic aromatic hydrocarbon emission. Spatial distributions and variability properties help corroborate the youthful nature of our sample. Most of the candidates are in regions with mid-IR nebulosity, associated with star-forming clouds, but others appear distributed in the field. Using Gaia DR2 distance estimates, we find groups of YSO candidates associated with the Local Arm, the Sagittarius-Carina Arm, and the Scutum-Centaurus Arm. Candidate YSOs visible to the Zwicky Transient Facility tend to exhibit higher variability amplitudes than randomly selected field stars of the same magnitude, with many high-amplitude variables having light-curve morphologies characteristic of YSOs. Given that no current or planned instruments will significantly exceed IRAC's spatial resolution while possessing its wide-area mapping capabilities, Spitzer-based catalogs such as ours will remain the main resources for mid-infrared YSOs in the Galactic midplane for the near future.
\end{abstract}

\section{Introduction}

The majority of young stellar objects (YSOs) in our galaxy are formed in massive star-forming complexes located near the Galaxy's midplane. The prevalence of star-forming regions in this part of the Galaxy is attested to by the spatially complex mid-infrared (mid-IR) nebulosity observed to permeate the entirety of the inner midplane and much of the outer midplane. For example, observations by the Spitzer Space Telescope \citep{Werner2004} and the Wide-field Infrared Survey Explorer \citep[WISE;][]{Wright2010} have identified more than a thousand interstellar medium bubbles in these regions, most of which are associated with star formation activity \citep{Churchwell2006,Churchwell2007,Simpson2012,Anderson2014,Bufano2018,Jayasinghe2019}. Nevertheless, apart from a few dozen well-studied star-forming regions, the YSOs in these regions remain either mostly or wholly unknown. This is a consequence of observational difficulties at low Galactic latitudes, including high dust column densities along many lines of sight, which limit optical studies, high stellar densities, which may produce source confusion and increase the number of contaminants in catalogs, and lines of sight that pass through multiple star-forming regions at different distances \citep{Feigelson18}.

There are many scientific applications for reliable lists of YSOs generated uniformly for large segments of the sky rather than on a region-by-region basis. For example, it remains an open question whether nearly all stars are formed in dense groups or whether there is a significant population of stars formed in low-density environments \citep[e.g.,][]{Carpenter2000,Bressert2010,Pfalzner2012,Gieles2012,Kuhn2015}. Hence, catalogs that sample YSOs from both types of environment help to address this question.
In addition, YSO catalogs, when combined with Gaia astrometric data, can be used to map out the kinematics of the youngest component of the Milky Way's thin disk. And, furthermore, with an increasing number of surveys searching large areas of the sky for transients, these catalogs would help in identifying outbursting YSOs and other YSO related variability \citep{Hodgkin2013,Rosaria2018,Graham2019}. 

Our goal here is to make optimum use of Spitzer survey data from the inner Galactic midplane 
(between $\ell\sim255^\circ$ and $110^\circ$ and $|b|<1^\circ$ to $3^\circ$) to identify YSOs out to several kpc in distance, using IR excess selection criteria that are independent of spatial clustering. Here, we focus on the 4-channel Infrared Array Camera \citep[IRAC;][]{Fazio2004} because this instrument provided the highest spatial resolution of any mid-IR imager with wide-area mapping capabilities over wavelengths from 3 to 9~$\mu$m. IRAC far exceeded the point-source sensitivity of {\it WISE} in the Galactic plane, because the latter was severely limited by both detector saturation and source confusion. The extensive IRAC observations of the Galactic plane were obtained as part of the Galactic Legacy Infrared Mid-Plane Survey Extraordinaire \citep[GLIMPSE;][]{Benjamin03,Churchwell09} along with several related Spitzer/IRAC programs that followed similar observing and data processing strategies. 
 
 \defcitealias{Povich11}{P11}
 \defcitealias{Povich13}{P13}
Spitzer has proven effective at identifying candidate YSOs \citep[e.g.,][and many others]{Allen2004,Hartmann2005,Harvey2007,Simon2007,Gutermuth2009,Povich11,Povich13}. However, these studies use differing criteria to select YSO candidates, ranging from simple cuts in color space to empirical probabilistic classification to fitting the spectral energy distributions (SEDs) with models of circumstellar dust actively infalling or accreting onto a central stellar object \citep[e.g.,][]{Robitaille2006,Robitaille2007,Robitaille17}. Our study employs a hybrid approach, which combines the strengths of SED fitting and principled statistical learning techniques.

\defcitealias{Robitaille08}{R08}
An earlier GLIMPSE study \citep[$\ell \sim 295^\circ$--$65^\circ$;][]{Robitaille08} identified $\sim$20,000 ``intrinsically red sources'' ($[4.5] - [8.0] \geq 1$), using strict photometric brightness and quality measures to guarantee that the infrared excesses they identify are real. However, they find that this selection criterion is sensitive not only to YSOs but also to intrinsically red contaminants, largely comprised of (post-)asymptotic giant branch stars (AGBs). They find that the 24~$\mu$m Spitzer/MIPS band is helpful for distinguishing between these cases. However, this band is not available for the vast majority of point sources detected in GLIMPSE \citep{Gutermuth2015}. In our study, we relax these criteria to identify significantly more YSOs candidates, but use patterns in the IRAC photometry to better distinguish between YSOs and contaminants. Our survey area also overlaps YSO catalogs for Cygnus~X \citep{Beerer10,Winston2020} and the Spitzer Mapping of the Outer Galaxy \citep[SMOG;][]{Winston19} survey; we use the latter as a benchmark to compare to our results.
 
 This paper is organized as follows.  \autoref{sec:data} describes the datasets. \autoref{sec:method} explains our statistical methodology. \autoref{sec:catalog} introduces our YSO catalog. Color-color and color-magnitude diagrams of candidate YSOs and probable contaminants are examined in \autoref{sec:color}. The next sections describe properties of YSO candidates related to environment (\autoref{sec:images}), spatial clustering and kinematics (\autoref{sec:clustering}), and variability (\autoref{sec:variability}). Comparisons with other catalogs are made in \autoref{sec:comparison}. \autoref{sec:conclusions} is the conclusion.

\section{Data}
\label{sec:data}
\subsection{IRAC catalogs}

\begin{figure*}
	\centering
	\includegraphics[width=0.45\textwidth]{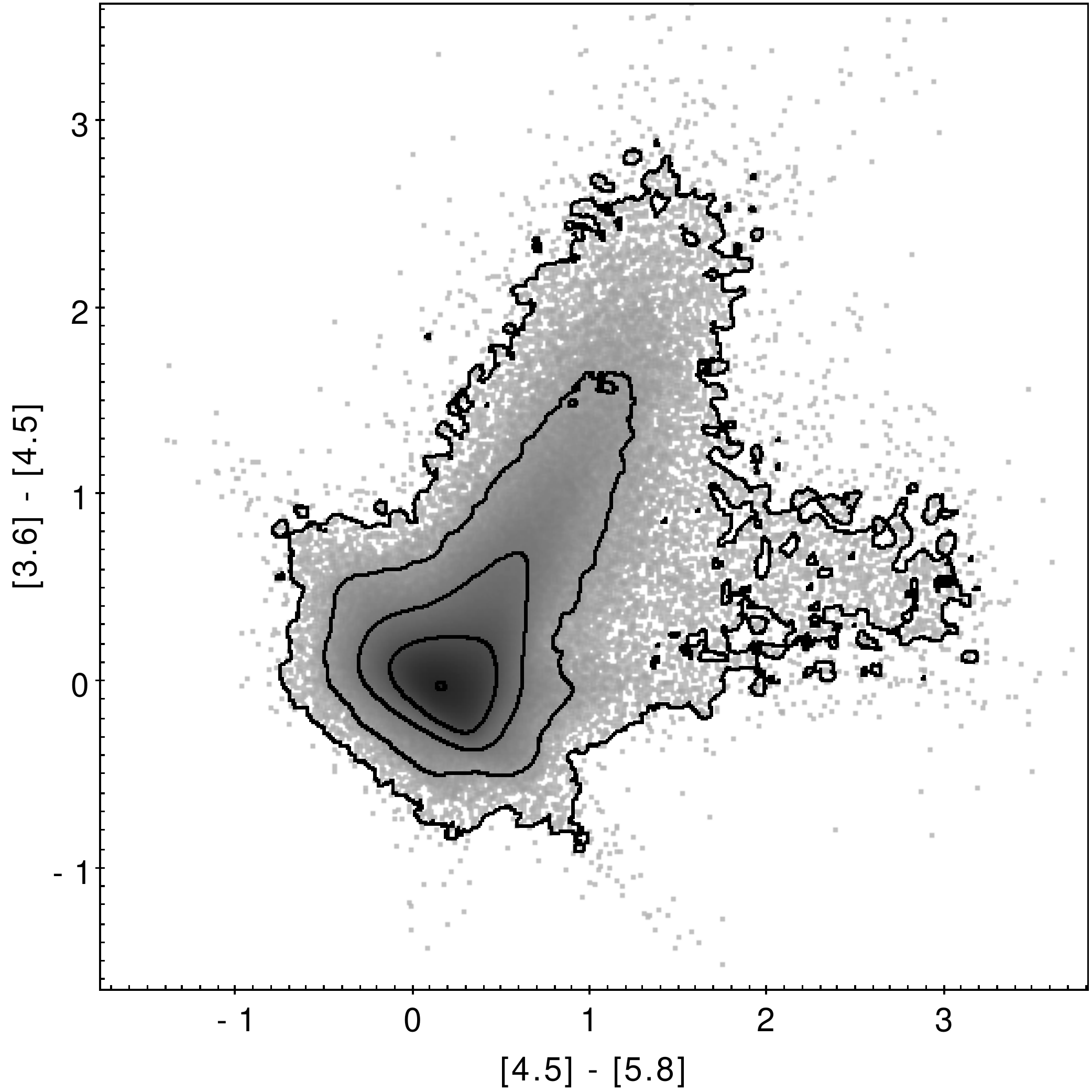}
	\includegraphics[width=0.45\textwidth]{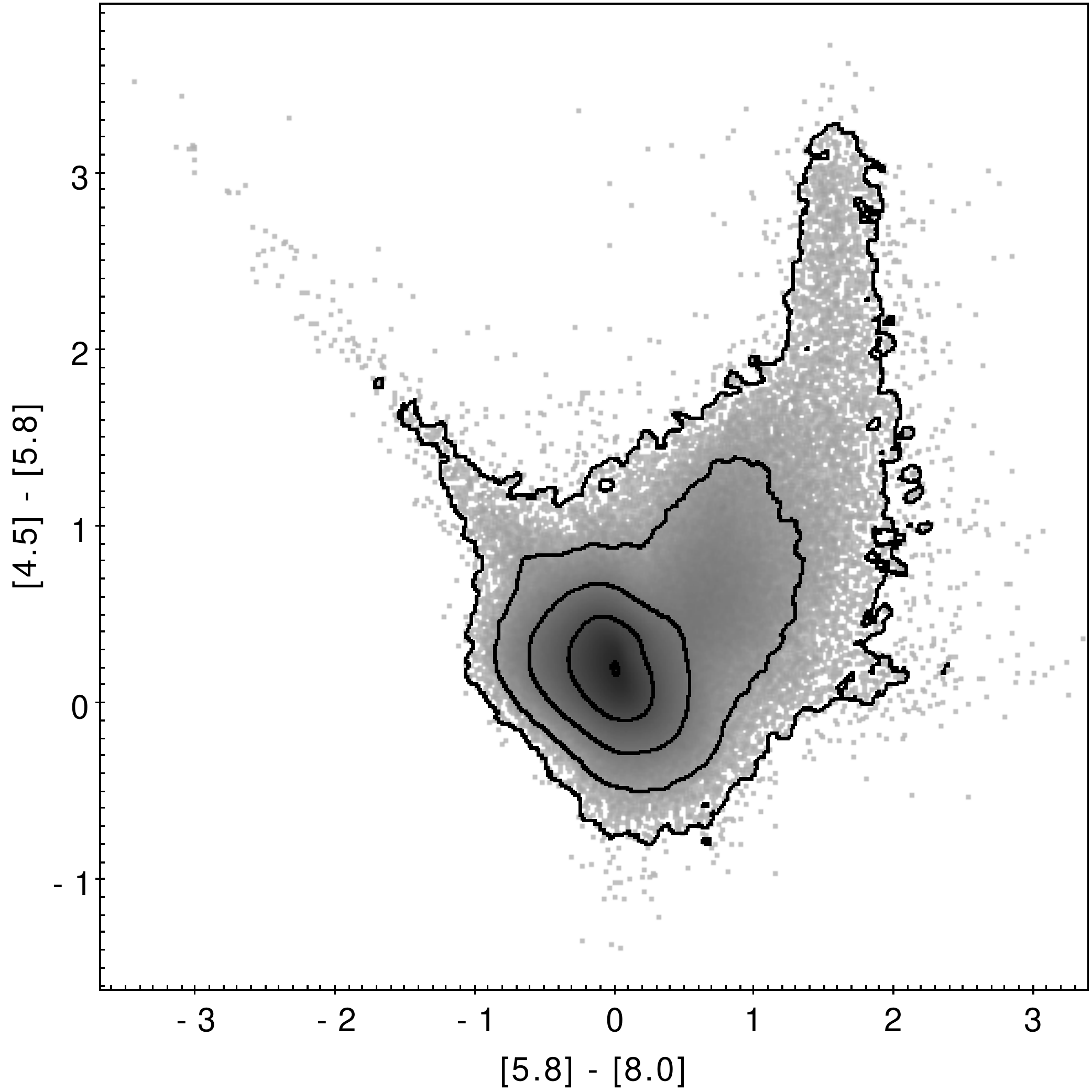}
    \caption{Colors of sources from the GLIMPSE Catalog. (Due to the high number of sources in the full tables, we display a random subsample for plotting convenience.) Contours are drawn at increases in density by a factor of 12. From these plots we see that the highest source density is at colors $\sim$0, but in both the $[3.6] - [4.5]$ vs.\ $[4.5] - [8.0]$ (left) and $[4.5] - [5.8]$ vs.\ $[5.8] - [8.0]$ (right) diagrams there is an excess of redder sources to the upper right. In the right panel, two additional features stand out. A streak from the origin to the upper left is an artifact resulting from source-extraction errors in the 5.8~$\mu$m band. To the upper right, there is a curved feature in the sources distribution, with $[4.5] - [5.8] \gtrsim 1.6$~mag and $[5.8] - [8.0] \approx 1.6\pm0.25$ mag. We argue that these colors are affected by PAH emission (\autoref{sec:PAH}).}
  \label{fig:GLIMPSE_COLORS}
\end{figure*}

The YSO selection is largely based on IRAC photometry from GLIMPSE \citep{Benjamin03,Churchwell09} and related surveys that used similar observing strategies and data reduction methodologies.\footnote{The IRAC point-source catalogs were obtained from the NASA/IPAC archive at \url{https://irsa.ipac.caltech.edu/data/SPITZER/GLIMPSE/overview.html}} These include GLIMPSE I (31,184,509~sources), GLIMPSE II (19,067,533~sources), and GLIMPSE 3D (20,403,915~sources), the Vela-Carina (2,001,032~sources) \citep{Majewski07,Zasowski2009}, Cygnus~X (3,913,559~sources) \citep{Beerer10}, and SMOG (2,512,099 sources) \citep{Winston19} surveys. Spitzer's observations of the Galactic Center \citep{Stolovy2006} were included in the GLIMPSE II Catalog. We use only the Spitzer photometry obtained during the cryogenic mission, which includes 4 mid-IR bands centered at 3.6, 4.5, 5.8, and 8.0~$\mu$m. We omit the GLIMPSE 360 data from the warm Spitzer mission that includes only the 3.6 and 4.5~$\mu$m bands.

The GLIMPSE team designed their reduction pipeline to provide reliable point-spread function (PSF) fitting photometry in crowded fields with spatially varying nebular emission \citep{Benjamin03,Kobulnicky2013} -- conditions that are common in IRAC images of star-forming regions. They provide two source lists for each survey, the ``Catalog'' which is more reliable, and the ``Archive'' which is more complete\footnote{\url{http://www.astro.wisc.edu/sirtf/docs.html}}. Following, \citet{Povich13}, we use the ``Catalog'' photometry. We make no additional cuts on quality flags, but certain flags are discussed in \autoref{sec:GLIMPSE_flags}. The Spitzer/IRAC images have PSFs with full widths at half maximum of 1.66$^{\prime\prime}$ at 3.6~$\mu$m, 1.72$^{\prime\prime}$ at 4.5~$\mu$m, 1.88$^{\prime\prime}$ at 5.8~$\mu$m, and 1.98$^{\prime\prime}$ at 8.0~$\mu$m. This is significantly better than the $\sim$6$^{\prime\prime}$ resolution provided by the WISE survey over a similar wavelength range \citep{Wright2010}, giving IRAC a distinct advantage in crowded fields in the Galactic midplane. The catalogs from the GLIMPSE team also include many stars that are missing from the Spitzer Enhanced Imaging Products (SEIP) due to GLIMPSE's better treatment mid-IR nebulosity. PSF photometry is also more accurate than SEIP aperture photometry in regions with variable backgrounds \citep[][]{Fang2020}.

The GLIMPSE I, II, 3D, Galactic Center, and Vela-Carina survey observations consisted of 2--$5\times1.2$~s integrations at each positions, while the Cygnus~X and SMOG surveys used a 0.4+10.4~s high-dynamics range mode. Given that Cygnus~X and SMOG are deeper than the rest of the data, we impose uniformity by omitting sources in these fields that are either brighter or fainter than the limits for the main GLIMPSE survey.\footnote{The bright limits for GLIMPSE are 7, 6.5, 4.0, and 4.0~mag in the 3.6, 4.5, 5.8, and 8.0~$\mu$m bands, respectively. The 3$\sigma$ detection limits are 15.5, 15.0, 13.0, and 13.0~mag in these bands, but completeness declines precipitously before reaching these limit (\url{http://www.astro.wisc.edu/sirtf/GQA-master.pdf}).
Completeness in the GLIMPSE Catalog is a strong function of both crowding and background sky level, with structure in the background playing a larger role than photon noise \citep{Kobulnicky2013}. In the main GLIMPSE survey, the magnitude distribution of Catalog sources peaks at $[4.5]\approx 13.6$~mag, before declining. We adopt a faint limit of $[4.5]=14.5$~mag (where density has decreased by a factor of $\sim$5) based on this magnitude distribution and because no source fainter than this is selected as a YSO candidate.} The magnitudes of our selected YSO candidates range from $[3.6]=8.3$--$14.9$~mag, $[4.5]=7.3$--$13.7$~mag, $[5.6]=6.4$--$12.9$~mag, and $[8.0]=5.5$--$12.2$~mag (1\%--99\% quantiles), and the median photometric uncertainties are 0.062, 0.073, 0.075, and 0.060~mag in these bands, respectively. 

Even in the full GLIMPSE catalogs, the presence of red sources and several catalog artifacts can be seen in color-color diagrams (\autoref{fig:GLIMPSE_COLORS}). The highest concentration of sources have colors close to 0 (expected for normal stars without IR excess), but numerous sources form a distribution extending to the upper right in these plots, which is made up of both YSOs and other red mid-IR sources (e.g., evolved stars, galaxies, etc.; \autoref{sec:contaminants}). In the $[4.5]-[5.8]$ vs.\ $[5.8]-[8.0]$ diagram, a streak can be seen extending from the origin to the upper left. This streak appears to be related to erroneous photometry in the 5.8~$\mu$m band for a low fraction of the GLIMPSE sources, and it extends from the origin because this is where the source density is highest. Another prominent feature in this diagram is a finger-like structure extending upward at $[5.8]-[8.0]\approx1.6$, which we attribute to polycyclic aromatic hydrocarbon (PAH) emission (\autoref{sec:PAH}).

\subsection{Cross-Matches to Near-IR Catalogs}

Near-infrared $JHK_s$ photometry from the Two Micron All Sky Survey \citep[2MASS;][]{Skrutskie06} is already included in the GLIMPSE (and extensions) data products.2MASS has a spatial resolution of $\sim$2$^{\prime\prime}$, which is comparable to the Spitzer/IRAC PSF. For our sample, 2MASS is nearly complete down to $J\sim15.4$~mag, $H\sim14.2$~mag, and $K_s\sim13.0$~mag, with median photometric uncertainties of 0.038, 0.040, and 0.035, respectively. While these limiting magnitudes correspond well with the limits of the GLIMPSE surveys, in practice YSOs are often found in regions of high interstellar reddening, where 2MASS may not be deep enough to detect NIR counterparts of red GLIMPSE sources.

Deeper NIR catalogs with higher spatial resolution are available from the United Kingdom Infra-Red Telescope (UKIRT) Infrared Deep Sky Survey \citep[UKIDSS;][]{Lawrence07}  and the Visible and Infrared Survey Telescope for Astronomy (VISTA) Variables in the V\'ia L\'actea survey \citep[VVV;][]{Minniti10} for the northern and southern portions of the Galactic plane, respectively, with overlap around the Galactic Center. These catalogs are deeper than 2MASS, but are saturated for brighter sources. We use the UKIDSS catalog from the Galactic Plane Survey \citep{Lucas08} and the averaged VVV photometry for multiple epochs from the VVV Infrared Astrometric Catalog \citep[VIRAC DR1;][]{Smith18}. For both deeper NIR surveys, the  photometry extends to $J\sim19$~mag, $H\sim18$~mag, and $K_s\sim16$~mag with formal photometric uncertainties $<$0.01~mag.

IRAC and UKIDSS/VVV were cross matched using a 1$^{\prime\prime}$ match radius in TOPCAT \citep{Taylor2005}. Photometric measurements from UKIDSS/VVV were omitted if they did not have the flag $mergedClass = -1$ (stellar) or if they had magnitudes $J<11$, $H<12$, $K<10.5$ (UKIDSS) or $J<12.5$, $H<13$, $K_s<11.5$ (VVV), for which saturation effects start to affect photometry. The higher spatial resolutions of the NIR catalogs mean that it is possible for multiple NIR sources to be associated with individual IRAC sources; however, examination of IRAC+UKIDSS matching by \citet{Morales2017} have found that the NIR flux is usually dominated by a single counterpart. 

In our analysis we perform the YSO candidate selection independently on cross-matches of IRAC+2MASS, IRAC+UKIDSS, and IRAC+VVV, and the results of these separate selections are merged. 

\subsection{Ancillary Data}

The Gaia mission \citep{2016A&A...595A...1G}, in its second data release \citep[Gaia DR2;][]{2018A&A...616A...1G},  has provided optical broad band photometry \citep{2018A&A...616A...4E} for the whole sky along with exquisite astrometric measurements \citep{Lindegren2018} for more than $1.3$ billion stars. These data on their own can be used for selecting possible pre--main-sequence stars \citep[e.g.,][]{Zari2018}, but for our study we use them as ancillary data to better understand the parallax ($\varpi$)  and proper motions ($\mu_{\alpha*}, \mu_{\delta}$),  distributions of the IR-excess selected YSO candidates.  From the YSOs candidate list (\autoref{sec:catalog}), 33\% have Gaia counterparts with the full 5-parameter astrometric solution \citep{Lindegren2018}. A match rate below 50\% is expected because many YSOs are enshrouded by dust and thus not optically visible. 

Longer wavelength photometry is available from both Spitzer's MIPS Galactic Plane Survey \citep[MIPSGAL;][]{Carey2009,Gutermuth2015} at 24~$\mu$m and the WISE All-Sky Data Release \citep{Wright2010} at 22~$\mu$m. In the Galactic plane AllWISE is affected by high numbers of spurious sources, particularly in the longer wavelength bands, so we follow the catalog cleaning recommendations from \citet{Koenig14} and apply the signal-to-noise and $\chi^2$ quality cuts on the profile-fit photometry given in their Equations~1--4. For MIPSGAL, we use the ``Catalog'' instead of the ``Archive.'' The WISE photometry, and to a lesser extent the MIPS photometry, is strongly affected by crowding and nebulosity in the regions around the Galactic plane that we are investigating, leading to fewer reliable source detections in these areas. The application of the quality cuts from \citet{Koenig14} leave visible holes in the spatial distribution of WISE sources surrounding the clusters of YSO candidates that we identify with the IRAC photometry. Only 30\% of our IRAC candidates have counterparts in the 24~$\mu$m MIPS band, and 8\% have reliable 22~$\mu$m WISE photometry. Because of the unavailability of longer wavelength data for the majority of our sample, we do not uses these bands for identifying candidates, but only for post-selection examination of the sample. For both catalogs we used a cross-match radius of 1.2$^{\prime\prime}$. 

\subsection{Published YSO Catalogs}\label{sec:mystix}

YSOs identified in earlier studies of star-forming regions within GLIMPSE (and extensions) can be used to train a classifier to find similar types of objects. We use YSOs identified as part of the Massive Young Star-Forming Complex Study in Infrared and X-ray \citep[MYStIX;][]{Feigelson13}, in addition to a similar, earlier study of the Carina Nebula \citep[][]{Townsley2011}. From the combined lists, we have included probable YSOs from the Carina Nebula, NGC~6611, M17, NGC~6530, M20, NGC~6357, NGC~6334, RCW~38, RCW~36, and DR~21, ranging from $\sim$0.7 to 2.7~kpc in distance. \citet{Povich11,Povich13} performed the IR excess detection for these projects using Spitzer data based on a strategy that included SED fitting of both reddened stellar atmospheres and the Robitaille YSO models, color cuts to remove certain types of contaminants, spatial filtering to remove objects that are not clustered, and visual examination of SEDs. 

 We select all objects from the MYStIX IR-excess catalogs classified as both a YSO ($Cl = 0$) and a probable member ($Mm = 1$). More recently, Gaia DR2 has become available, so, for the subset of sources with Gaia parallaxes ($\sim$30\% of the sample), we refine the sample further by removing any source with a parallax that is discrepant from the median parallax of the group by $>$2 times the reported parallax error. 
 
In addition to the aforementioned studies of multiple regions and large areas, GLIMPSE has been used in hundreds of papers about individual (or several) star-forming regions. A few representative examples of these include \citet{Zavagno2006}, \citet{Watson2008}, \citet{Povich09}, \citet{Dewangan2013}, \citet{Samal2014}, \citet{Mallick2015}, and \citet{Povich16}. 

\section{Methodology}
\label{sec:method}

YSOs make up a minuscule fraction of the nearly fifty million sources detected in the Spitzer/IRAC surveys included in this project. This means that selection of YSO candidates requires rejection of numerous contaminants (mostly field stars) along similar lines of sight. The first steps in our procedure, in which we reject sources that can be explained without IR excess, are nearly identical to those from \citet{Povich13}. 
These steps greatly reduce the sample size and are based on well established stellar atmosphere models.
However, in the next steps -- classification of the remaining sources -- rather than fitting models of YSO SEDs as \citet{Povich13} do, we use their resulting MYStIX YSO sample to train our random forest classifier. 

A data-driven approach offers some advantages. For instance, we use IRAC photometry of actual stars as a training set instead of artificial photometry generated from theoretical YSO models. This means that the method will tend to avoid classifying an object with unusual colors as a YSO even if these colors can be reproduced by a physically unrealistic configuration of a star, disk, and envelope that exists in a grid of theoretical YSO models. Furthermore, it takes significantly less computational time to apply a trained classifier to millions of stars than it does to fit each of them with several categories of parametric YSO model. Nevertheless, the YSO SED fitting method does play an important role in generating training sets for the classifier (\autoref{training.sec}).

\subsection{Removing Sources without Significant IR Excess}\label{sec:sed_reddened_photosphere}

In the Galactic midplane, many background stars are affected by high levels of foreground extinction, so any source that is either insufficiently red or whose red colors can plausibly be explained by reddening alone is dropped from further scrutiny. 

Cuts on IRAC colors and color uncertainties can remove many objects that have no chance of being selected as reliable IR excess objects. We apply the rules recommended by \citet{Povich11,Povich13}, decreasing the number of sources in our sample by a factor of $\sim$10. All retained sources must be detected in at least 4 out of the 7 IR bands, two of which must be 3.6 and 4.5~$\mu$m. Sources are kept if there is the suggestion of IR excess in the $[3.6]-[4.5]$ color using the criterion
\begin{equation}
    [3.6] - [4.5] - 0.408 > \mathrm{error}([3.6]-[4.5]),
\end{equation}
where ``error'' denotes uncertainty in color, calculated by adding the photometric uncertainties for the two bands in quadrature. The value 
0.408 is the expected reddening of this color with $A_V\approx30$~mag of extinction.
Sources are also kept if they have have photometric measurements in the 5.8 and 8.0~$\mu$m bands and either meet both the criteria
\begin{eqnarray}
\left|[4.5]-[5.8]\right|&>&\mathrm{error}([4.5]-[5.8])\\
~[5.8]-[8.0]&>&\mathrm{error}([5.8]-[8.0]),
\end{eqnarray}
or 
\begin{eqnarray}
\left|[4.5]-[5.8]\right|&\leq&\mathrm{error}([4.5]-[5.8])\\
~\left|[5.8]-[8.0]\right|&\leq&\mathrm{error}([5.8]-[8.0]).
\end{eqnarray}
These rules, optimized from experience with GLIMPSE data, ensure that determination of IR excess is based on more than just the 8.0~$\mu$m band, which can occasionally give a spuriously bright measurement.

We fit the $JHK$+IRAC SEDs of the remaining sources with reddened \citet{CastelliKurucz2004} stellar atmosphere models, using the \citet{Indebetouw2005} extinction law. The fitting procedure takes into account the statistical photometric uncertainties on the data, which happen to be of similar size for both 2MASS and IRAC photometry. The UKIDSS and VVV datasets provide $JHK$ photometry for many objects that were not detected in 2MASS, allowing many more sources to be included. However, the statistical measurement uncertainties for most UKIDSS and VVV sources are far more precise than for 2MASS or IRAC. Given that we are mostly interested in detecting deviations from a reddened stellar atmosphere model in the IRAC bands and we want similar selection performance for each dataset, we re-scale all UKIDSS and VVV error bars that are smaller than the median 2MASS error bars to be equal to the median 2MASS error bars. The sources that are poorly fit by the reddened stellar photosphere, with $\chi
^2$ per data point $>$4, comprise the target set for our random forest classifier. 

Overall, these pruning steps leave 319,251 2MASS+IRAC, 188,701 UKIDSS+IRAC, and 257,334 VVV+IRAC sources with possible IR excess as inputs to our classification step below.

\subsection{Training Sets}
\label{training.sec}

The training data includes both MYStIX IR-excess sources (\autoref{sec:mystix}) that we label ``YSO'' and sources unlikely to be YSOs that we label ``contaminant'' (see discussion of contaminants below). 
Although lists of members are more complete in some of the nearest star-forming regions \citep[e.g., Ophiuchus or Taurus;][]{Evans2009,Luhman2018_Taurus},
we choose to use MYStIX because these massive star-forming complexes may better represent the regions we expect to probe in the Galactic midplane, i.e.\ at greater distances, with higher extinction, and in more extreme environments. Furthermore, many of the MYStIX regions lie within the survey region of GLIMPSE and its extensions, meaning that homogeneous data products are available for both the training and target sets. 

Contaminants can include both sources that occur in star-forming regions (e.g., non-stellar sources such as nebular knots and shocked emission) and sources that are smoothly distributed on the sky \citep[e.g., AGB stars and galaxies;][]{Robitaille08,Gutermuth2009} -- see \autoref{sec:contaminants} for discussion of these objects.  \citet{Povich11,Povich13} used a variety of techniques to remove these objects from their catalogs, including SED fitting of \citet{Robitaille2007} models, color cuts, and visual inspection. We label any object within the field of view of these studies that was not classified as a probable young star by either IR or X-ray criteria as a field object. 

To enlarge our sample of contaminants, we identify several fields near the Galactic plane that have no signs of star formation (Appendix~\ref{sec:field}) and label objects in these fields as non-YSOs. These fields were selected to include lines of sight at multiple Galactic longitudes, in the midplane and up to several degrees above or below it, and with different amounts of Galactic extinction.

Training sets are generated separately for each combination of NIR+IRAC data due to the differences in NIR filters. For 2MASS+IRAC, the training set contains 2,865 YSOs, 3,436 field objects in the MYStIX fields, and 7,718 other field objects for the 2MASS+IRAC dataset. For UKIDSS+IRAC these numbers are 919, 2,000, and 1,128, and for VVV+IRAC they are 1,266, 1,459, and 2,595, respectively. 

The distributions of training-set object in color space is discussed in \autoref{sec:training_ccds}. The full IRAC catalog of $\sim 5\times10^7$ sources includes a few outliers in region of color space that are not well sampled by either the labeled YSOs or labeled non-YSOs in the training set. (The limits used to identify these outliers are given in \autoref{sec:training_ccds}.) Given that we have little basis to assign such objects to either category, we are cautious and do not include these objects in our final YSO list.

\subsection{Missing data}

\begin{figure}
	\centering
    \includegraphics[width=0.48\textwidth]{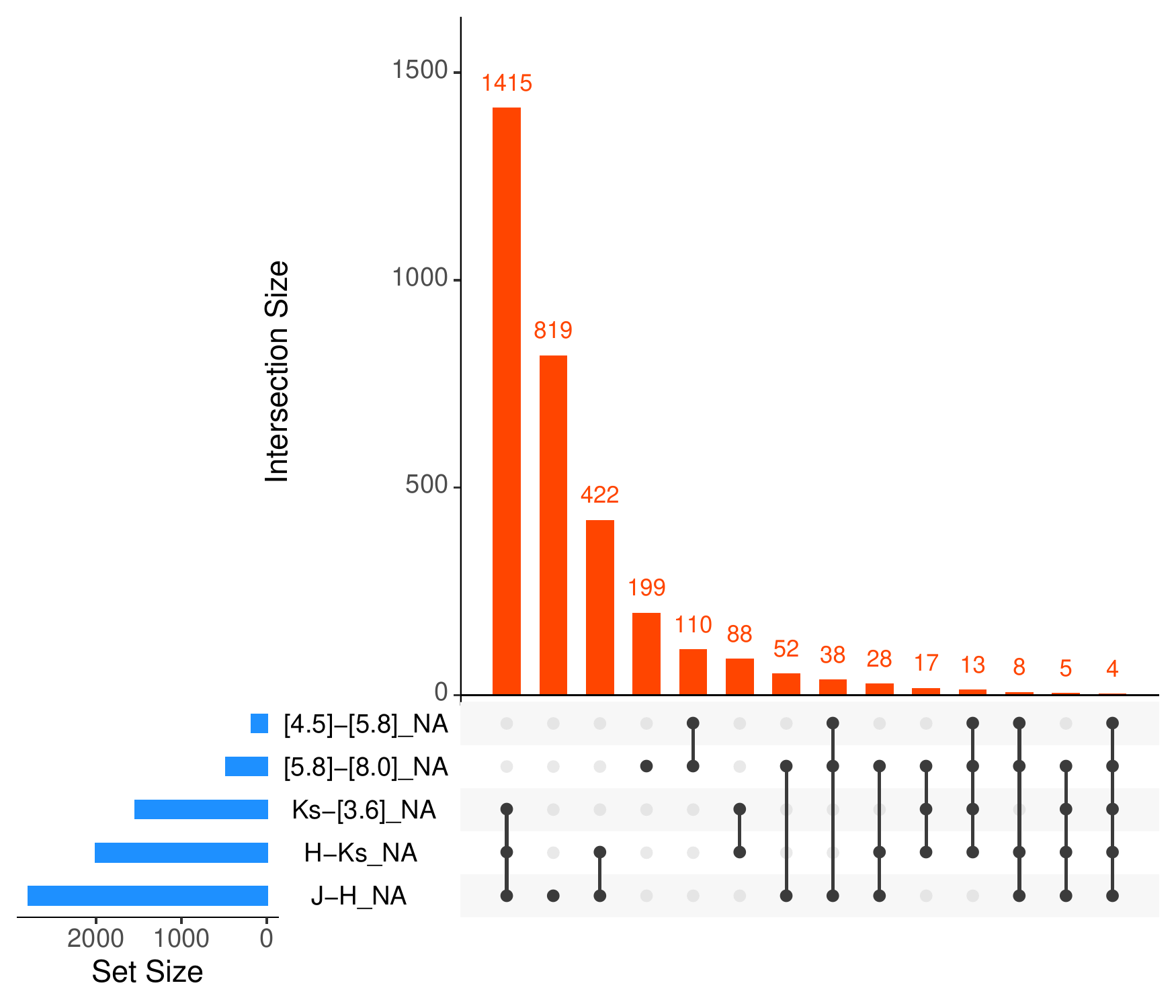}
    \caption{Missing data pattern the labeled 2MASS+IRAC training set. Blue bars are the number of missing colors, the connected black dots indicate combinations of missing colors, and the red histograms indicate the number of instances these combinations are missing. The UKIDSS and VVV diagrams are provided in the figure set.\\
    (The complete figure set (3 images) is available in the online version.)}
  \label{fig:2MASS_NA}
\end{figure}

When data are combined across multiple catalogues, it is almost certain that missing data will occur, as is the case here. \autoref{fig:2MASS_NA} depicts the missing pattern for the 2MASS+IRAC, UKIDSS+IRAC, and VVV+IRAC training sets, from which only 57\%, 46\% and 29\% of objects, respectively, have complete information. About 20\% of the 2MASS+IRAC objects are missing three colors at once, and $JHK_s$ are often missing together. The VVV+IRAC dataset has the most missing data, with 57\% of the objects missing at least three bands. While UKIDSS+IRAC is the most complete, more than 35\% of their rows have at least two missing colors. Thus, a naive removal of rows presenting missing values would throw away a non-negligible amount of valuable information. 

As a final pre-processing step before training our YSO classifier we employed a multiple copula imputation. This decomposes joint probability distributions into their marginal distributions and a function, the copula, that couples them \citep{Nelsen10}.  Copulas have been used previously in astronomy, for example, to construct likelihood functions for weak lensing analysis \citep{Sato2011,Lin2016} and to infer bivariate luminosity and mass functions \citep{Andreani2018}.  Previous tests suggest that this method outperforms other popular approaches, such as multiple imputation via chained equations \citep{Stef2011} and Amelia \citep{Amelia2011}, in terms of bias and coverage, especially in cases where the variables are not normally distributed \citep{hoff2007}.  
The underlying idea of copula imputation is to derive conditional density functions of the missing variables given the observed ones through the corresponding conditional copulas, and then impute missing values by drawing observations from them.  Finally, the choice of performing imputation before training the random forest models has been previously assessed by other studies \citep{jaeger2020impute}, which have shown it to reduce the variance in model error estimate, without any detectable change in precision. The imputation method was implemented using the {\sc sbgcop} package \citep{sbgcop} within the {\sc r} language \citep{rcore19}. Copulas were fit simultaneously to both training and target datasets.

Overall, the imputed data preserves the coverage of the original dataset (\autoref{sec:training_ccds}). Nevertheless, we do not advocate for the use of these colors in other contexts. They are treated here as nuisance parameters to enable classification of the entire dataset.

\subsection{Tree-based Classification}\label{tree.sec}

\begin{figure*}
	\centering
    \includegraphics[width=0.24\textwidth]{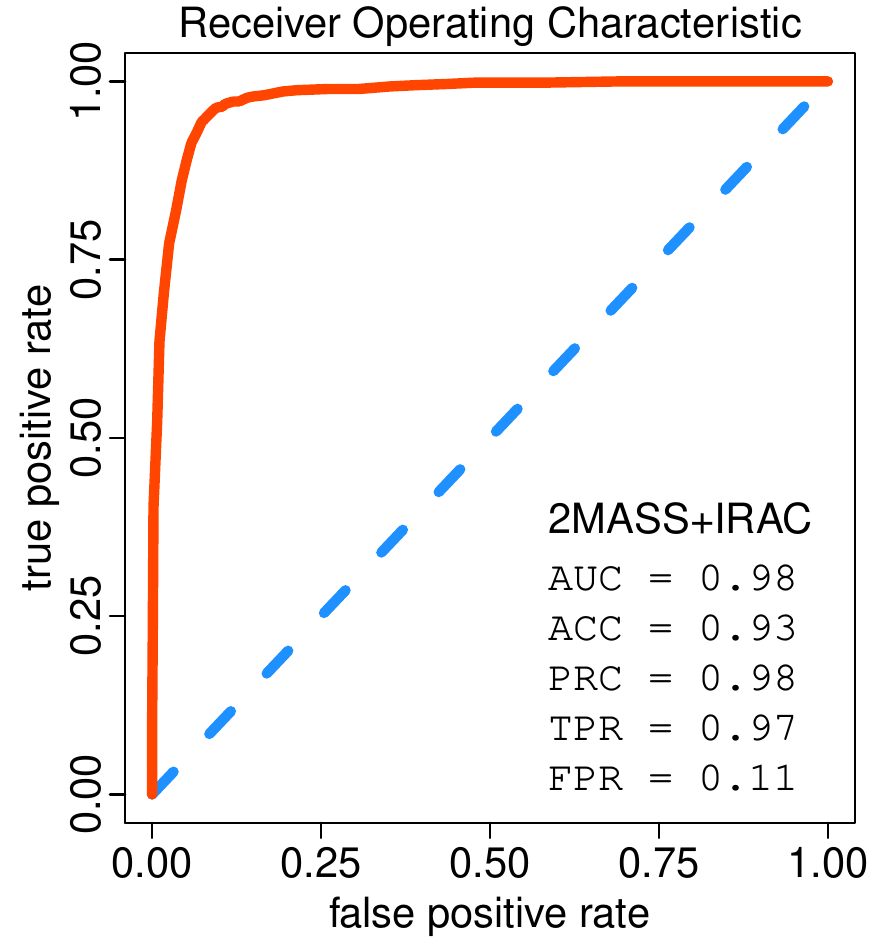}
	\includegraphics[width=0.24\textwidth]{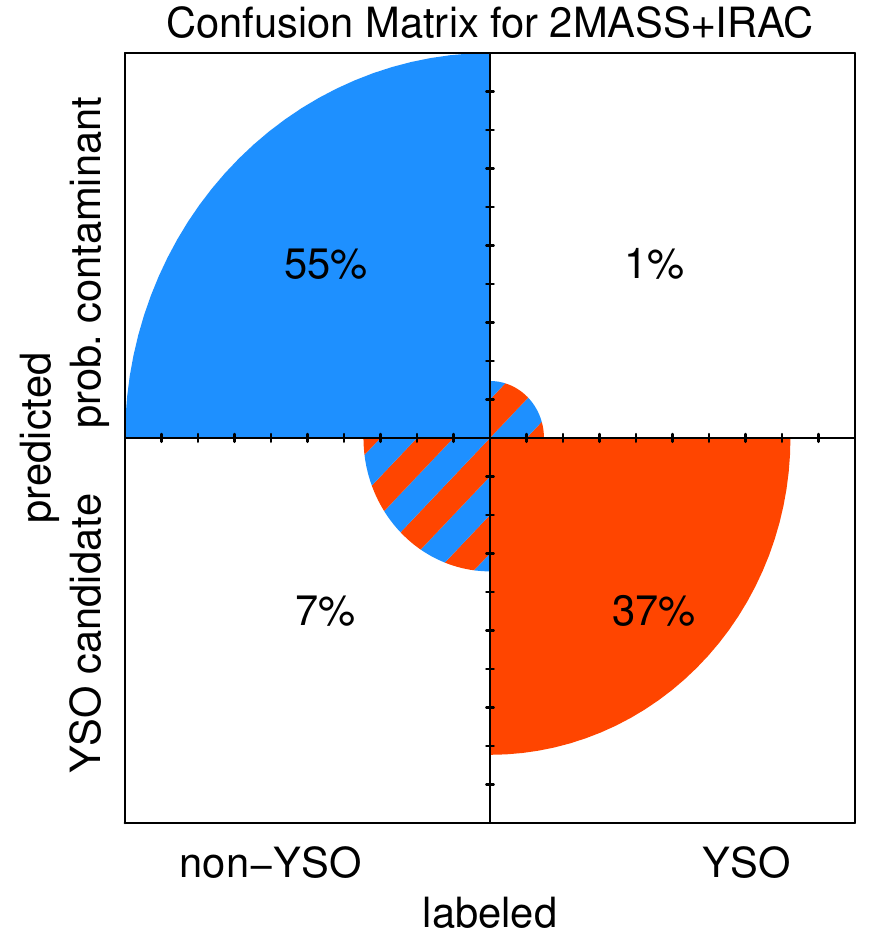}
	\includegraphics[width=0.24\textwidth]{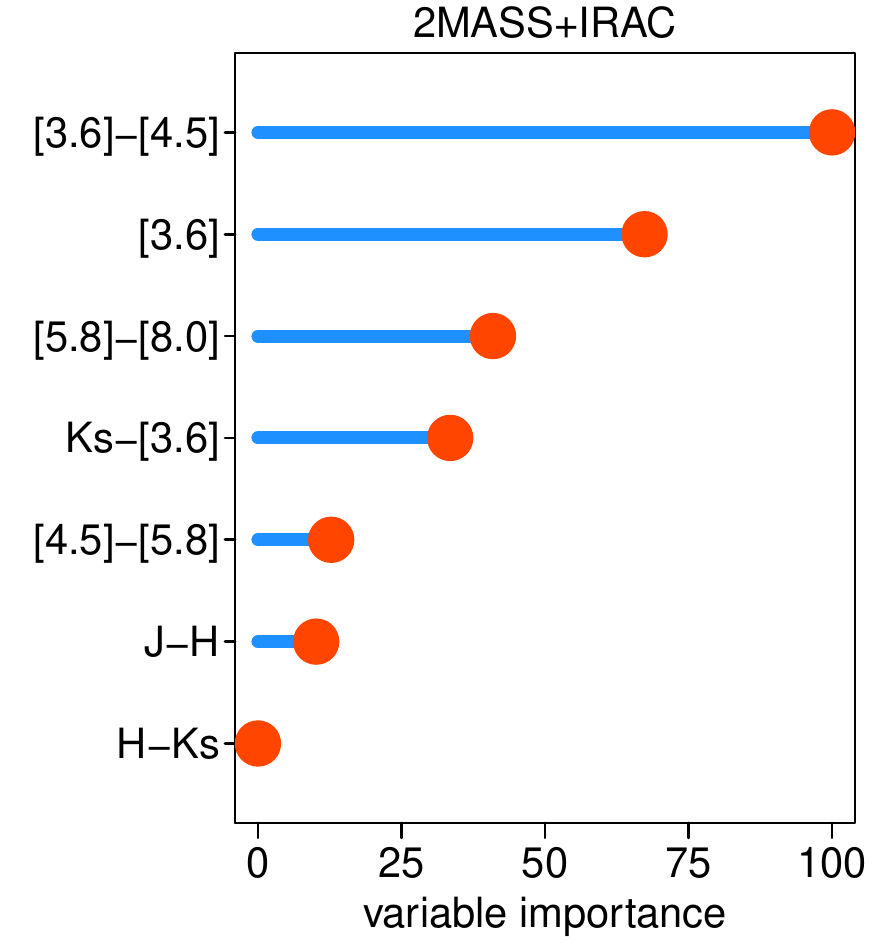}
	\includegraphics[width=0.24\textwidth]{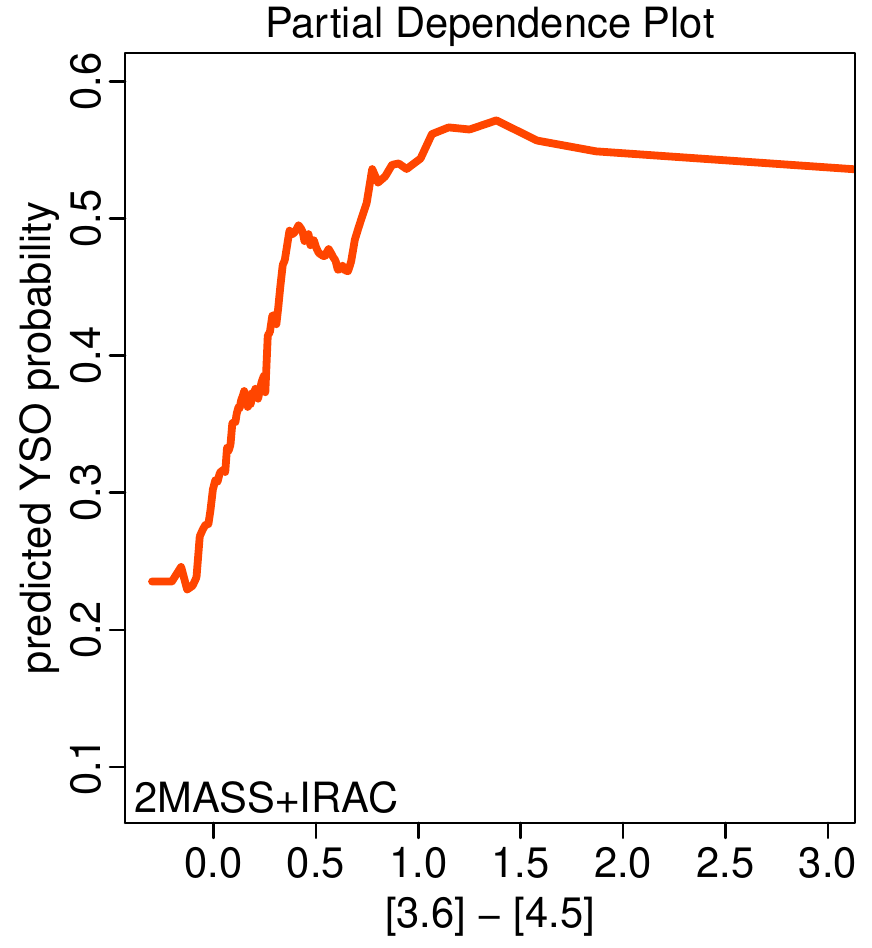}
    \caption{Diagnostic plots for the 2MASS+IRAC classifier (plots for additional variables and for the UKIDSS and VVV classifiers are included in the figure set). 
    Far left: ROC curve. Values are provided for area under the curve (AUC), accuracy (ACC), precision (PRC), true positive rate (TPR), and false positive rate (FPR). Center left: Confusion matrix. Wedge areas are proportional to the numbers of objects in each category. Center right: Estimated importance of colors and magnitudes in the random forest model. Far right: Partial dependence plot for the $[3.6]-[4.5]$ color.\\
    (The complete figure set (30 images) is available in the online version.)
    }
  \label{fig:RF_pdp}
\end{figure*}

Decision trees are learning algorithms that resemble the natural flow of human decision making. At each node of the tree, the algorithm randomly selects one feature and, based on the distribution of training data, determines the decision boundaries that best separate different classes. Training objects are then propagated along their branch of the tree to the next node, where a new feature is selected. The process is repeated until the tree reaches a pre-determined depth or until all objects in a leaf belong to the same class \citep[for a detail description see][]{rokach2014}. This basic concept has given rise to successful algorithms in many different fields. However, a single decision tree trained on an entire dataset is prone to overfitting, presenting low accuracy results whenever faced with data not used in training. This problem can be overcome by randomizing different stages of the tree construction and combining many independent estimators in a more robust classifier. This type of approach belongs to the wider class of ensemble models. 

Ensemble methods \citep[e.g.][]{sagi2018} are regression algorithms, constructed from the combination of many weak classifiers that, when considered together, provide a robuster estimate than any of their constituents. Random forests \citep{ho1995, breiman2001random} are one such algorithm, composed of many decision trees, each constructed independently. The final classification is determined via majority vote, considering all trees in the forest. In this context, the probability of being a YSO is approximated by the percentage of trees in the ensemble voting for a YSO candidate -- we call this probability estimate the ``YSO score.'' Random forests have been successfully used to classify YSOs in smaller scale studies, including with missing data imputation  \citep[e.g.,][]{2017A&A...597A..90D,Melton2020}.

We construct the YSO random forest classifiers using the following covariates: $J-H$, $H-K_s$, $K_s-[3.6]$, $[3.6]-[4.5]$, $[4.5] - [5.8]$, $[5.8]-[8.0]$, and the 3.6~$\mu$m band magnitude. To mitigate the class imbalance influence on tree optimization,  we employed an up-sampling method during the training process. Overall this does not have a perceptible effect on the classification performance but pushes the decision boundary probability between classes closer to 0.5. Each model was independently trained for 2MASS + IRAC, UKIDSS + IRAC, and VVV + IRAC datasets. The random forest was employed using the {\sc caret} R package, with 1500 trees, which was sufficient to guarantee a stable solutions. As a sanity check, we tested few other regression models (including generalized additive models, support vector machine, gradient boosting machines, and conditional random forest), but no significant difference in the final YSO candidate set was found. This suggests that random forest (or other typical non-linear classifiers) captures the data complexity well enough without the need for highly complex models.

\subsection{Classifier Performance}
\label{sec:diagnostics}

The performances of the classifiers were assessed with validation tests in which we partitioned the labeled objects into a training set (80\%) and a test set (20\%).

\autoref{fig:RF_pdp} (far left) shows true positive rate (TPR) as a function of false positive rate (FPR), a curve often called the receiver operating characteristic (ROC). The area under the curve (AUC) can range from 0 to 1, where random guessing gives an area of 0.5, and higher values indicate better performance. Our classifiers all have $\mathrm{AUC}>0.9$. A threshold on YSO score of $p=0.5$ gives high TPRs and low FPRs for each classifier, meaning that this threshold is effective for distinguishing between the YSO and non-YSO classes. Other performance measures are listed on the plot, include accuracy (ACC), the number of true positives plus true negatives divided by the total population, and precision (PRC), the number of true positives divided by the number of true positives plus false positives. The confusion matrix (\autoref{fig:RF_pdp}, center left) shows few misclassifications. 

Variable importance (\autoref{fig:RF_pdp}, center right) is evaluated via out-of-bag samples, which consists of random samplings of the data that are left out of each tree. This is calculated by measuring variations in the prediction error when the out-of-bag data are permuted solely among a specific color, leaving the others unchanged. The process is then repeated across all trees. The final result is a measure of the incremental error for a given color when compared with the unperturbed colors for all the 1500 trees over the entire forest.   

\autoref{fig:RF_pdp} (far right) displays a partial dependence plot \citep[PDP;][]{Brandon2017} for $[3.6]-[4.5]$ color (plots for other variables are included in the figure set). PDPs are useful for visualizing the relationship between individual features and the response while accounting for the average effect of the other predictors in the model. The shape and steepness of the curves are indicators of the predictor's relative influence. Note the sharp behaviour of $[3.6] - [4.5]$, one of the best indicators of YSOs candidates.  

\section{Catalog}
\label{sec:catalog}

\begin{sidewaysfigure}
	\centering
	\includegraphics[width=0.995\textwidth]{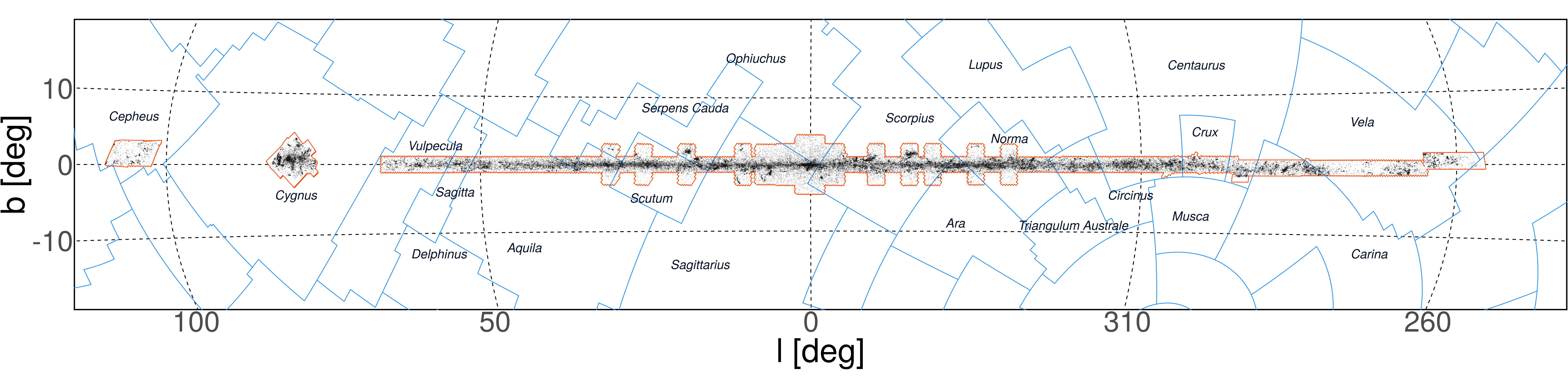}
    \caption{Spatial distribution of the candidate YSOs (black points) within the Spitzer/IRAC survey regions (outlined in red). The data are plotted in Galactic coordinates and the constellation boundaries are shown in blue. Candidate YSOs tend to be concentrated toward the midplane and/or in spatial clusters.}
  \label{fig:skymap}
\end{sidewaysfigure}

\begin{figure*}
	\centering
	\includegraphics[width=0.995\textwidth]{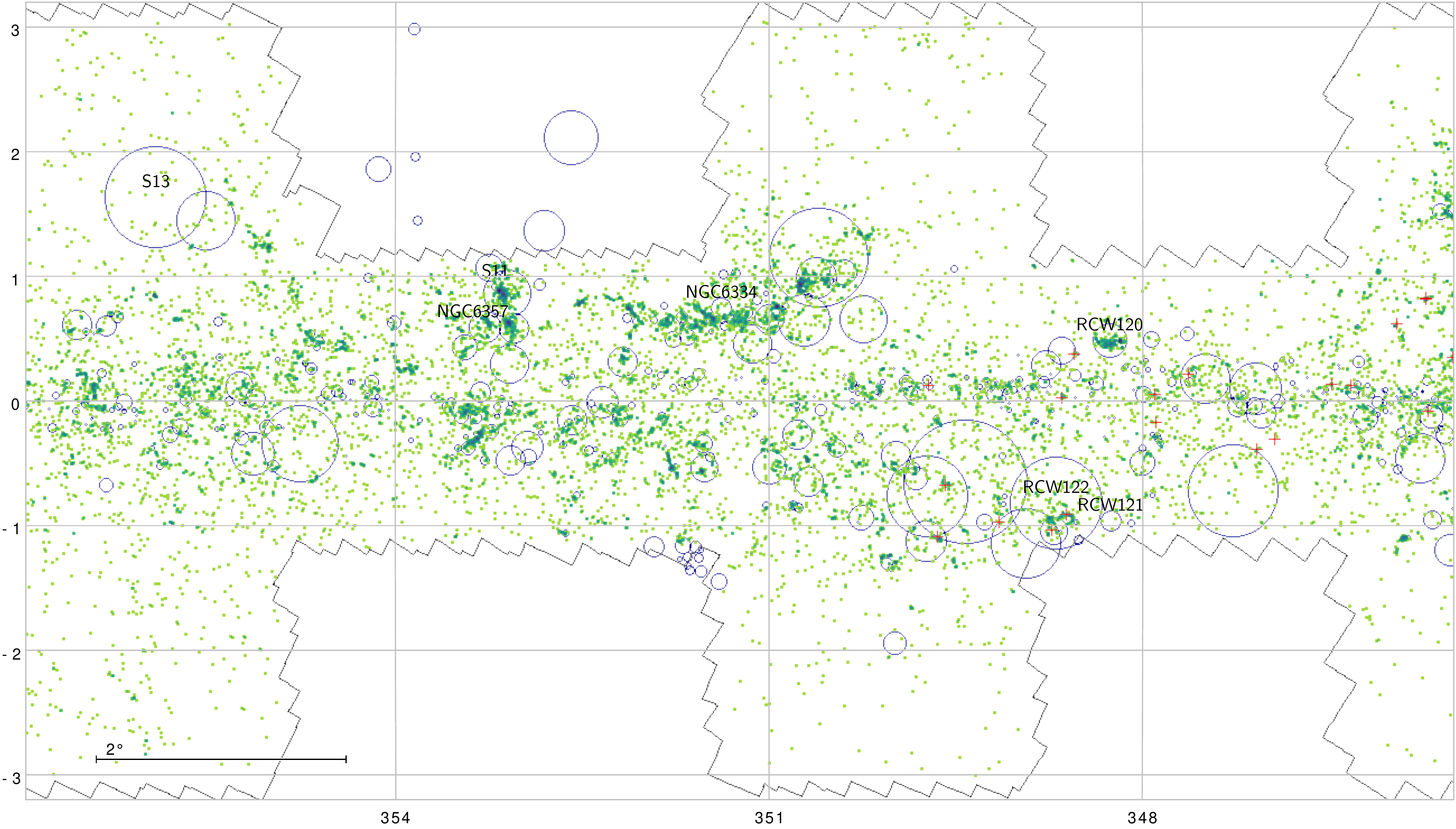}
    \caption{This figure set (19 components) provides an atlas of the Galactic midplane with locations of the YSO candidates (green points) and the boundaries of the IRAC surveys (black lines). Overlapping points produce darker shades of green, using a square-root scale and the ``viridis'' color pallet. For context, we also show outlines of H\,{\sc ii} bubbles from WISE \citep[blue circles;][]{Anderson2014}, massive YSOs from MSX \citep[red crosses;][]{Lumsden2013}, and labels of select star-forming regions. }
  \label{fig:atlas}
\end{figure*}

\begin{figure*}
	\centering
	\includegraphics[width=0.45\textwidth,angle=270]{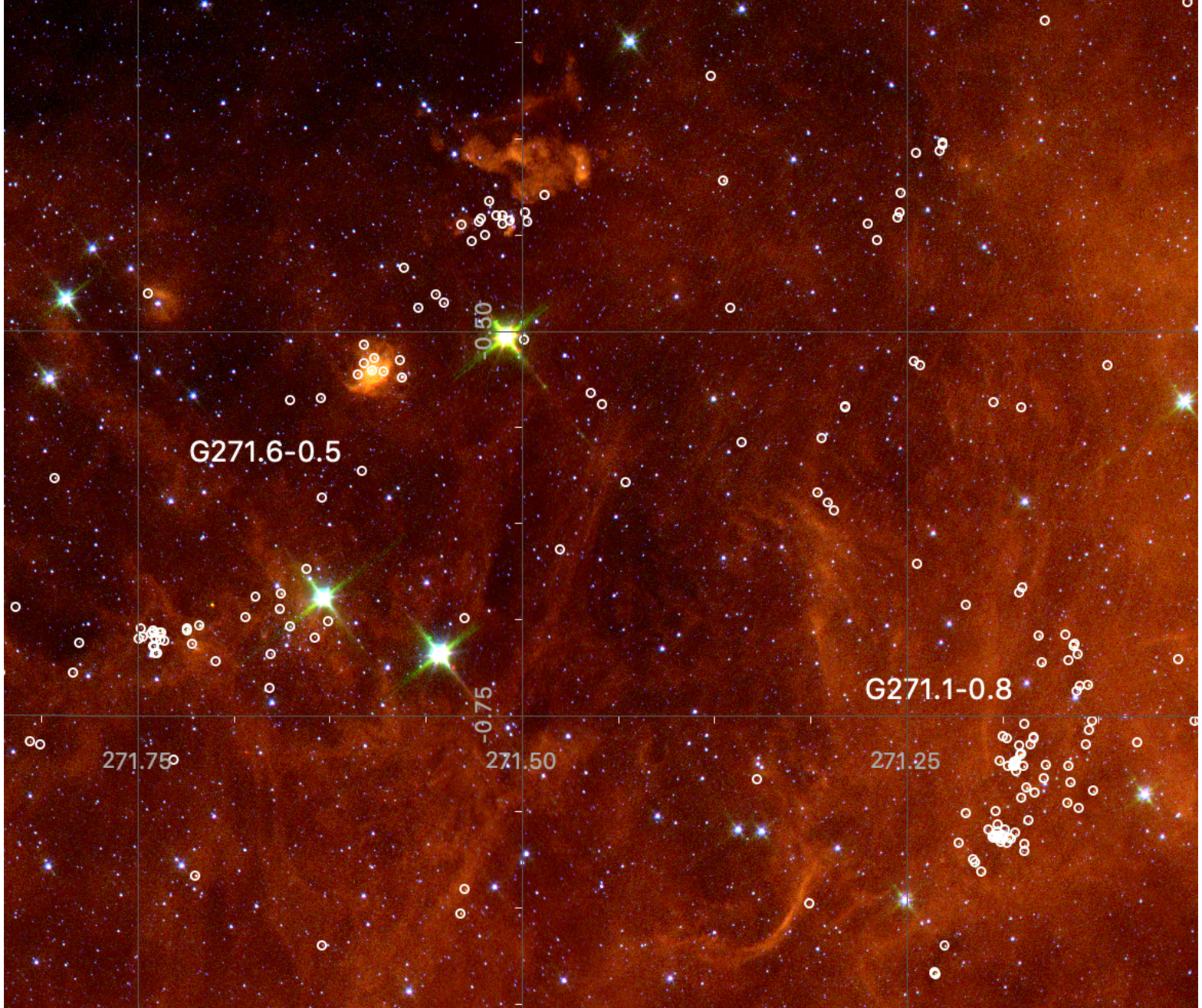}
    \caption{Spitzer/IRAC image  (Vela-Carina survey) with our YSO candidates marked by the white circles. The image is composed of the 3.6~$\mu$m (blue), 5.8~$\mu$m (green), and 8.0~$\mu$m (red) images. The image captures two groups of stars, the previously un-studied group G271.6-0.5 and a neighboring group G271.1-0.8.}
  \label{fig:COIN_cluster}
\end{figure*}

All objects with YSO scores $>$50\% from any of the three random forests are classified as candidate YSOs, while other sources are regarded as probable contaminants. Among the Spitzer sources with IR excess, there are 117,446 candidate YSOs and 180,997 probable contaminants. The candidates are listed in \autoref{tab:spicy} with the designation Spitzer/IRAC Candidate YSO (SPICY).

\autoref{fig:skymap} shows how the candidates are distributed within the footprints of the Spitzer surveys. Many of the candidate YSOs are concentrated toward the Galactic midplane while others form prominent clumps. 

More detail is visible in the zoomed-in maps from the atlas (\autoref{fig:atlas}), which consist of 19 panels ($\sim$6$^\circ\times$12$^\circ$) covering the entire survey area. In this atlas, we have also marked other features possibly connected to star formation, including H\,{\sc ii} bubbles from WISE \citep{Anderson2014} and massive YSOs from MSX \citep{Lumsden2013}. These maps show that the YSO candidate distribution can be resolved into stellar clusters and associations, along with a non-negligible number of widely distributed objects. It also appears that dense groups of YSOs tend to coincide with the locations of bubbles as seen in the plane of the sky.

With the new YSO candidates, some previously unrecognized stellar groups become apparent. In \autoref{fig:COIN_cluster}, we show an image containing one such group located in the Vela-Carina portion of the survey and designated G271.6-0.5 (left side of the image). To the south west of this group, a previously identified, but little studied, star-forming region, G271.1-0.8 can also be seen. 

The SPICY catalog is the largest homogeneous sample of YSO candidates available to date for the inner regions of the Milky Way. It seems unlikely that this mid-IR list of YSOs will be superseded in the near future given that no existing or planned mid-IR instrument exceeds Spitzer's spatial resolution in tandem with its wide-area mapping capabilities. The catalog is intended for both use in addressing questions about star formation on Galactic scales and assistance in searches for interesting individual YSOs. 
However, some contaminants inevitably remain, and formal assessment of contamination requires followup observations (e.g., spectrographic surveys). Nevertheless, the properties of these stars, including their colors, the environments in which they are found, their spatial and kinematic distributions, and their photometric variability (discussed in Sections~\ref{sec:color}--\ref{sec:variability}), are useful for corroborating the results of the random forest classifier and may give a qualitative sense of the level of remaining contamination. 

\section{Color and Magnitude Distributions}\label{sec:color}

\begin{figure}
	\centering
	\includegraphics[width=0.5\textwidth]{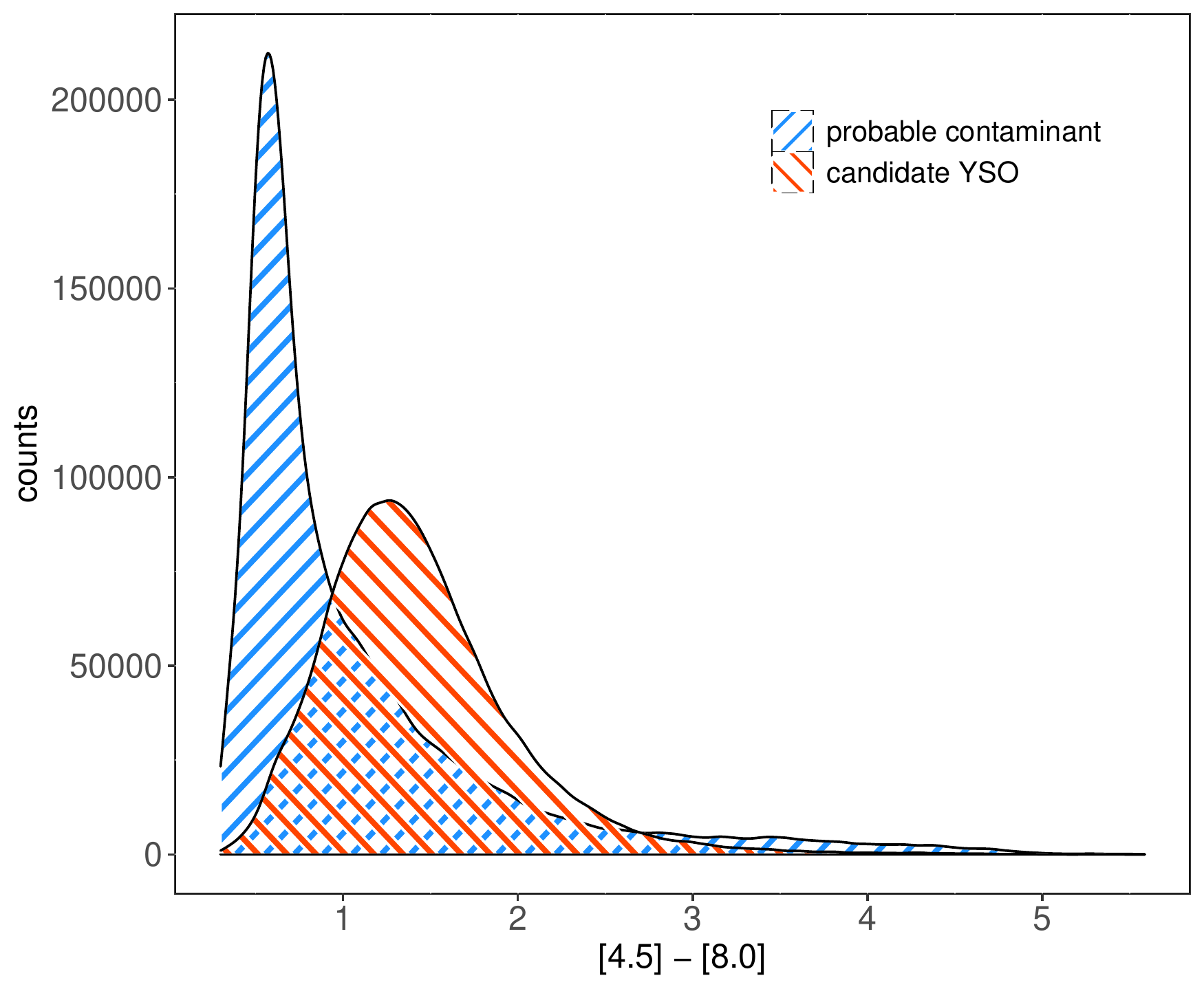}
    \caption{Distributions of $[4.5]-[8.0]$ color for the candidate YSOs (red stripes) and the probable contaminants (blue stripes). Overall, the candidate YSOs tend to be redder than the probable contaminants.
    Densities of sources in both samples are approximately equal at $[4.5] - [8.0] \approx 1$~mag, the limit imposed in the study by \citet{Robitaille08}, but in our sample 18\% of the YSO candidates are bluer than this limit and 25\% of the probable contaminants are redder. Probable contaminants also outnumber candidate YSOs at colors $[4.5] - [8.0] \gtrsim 3.5$~mag.}
  \label{fig:45_80_hist}
\end{figure}

\begin{figure*}
	\centering
	\includegraphics[width=1\textwidth]{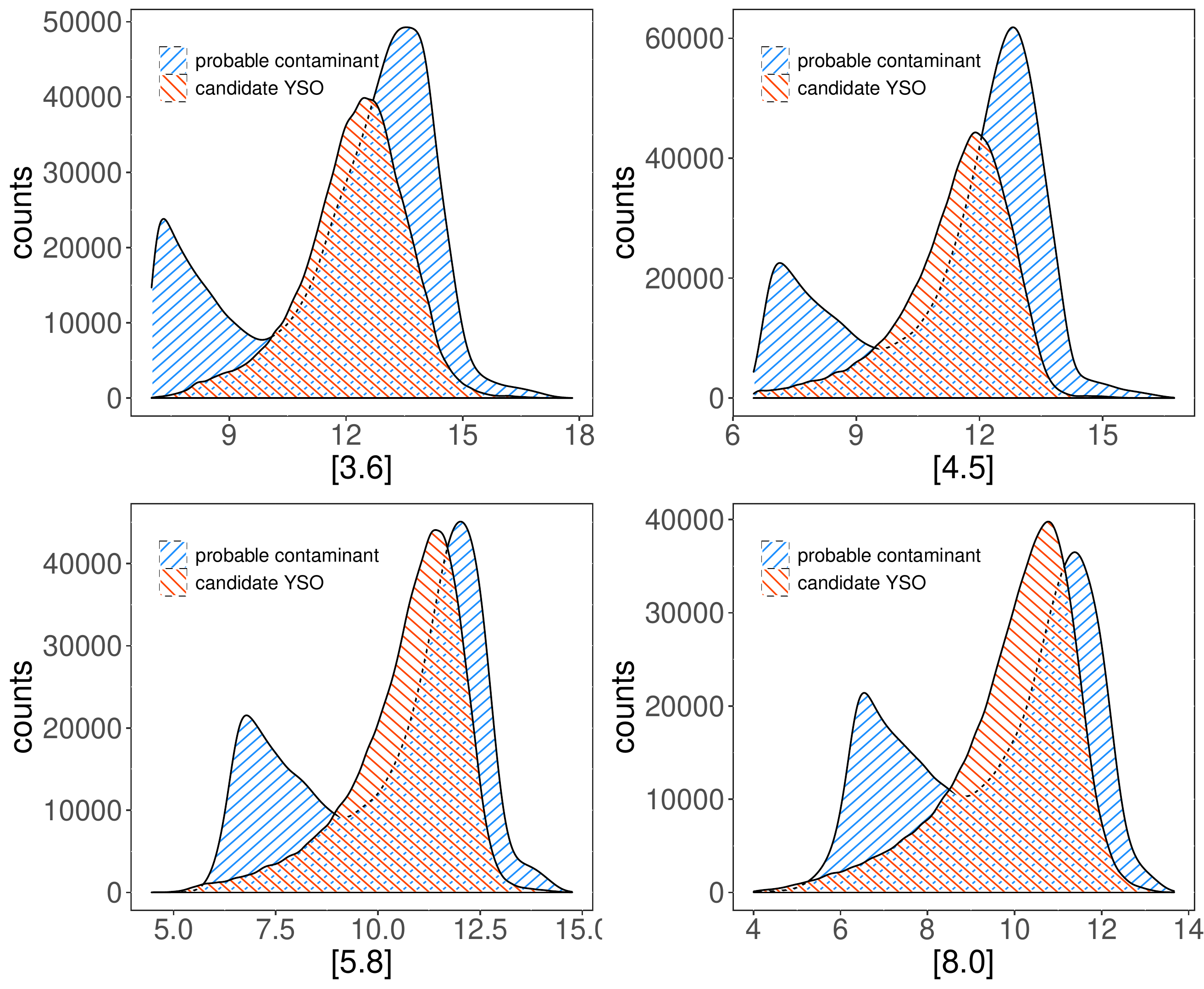}
    \caption{Distributions of IRAC magnitudes for the candidate YSOs (red stripes) and the probable contaminants (blue stripes). The distributions for probable contaminants are all multimodal, while the YSO candidates each have a single mode toward the fainter end of the distribution, and a heavy tail consisting of brighter sources.}
  \label{fig:hist_filters}
\end{figure*}

By examining the IR color and magnitude distributions for classified objects, we gain insight into how the classifier makes its decisions and how it compares to other selection criteria used in previous studies.

\autoref{fig:45_80_hist} shows the distribution of $[4.5]-[8.0]$, one of the main features used in the earlier study by \citet{Robitaille08}. The sources we input into the classifier have a bimodal distribution in this color, but each of the output classes has a unimodal distribution, with the probable contaminants making up the bluer peak and the YSO candidates making up the redder peak. The densities are approximately equal at $[4.5]-[8.0]\approx1$, the threshold used by \citet{Robitaille08}, but we also find a substantial number of objects of both classes (but particularly the contaminants) crossing the threshold.

The four panels of \autoref{fig:hist_filters} show the magnitude distributions of the classified sources in each IRAC band. The input distributions are bimodal, with lower peaks near the brightness limits and higher peaks at fainter magnitudes. The peaks at bright magnitudes can be attributed to an artifact of the $\chi^2$ fitting step because bright sources tend to have smaller magnitude uncertainties, and thus a smaller deviation is capable of leading to a formally ``bad fit.'' The classifier has identified the majority of sources associated with the bright peaks as probable contaminants. The distributions of the candidate YSOs are all unimodal, with peaks at fairly faint magnitudes, and heavy tails extending to brighter magnitudes. The contaminants also exhibit a second peak at magnitudes slightly fainter than the peak for YSOs. 

The YSO magnitude distributions appear reasonable, given that we would expect most of them to be low-to-intermediate mass objects at distances of one to several kpc, with a low number of brighter objects that could either be massive YSOs or nearby objects. The tendency to classify the faintest objects as probable contaminants may inherit a bias from the MYStIX training set which only includes YSOs out to $\sim$3~kpc. However, the differences in colors of the faintest objects (examined below) imply that they may be intrinsically different.  

Figures~\ref{fig:ir_cmds}--\ref{fig:nir_mir_ccds} show various $JHK$+IRAC color-magnitude and color-color diagrams. Candidate YSOs (red points) overlap probable contaminants (blue points) in each of these projections. Nevertheless, the locations in these diagrams with greatest source density are different for the two classes. We show reddening vectors indicating the effect of $A_K\approx1$~mag ($\sim$9~mag in the $V$ band) of extinction, adopting the reddening law from \citet{Rieke1985} for $JHK$ and \citet{Indebetouw2005} for the IRAC bands. We also plot curves for the near-IR stellar colors for stellar models without additional IR excess. For graphical display, we have merged the 2MASS, UKIDSS, and VVV photometry, converting UKIDSS and VVV to the 2MASS system using the first-order transformations from \citet{Hodgkin2009} and \citet{Soto2013}, and picking the most reliable photometry for each source.  

Some of these color spaces have been used in previous studies for selecting YSOs based on cuts on color. For example, the selection boundaries between YSOs and contaminants used by \citet[][]{Gutermuth2009} are depicted as gray lines in several of the diagrams, including $[4.5]$ vs.\ $[4.5]-[8.0]$ (\autoref{fig:ir_cmds}, left panel), $[3.6]-[4.5]$ vs.\ $[4.5]-[5.8]$, and $[4.5]-[5.8]$ vs.\ $[5.8]-[8.0]$ (\autoref{fig:mir_ccds}, upper panels). 

In the following subsections, we examine the IR criteria used for classification, evidence from Gaia that stars are pre--main-sequence, properties of the stars at 24~$\mu$m, YSO evolutionary classes, and the effects of various IR absorption and emission features. 

\subsection{Color-Magnitude Diagrams}

\begin{figure*}
	\centering
	\includegraphics[width=0.45\textwidth]{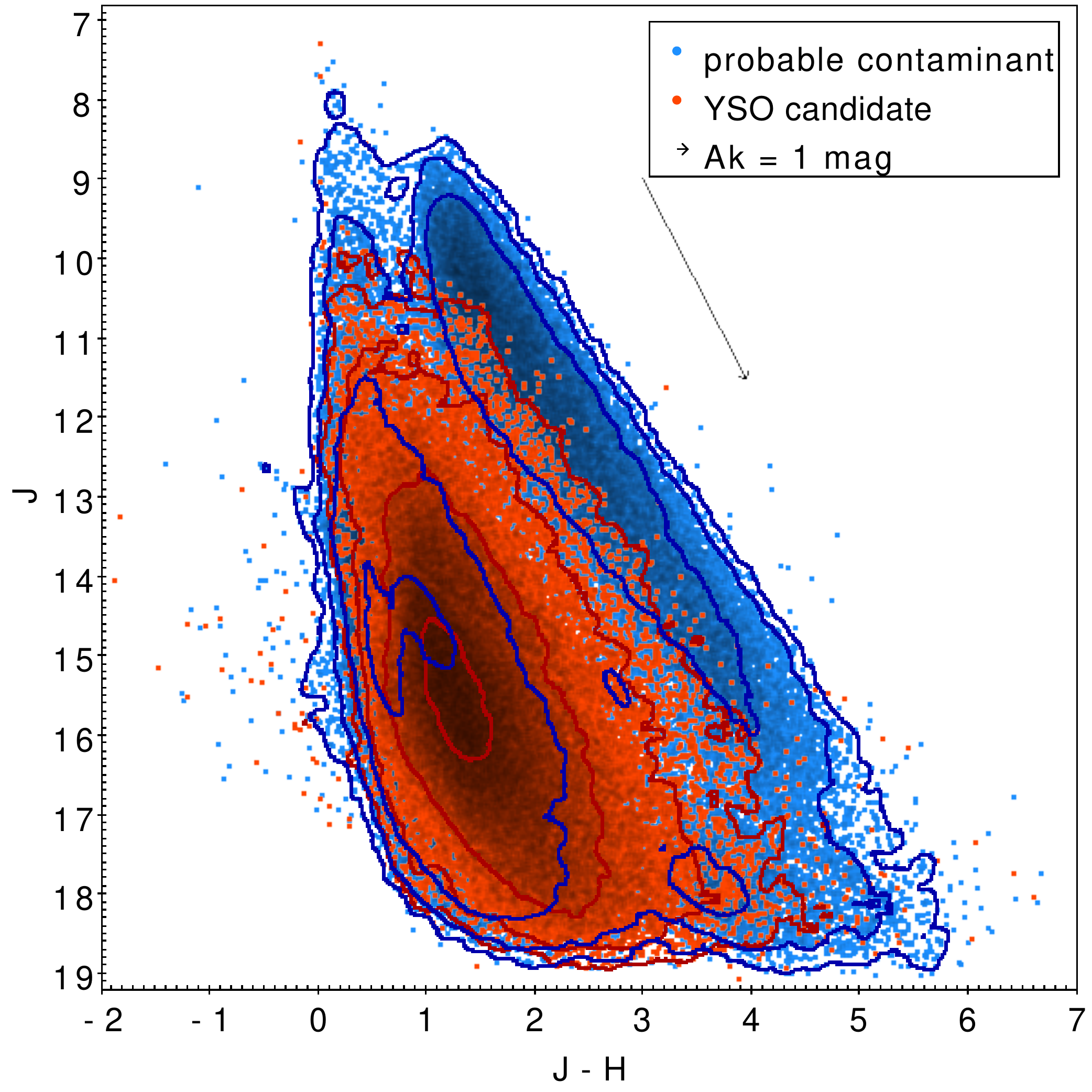}
	\includegraphics[width=0.45\textwidth]{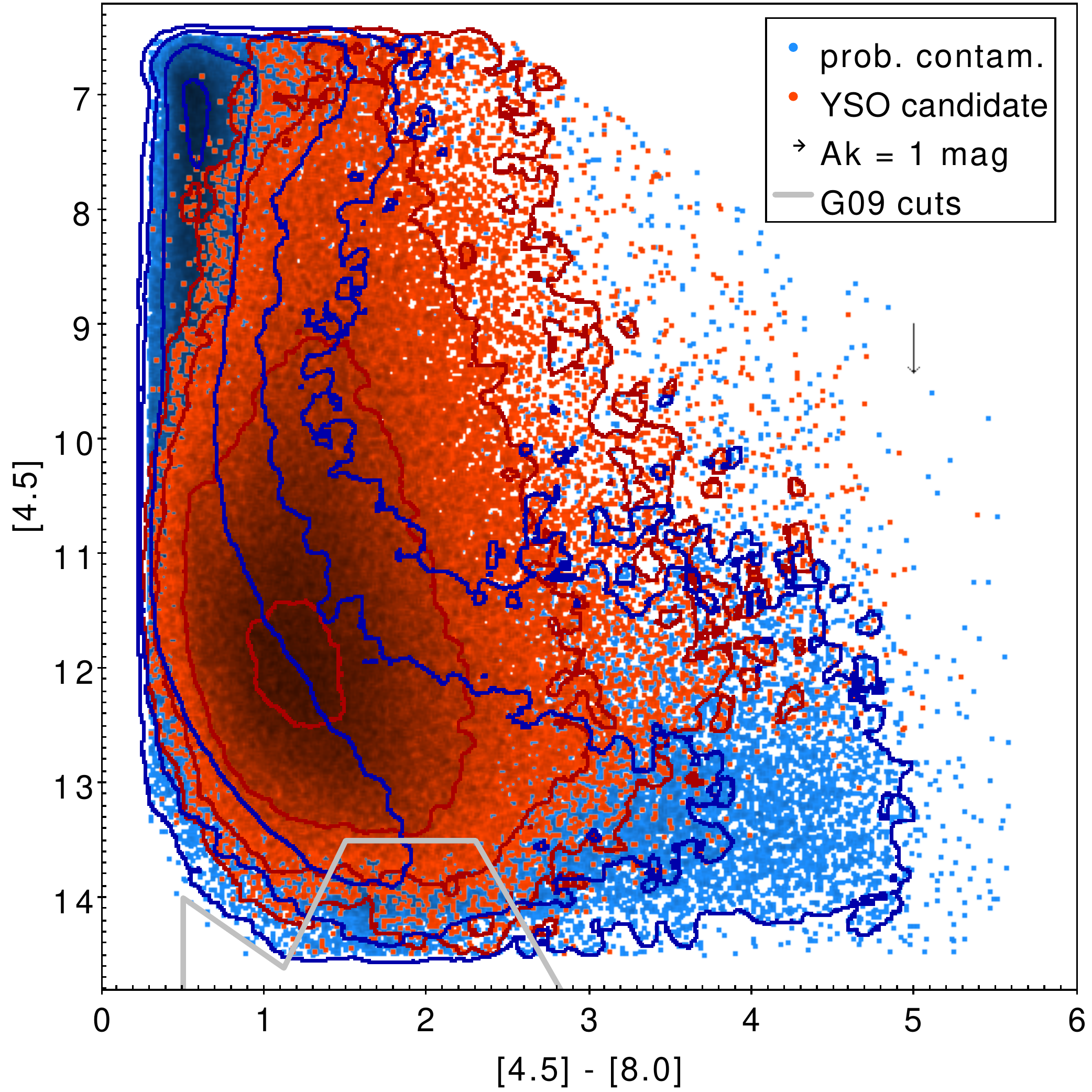}
    \caption{Infrared color-magnitude diagrams, $J$ vs.\ $J-H$ (left) and $[4.5]$ vs.\ $[4.5]-[8.0]$ (right), with candidate YSOs (red) and probable contaminants (blue). In low-density parts of the scatter plot, individual points are drawn, but in areas with overlapping points, darker colors indicate higher density. We also include contours at evenly spaced logarithmic increases in density. The arrow indicates the approximate shift produced by extinction of $A_K=1$~mag assuming the \citet{Indebetouw2005} reddening law. The gray polygon demarcates the region used by \citet{Gutermuth2009} to select contaminants.}
  \label{fig:ir_cmds}
\end{figure*}

On the $J$ vs.\ $J-H$ diagram (\autoref{fig:ir_cmds}, left), both candidate YSOs and probable contaminants occupy a triangular region of color--magnitude space, where the upper edge of the triangle is approximately parallel to the reddening vector. The YSO candidates are densest around $J\sim15.5$~mag and $J-H\sim1.3$~mag, whereas the probable contaminant distribution is multi-modal, with one peak just blueward of the peak of the YSO candidates, and another strip of stars along the upper right edge of the triangle. The stars in this strip, which are more luminous than the typical YSO candidate with the same $J-H$ color, lie in the region of the diagram that would be occupied by reddened post--main-sequence stars.

On the $[4.5]$ vs.\ $[4.5]-[8.0]$ diagram (\autoref{fig:ir_cmds}, right), the YSO candidates form a smooth distribution ranging from the bright limit at $[4.5]=6.5$~mag to $\sim$14~mag, where sensitivity declines, with the peak of the distribution at $[4.5]\sim12.2$~mag and $[4.5]-[8.0]\sim1.2$. A $\sim$1~Myr old (pre-)main-sequence star with a mass in the range 0.4--10~$M_\odot$ at a distance of $\sim$1-2~kpc would have an unreddened photospheric magnitude  $9\lesssim[4.5]\lesssim 14$~mag \citep{Bressan2012} -- approximately where were find the bulk of the YSO candidates. The probable contaminant distribution peaks at both bright and faint magnitudes. The bright contaminants form a band that tends to be bluer than the YSOs in $[4.5]-[8.0]$, while the faint contaminants tend to have redder $[4.5]-[8.0]$ colors. 

The gray lines on the $[4.5]$ vs.\ $[4.5]-[8.0]$ diagram were defined by \citet{Gutermuth2009} to separate dusty AGNs from YSOs in their studies of nearby star-forming regions. Although many of the faintest 4.5~$\mu$m sources in our sample have been classified as probable contaminants, the region defined by \citet{Gutermuth2009} for selecting AGNs  does not appear to separate our classes well. This apparent discrepancy may arise because \citet{Gutermuth2009} examined deeper Spitzer surveys of relatively nearby star-forming clouds at higher Galactic latitudes, where more AGN are expected to be detected, whereas GLIMPSE is less sensitive to this type of contaminant. Furthermore, GLIMPSE includes more distant star-forming regions in which legitimate YSOs will present fainter observed [4.5] magnitude distributions.

\subsection{Color-Color Diagrams}

\begin{figure*}
	\centering
	\includegraphics[width=0.45\textwidth]{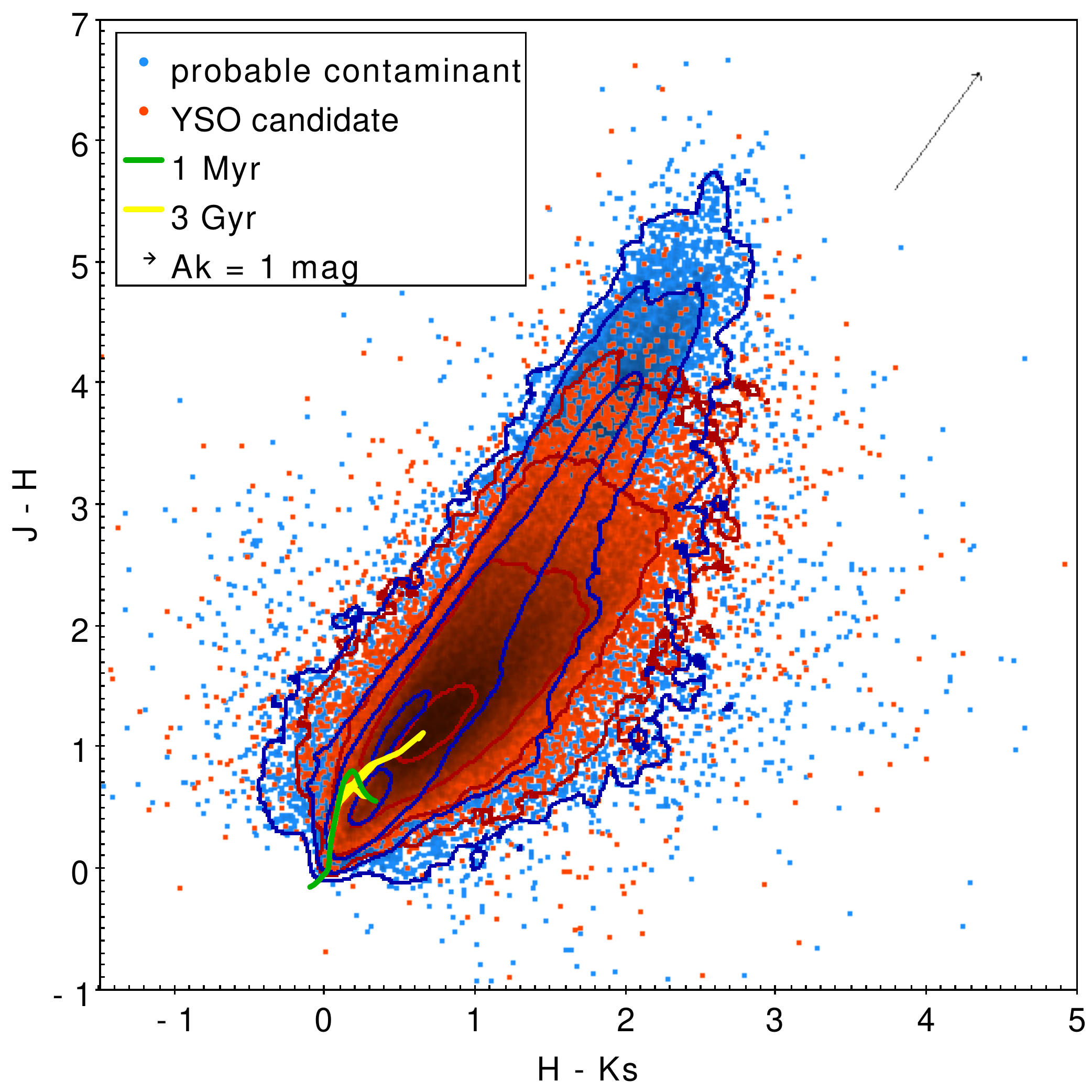}
	\includegraphics[width=0.45\textwidth]{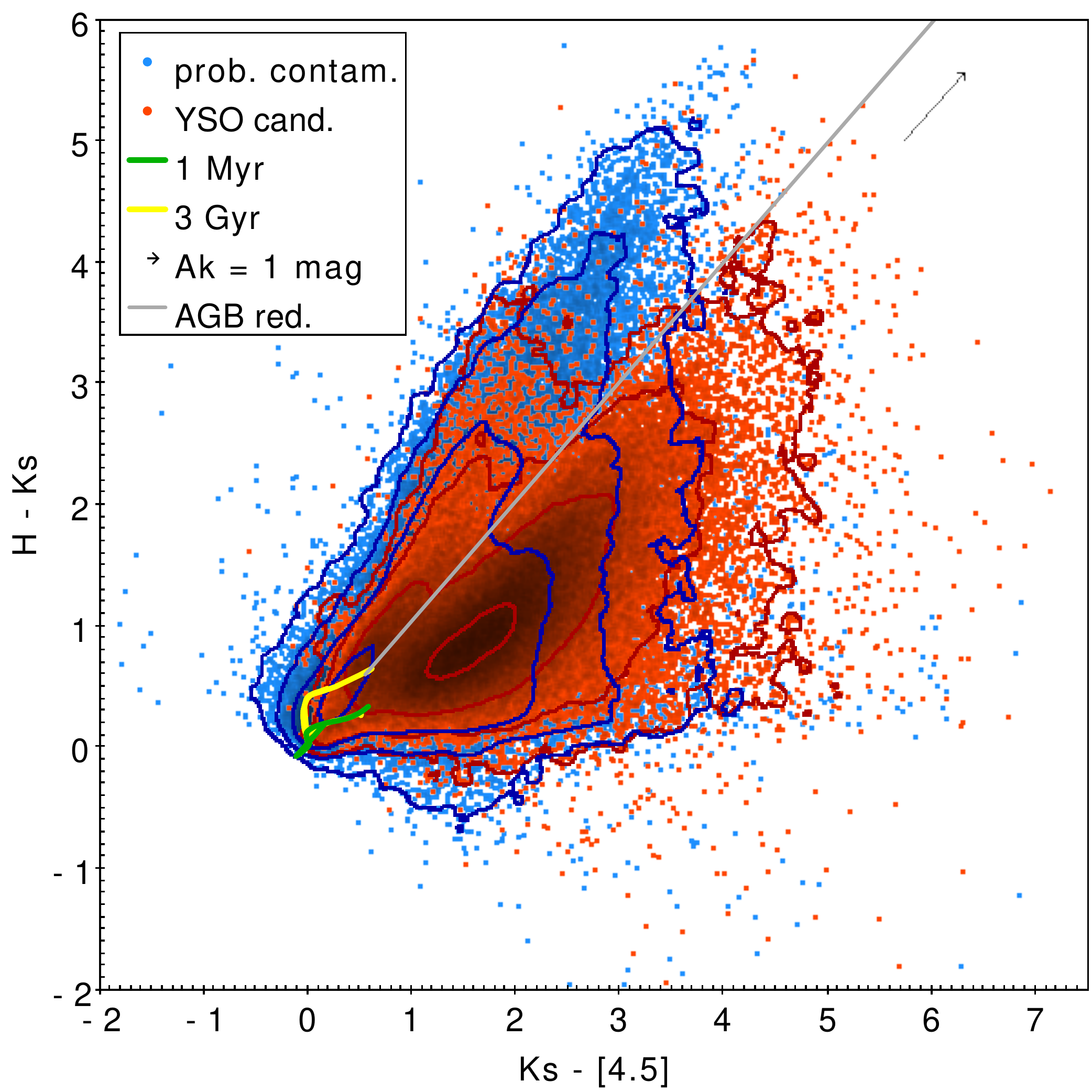}
    \caption{Color-color diagrams for $J-H$ vs.\ $H-K_s$ (left) and $H-K_s$ vs.\ $K_s - [4.5]$ (right) with candidate YSOs (red) and probable contaminants (blue). Curves indicate 1~Myr (green) and 3~Gyr (yellow) isochrones \citep{Bressan2012}, with models for the AGB phase \citep{Marigo2013} included in the 3~Gyr isochrone. In the right panel, a gray line parallel to the \citet{Indebetouw2005} reddening vector extends from the tip of the AGB. In our sample, objects classified as contaminants predominate above this line while objects classified as candidate YSOs are more abundant below.} 
  \label{fig:nir_ccds}
\end{figure*}

\begin{figure*}
	\centering
	\includegraphics[width=0.45\textwidth]{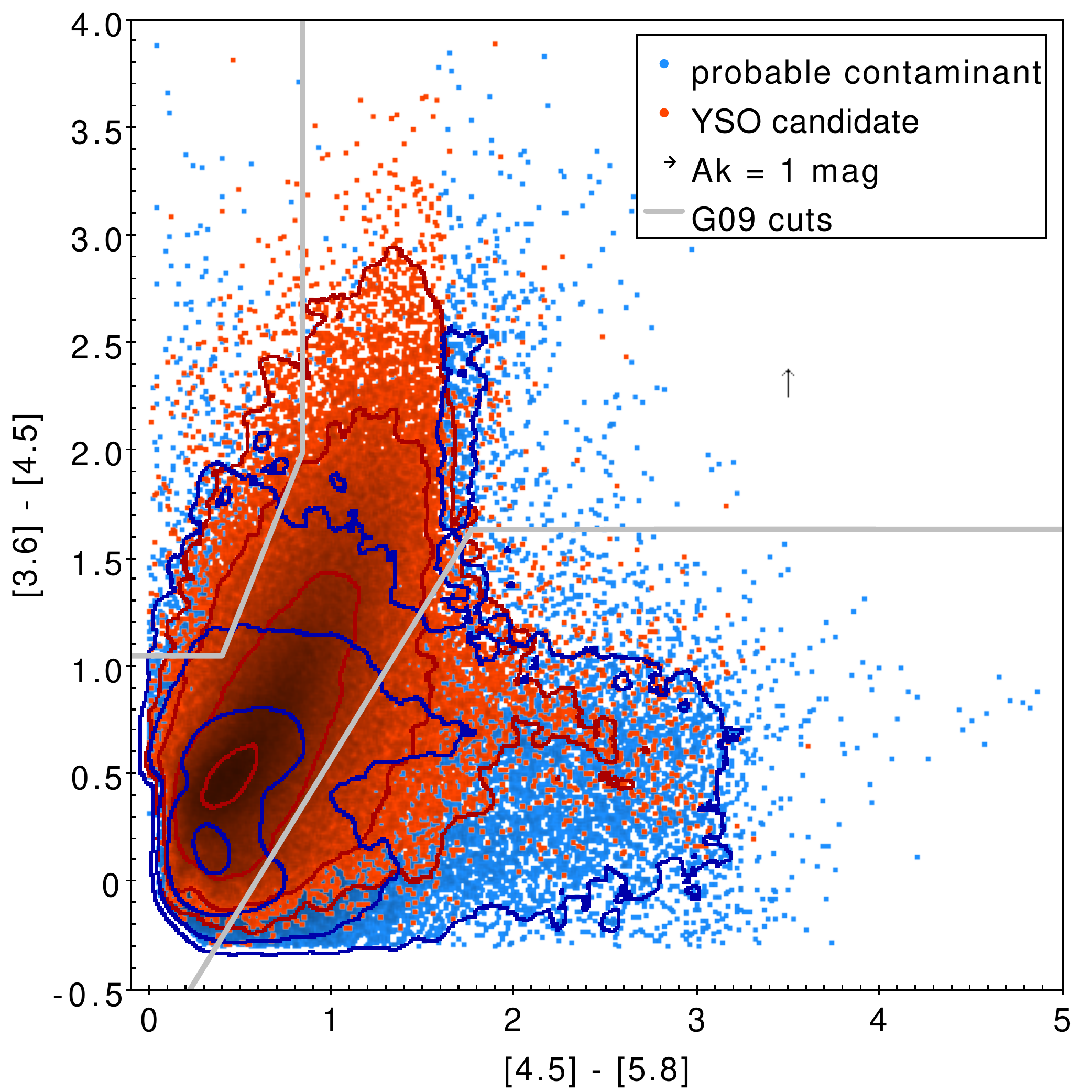}
	\includegraphics[width=0.45\textwidth]{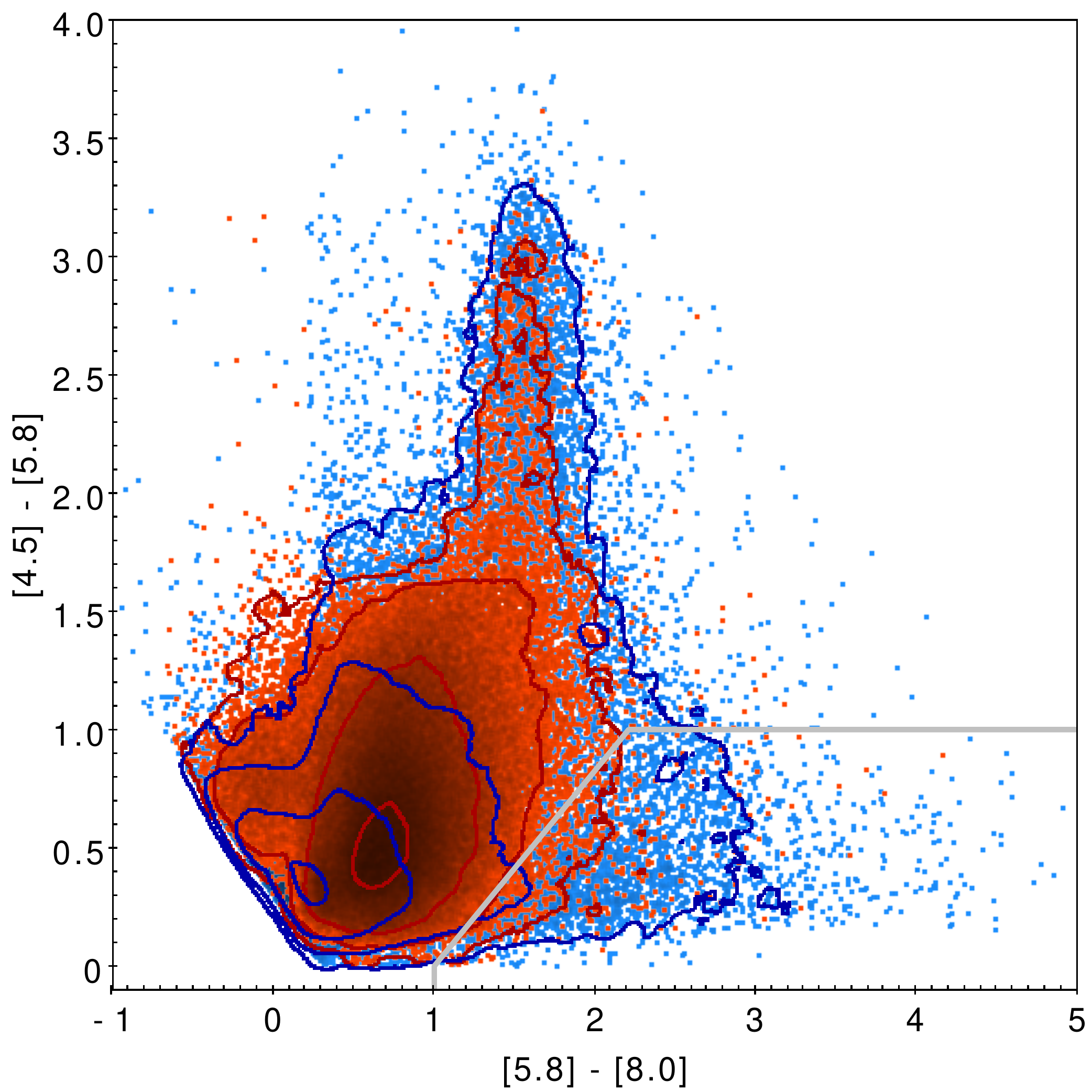}
	\includegraphics[width=0.45\textwidth]{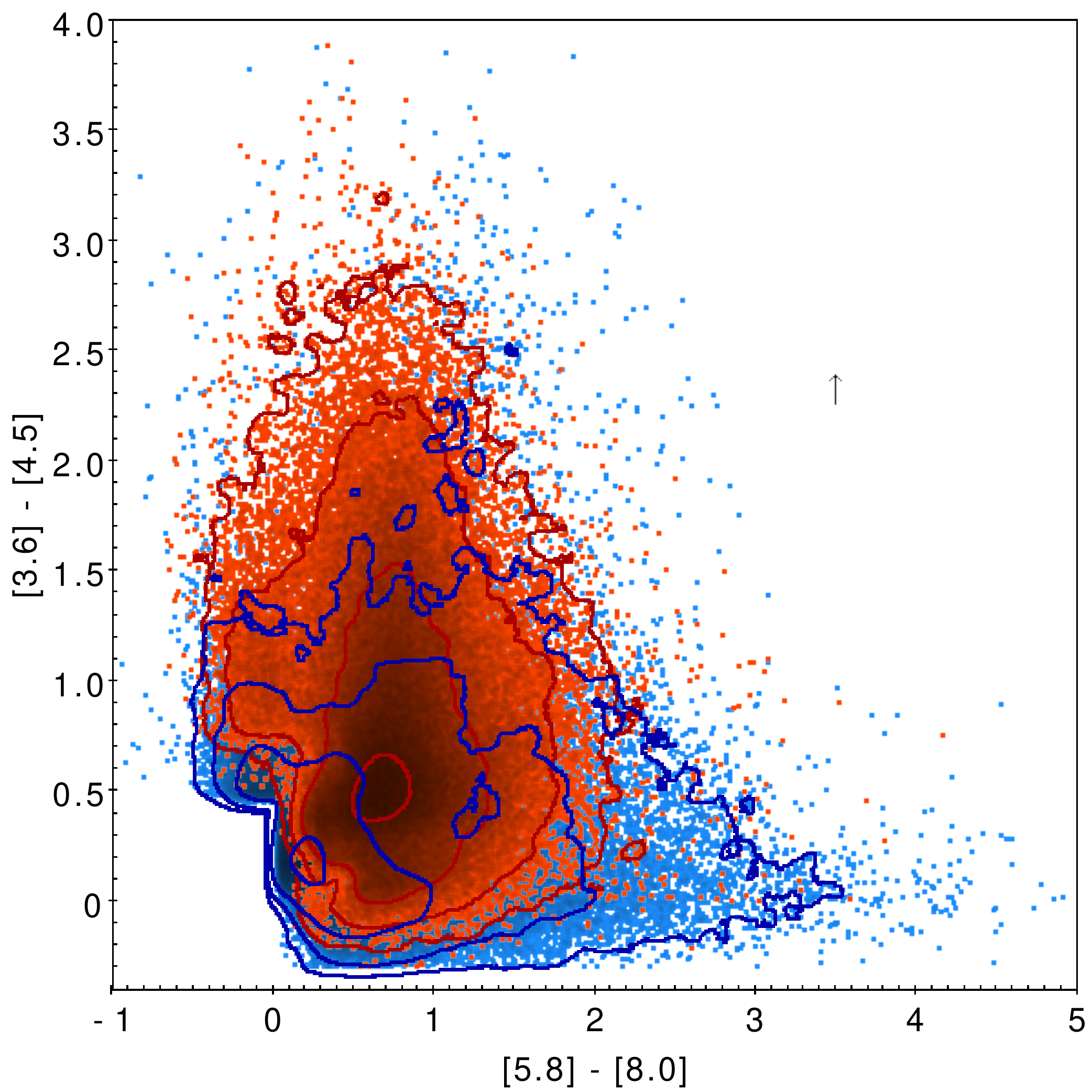}
	\includegraphics[width=0.45\textwidth]{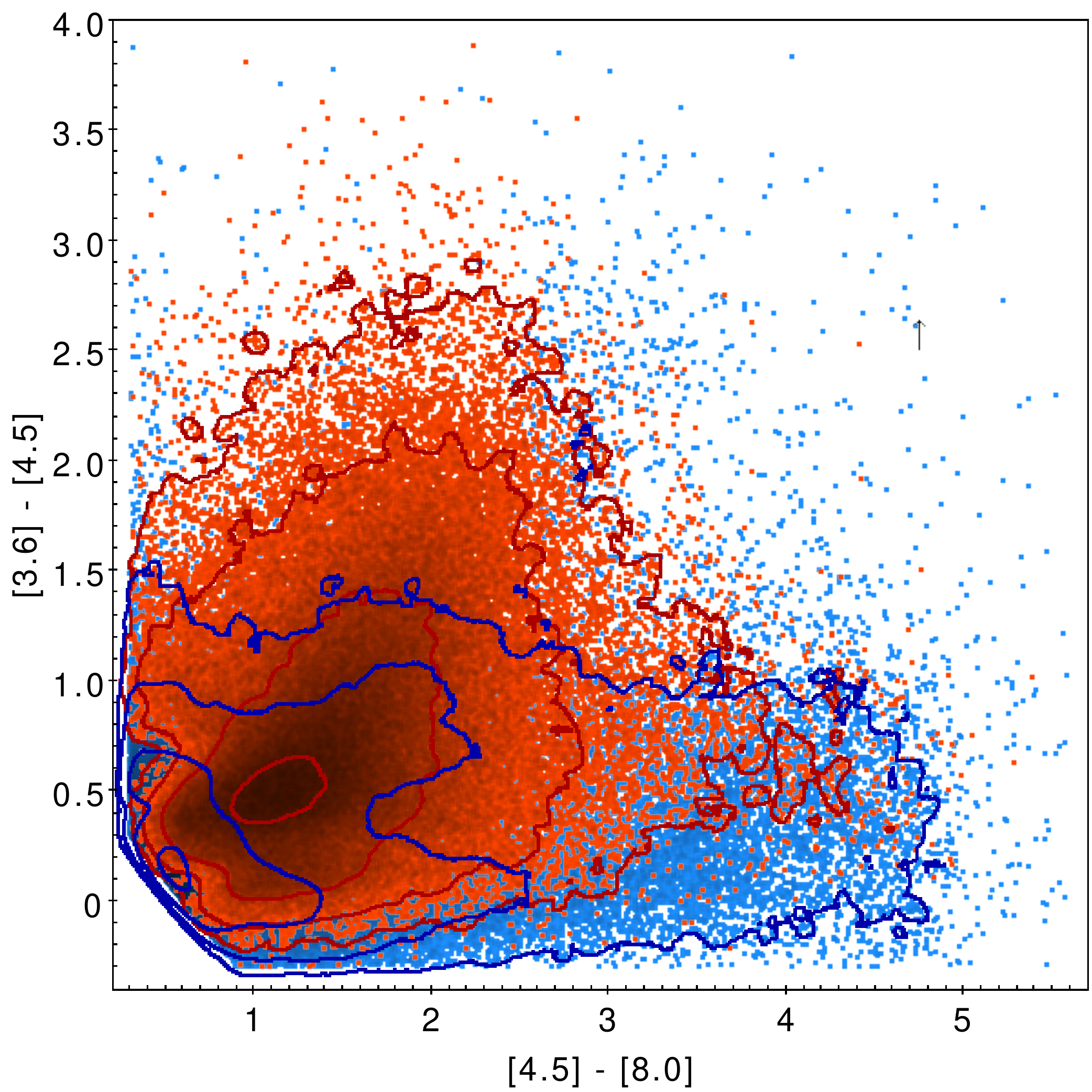}
    \caption{Color-color diagrams in the IRAC bands. The YSO candidates (red points) and probable contaminants (blue points) partially overlap in each of these projections, but differences are visible in their distributions. The short length (or absence) of the $A_K=1$~mag reddening vectors (black arrows) implies that extinction would need to be extreme to significantly change these distributions. The PAH feature is distinctly visible on the $[4.5]-[5.8]$ vs.\ $[5.8]-[8.0]$ diagram. The \citet{Gutermuth2009} criteria are indicated by gray lines for comparison. }
  \label{fig:mir_ccds}
\end{figure*}

\begin{figure*}
	\centering
	\includegraphics[width=0.45\textwidth]{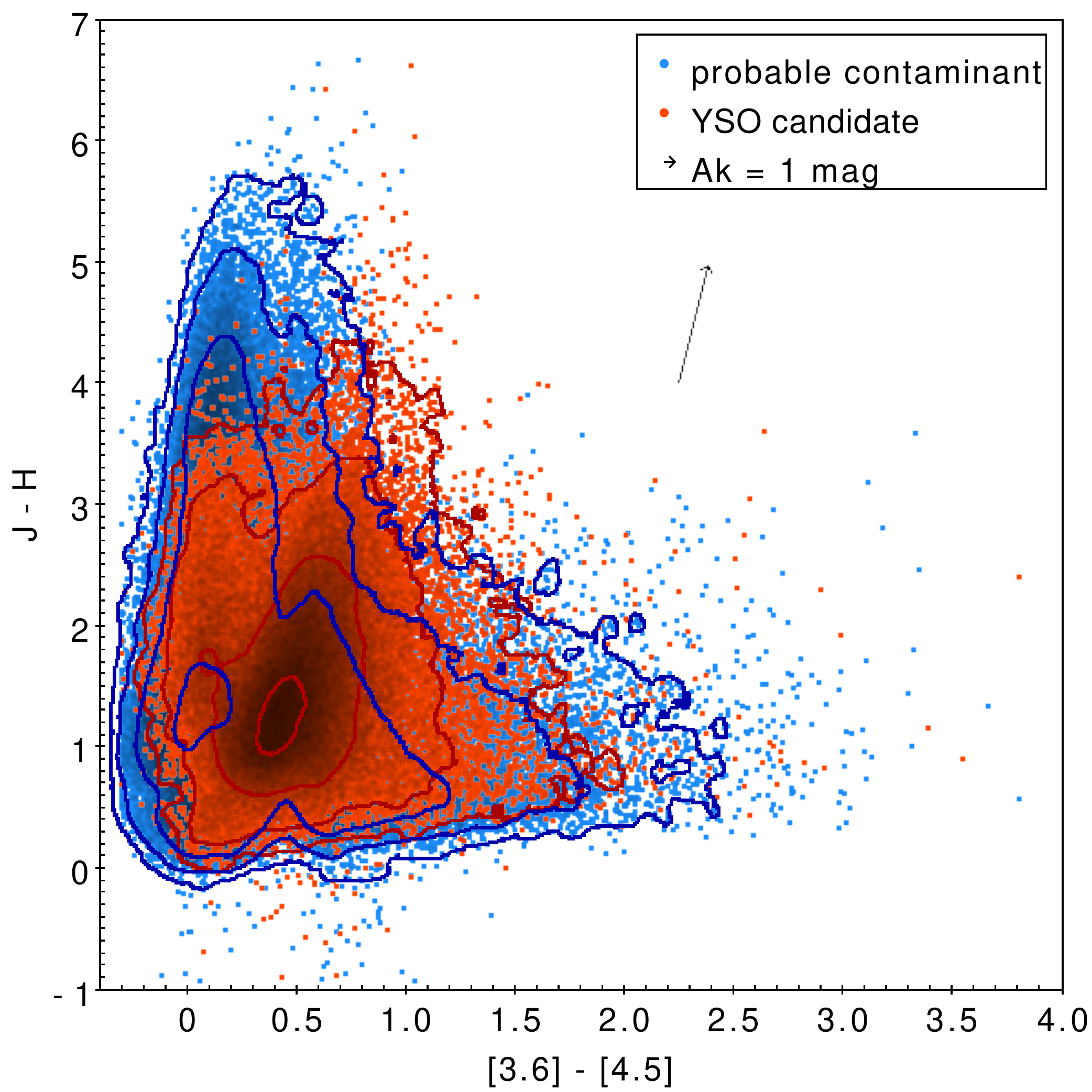}
	\includegraphics[width=0.45\textwidth]{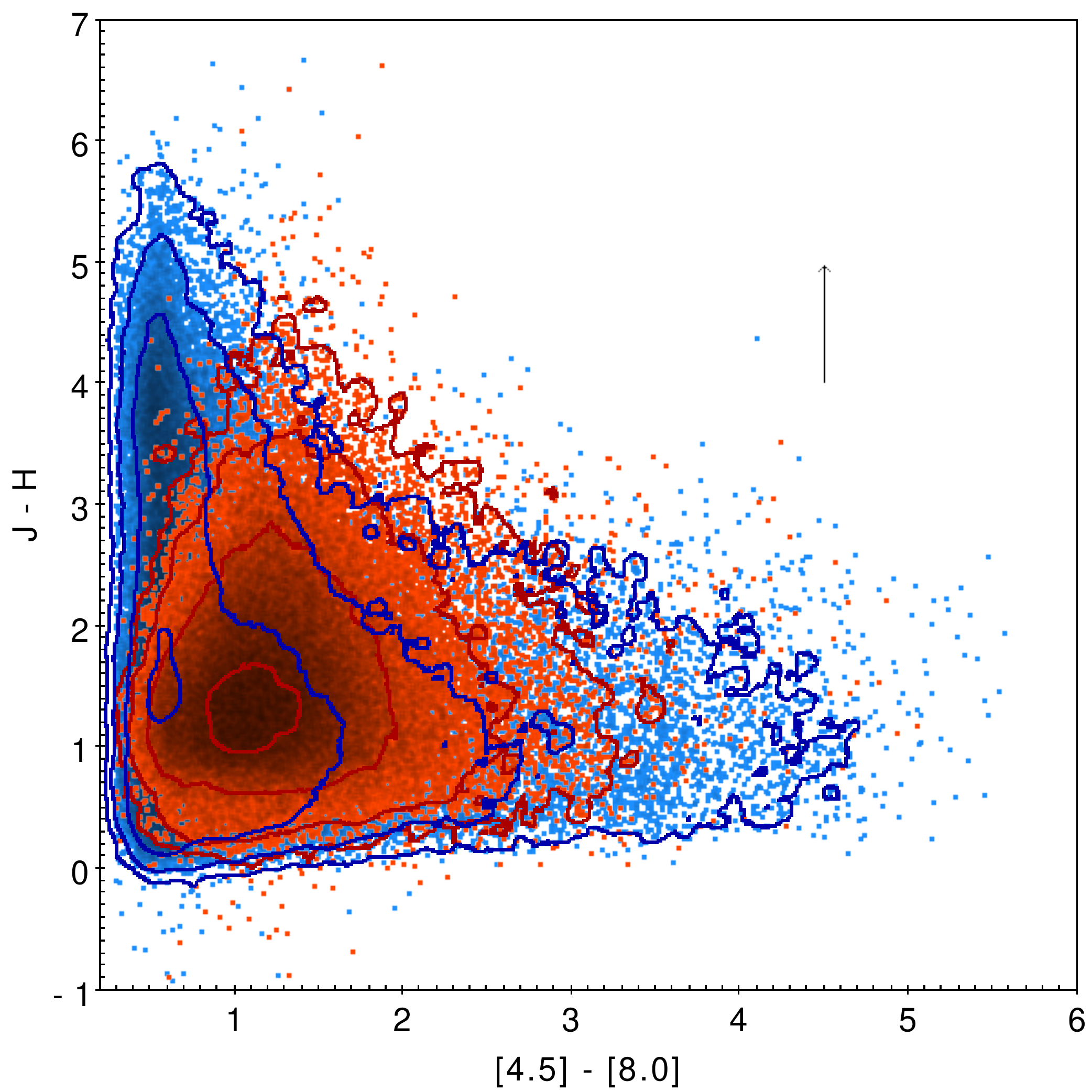}
    \caption{Color magnitude diagrams showing $J-H$ (the color most sensitive to reddening) vs.\ $[3.6]-[4.5]$ (left) and $[4.5]-[8.0]$ (right), which are both useful for selecting YSOs. Symbols and lines are the same as in \autoref{fig:ir_cmds}.}
  \label{fig:nir_mir_ccds}
\end{figure*}

\begin{figure}
	\centering
	\includegraphics[width=0.5\textwidth]{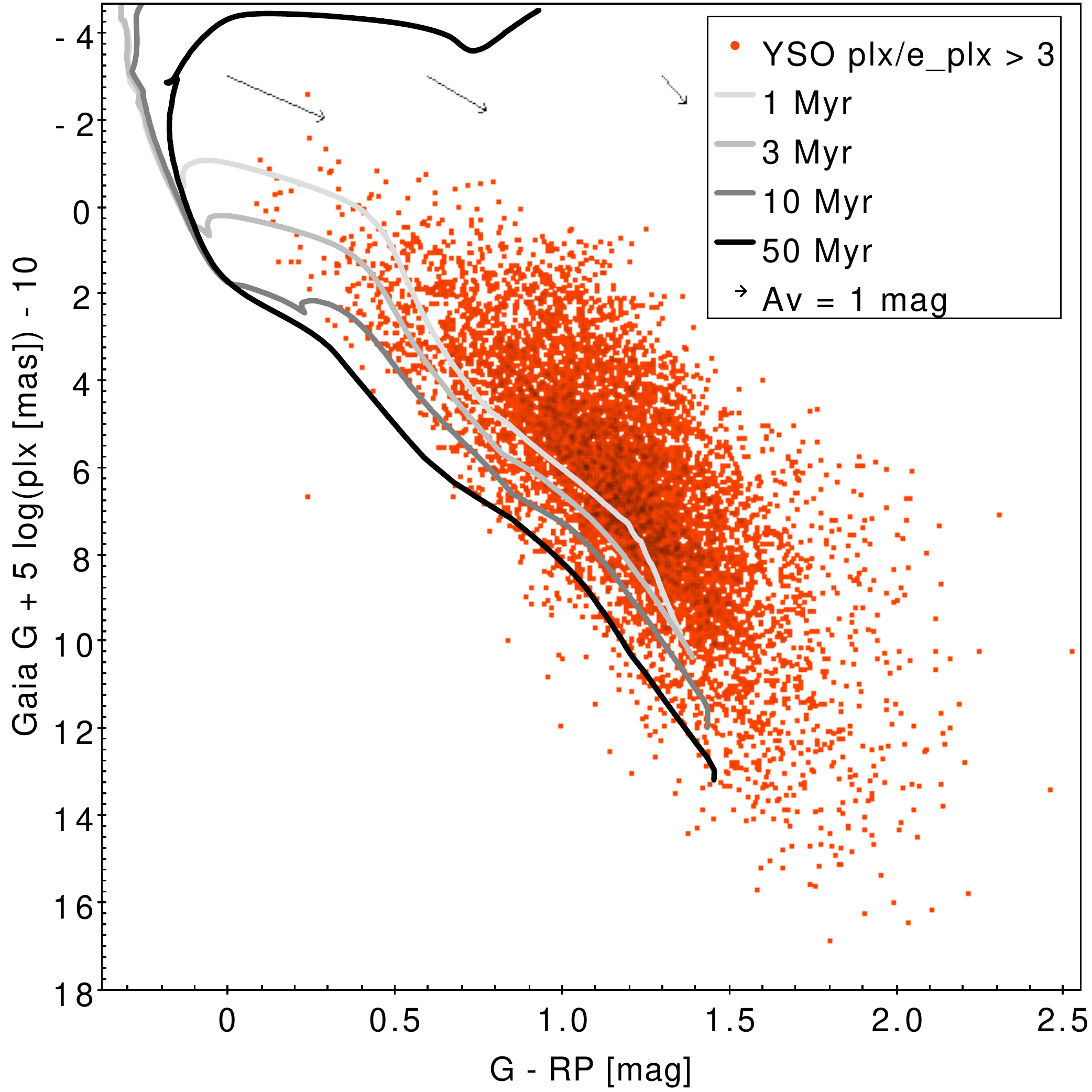}
    \caption{Absolute Gaia $G$-band magnitude vs.\ $G-RP$ color for candidate YSOs with $\varpi/\sigma_\varpi > 3$. The curves are unreddened isochrones with ages of 1, 3, 10, and 50~Myr from \citet{Bressan2012}. The arrows indicate approximate Gaia reddening vectors using the \citet{Cardelli1989} and \citet{ODonnell1994} extinction curves with $R_V=3.1$. The broad Gaia bands mean that these vectors vary with color, so we show three vectors estimated using stellar spectra with intrinsic colors of $G-RP=0$, 0.6, and 1.3. Nearly all of these candidate YSOs are in the region of this color--magnitude diagram consistent with the pre--main sequence.} 
  \label{fig:fig_Gaia_CMD}
\end{figure}

\begin{figure}
	\centering
	\includegraphics[width=0.45\textwidth]{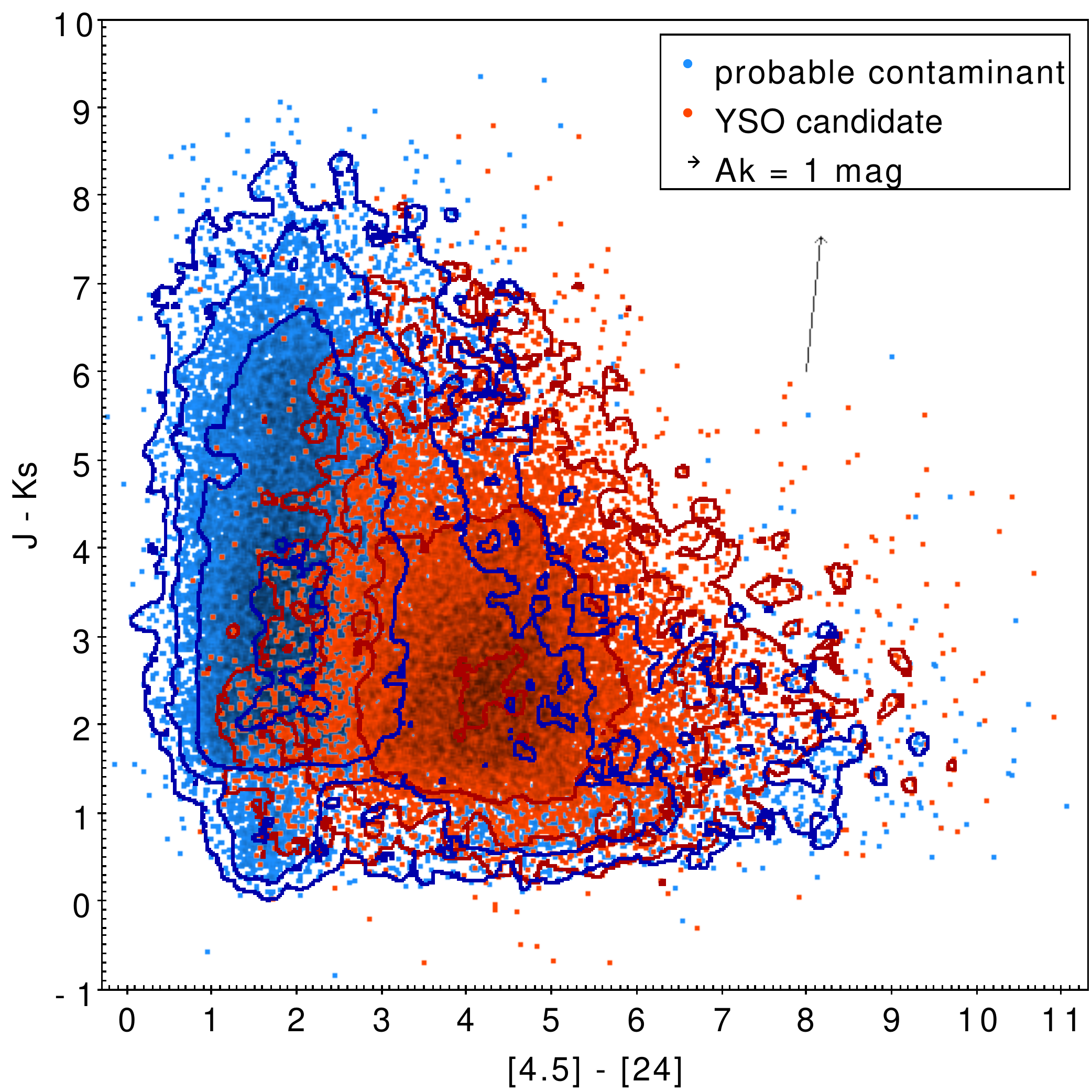}
    \caption{The $J-K_s$ vs.\ $[4.5]-[24]$ color-color diagram for candidate YSOs and probable contaminants. This diagram may be useful for verifying separation between AGB stars and YSOs. AGB stars typically have steep red SED shapes in the near-IR, but turn over to a Rayleigh-Jeans tail around 24~$\mu$m. We find the sources with reddest $J-H$ colors, but not as red $[4.5]-[24]$ colors are mostly classified as probable contaminants, consistent with being AGB stars.}
  \label{fig:explanatory24micron}
\end{figure}

\begin{figure}
	\centering
	\includegraphics[width=0.45\textwidth]{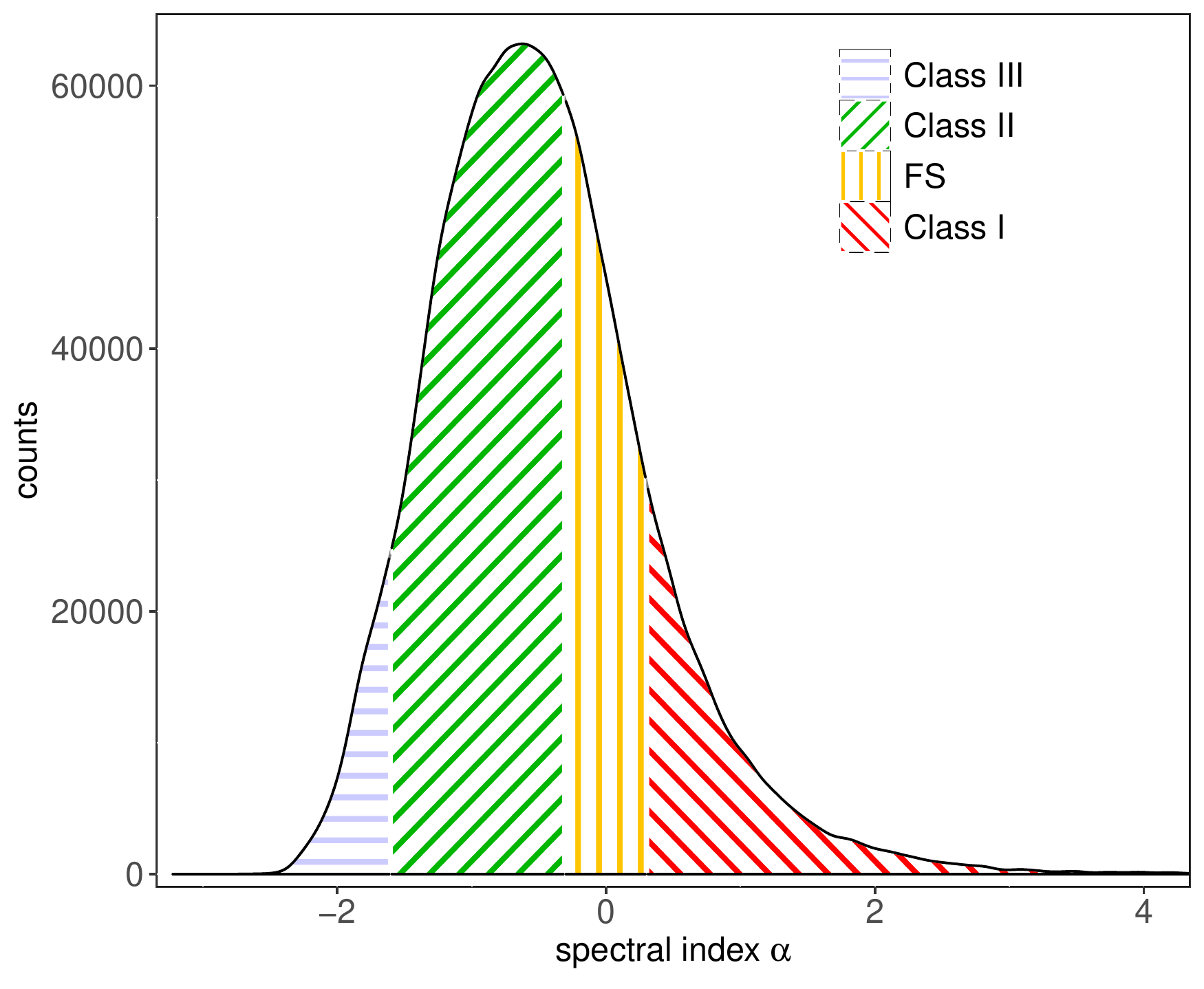}
    \caption{Distribution of spectral index $\alpha$ for YSO candidates, subdivided into YSO class using the customary demarcations at $\alpha=-1.6$, $-0.3$, and $0.3$. The shape of the distribution will be the product of the prevalence of the YSO classes, with Class~II/III YSOs being more common than Class~I/flat SED YSOs due to the longer lifetimes of the later evolutionary stages \citep[e.g.,][]{Evans2009}, and our sensitivity to each class, which may be lower for YSOs with smaller IR excesses (e.g., Class~III) and for deeply embedded YSOs (e.g., Class~I).}
  \label{fig:alpha}
\end{figure}

\begin{figure*}
	\centering
	\includegraphics[width=0.45\textwidth]{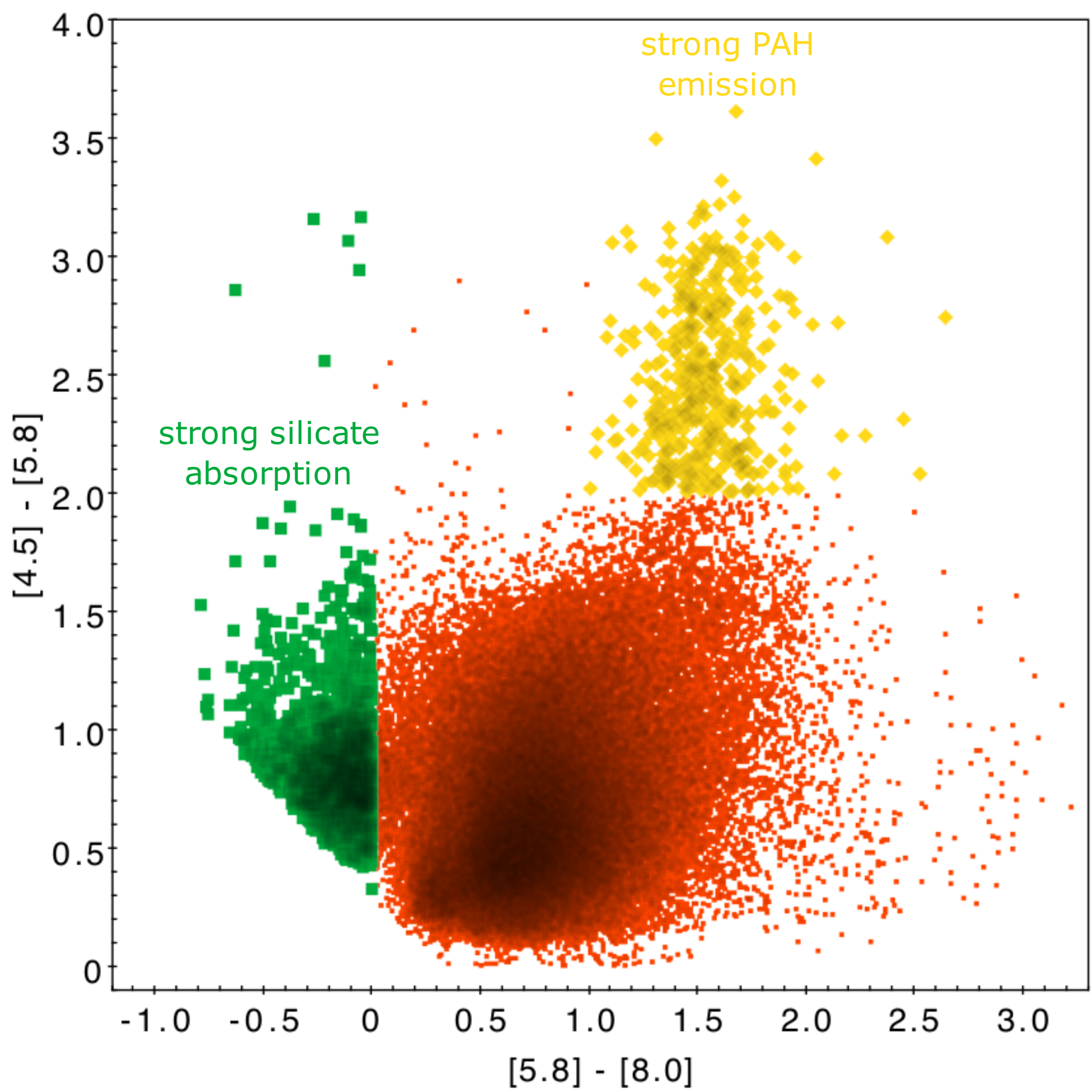}
	\includegraphics[width=0.45\textwidth]{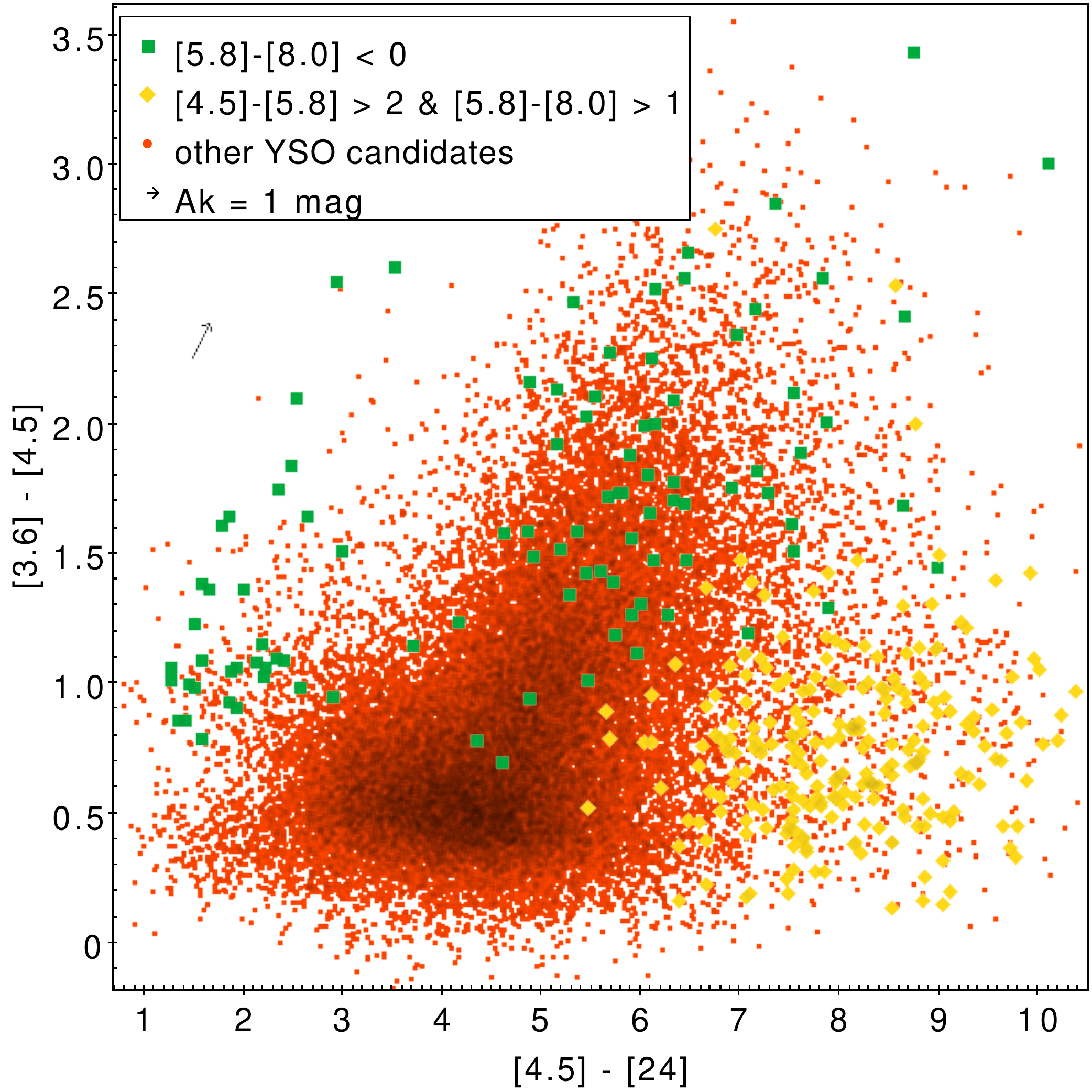}
    \caption{Left: IRAC color-color diagram with YSO candidates with possible strong silicate absorption and PAH emission labeled.
    Right: YSO candidates on the $[3.6]-[4.5]$ vs.\ $[4.5]-[24]$ diagram, with several subcategories selected. YSO candidates with low values of $[5.8]-[8.0]$ colors are shown as green squares and YSO candidates with suspected PAH emission are yellow-orange diamonds. Both groups have redder than average $[4.5]-[24]$ colors, which is consistent with these classes of sources being deeply embedded.  }
  \label{fig:explanatory}
\end{figure*}

\begin{figure*}
	\centering
	\includegraphics[width=0.95\textwidth]{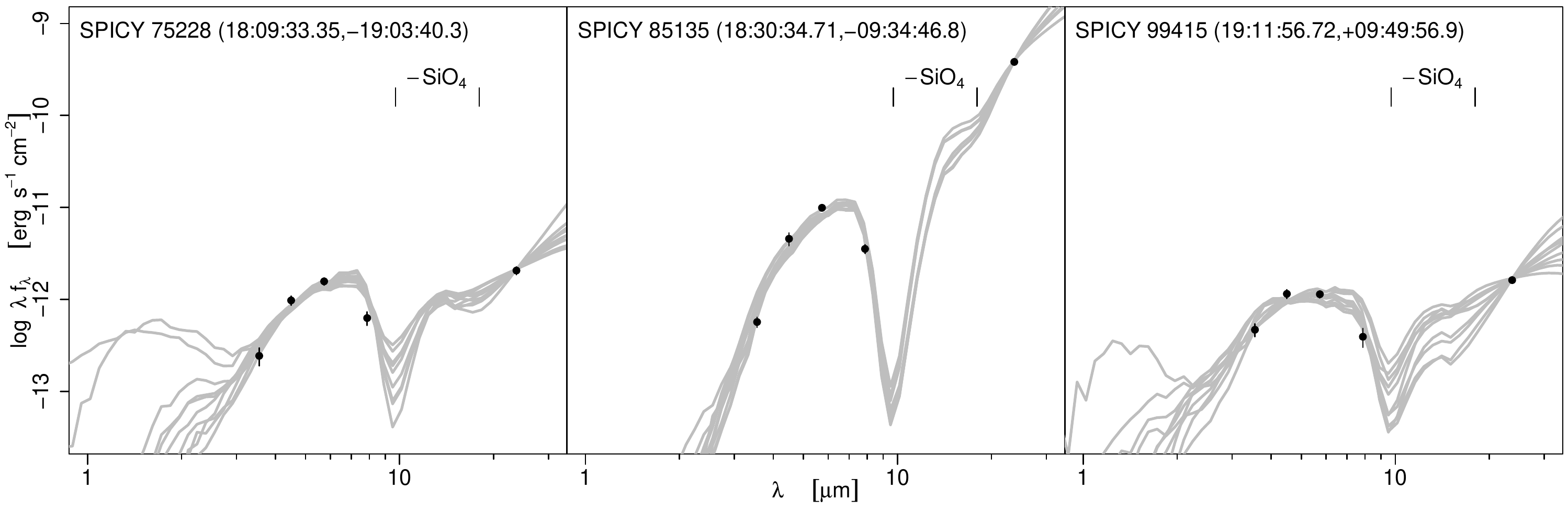}
    \caption{SEDs for three YSO candidates with $[5.8]-[8.0]< 0$. The black points represent the photometry in the IRAC bands, with 1$\sigma$ error bars. The gray lines are the 10 best-fitting convolved YSO models from \citet{Robitaille17}. Although not all of these models provide formally good fits, they represent the SED shapes necessary to produce the observed IRAC colors. These models demonstrate that strong silicate absorption features (centers indicated by tic marks) are necessary to reproduce observed photometry.}
  \label{fig:SED}
\end{figure*}

\autoref{fig:nir_ccds} shows the distributions of sources in $J-H$, $H-K_s$, and $H-[4.5]$. On the $JHK_s$ diagram, we include a representative isochrone for $\sim$1~Myr unreddened stellar models. Most of the objects are shifted to the upper right from this curve, in the approximate direction of the reddening vector. However, the distribution of the YSO candidates spreads to redder $H-K_s$ colors, which would be expected for stars with $K_s$-band excess. Objects with very red $J-H>5$ colors are largely classified as contaminants. 

On the $H-K_s$ vs.\ $K_s-[4.5]$ diagram, we show both a 1~Myr isochrone for (pre--)main-sequence stars and a 1~Gyr isochrone that also includes post--main-sequence stars. The red-giant branch extends upward to stars with redder $H-K_s$ colors than the (pre--)main-sequence, allowing these groups to be better separated. On this plot, the reddening vector points to the upper right. If we consider a line parallel to the reddening vector, starting from the tip of the asymptotic giant branch (as shown by the gray line in the figure), we would expect that many of the stars lying above this line could be evolved stellar contaminants. This is consistent with what the classifier finds; most objects above this line are classified as probable contaminants, while the candidate YSOs are more abundant below this line. The slope of the \citet{Indebetouw2005} reddening vector is not precisely parallel to the upper edge of the source distribution; this may arise due to systematic uncertainties in the reddening law or could be a property of IR colors of highly obscured evolved stars. 

\autoref{fig:mir_ccds} shows four projections of sources in IRAC color-color space. On the $[3.6]-[4.5]$ vs.\ $[4.5]-[5.8]$ diagram, the YSO candidates are smoothly distributed, with a peak in density around $[3.6]-[4.5]\sim0.5$ and $[4.5]-[5.8]\sim0.4$, and a tail that extends up and to the right. In contrast, the contaminant distribution peaks slightly bluer in $[3.6]-[4.5]$, and the distribution appears bifurcated, with some sources being redder in $[3.6]-[4.5]$ while others are redder in $[4.5]-[5.8]$. This bifurcation may be related to the types of contaminants. For example, the contaminants to the upper left roughly correspond to a region of color space identified by \citet{Gutermuth2009} (and indicated by the gray boundary) as containing sources produced by knots of shocked emission from H$_2$, while the contaminants to the lower right were associated with knots of PAH emission.  Both areas outlined by Gutermuth et al.\ are dominated by objects that we classify as probable contaminants, but the edges of the YSO distribution also overlaps these boundaries. 

On the $[4.5]-[5.8]$ vs.\ $[5.8]-[8.0]$ diagram the peak density of YSO candidates is redder in both colors than the peak density of probable contaminants. The objects with the most extreme $[5.8]-[8.0]$ colors are nearly all classified as contaminants. These lie in a region of the diagram identified by \citet{Gutermuth2009} (gray boundary lines) as being dominated by unresolved star-forming galaxies. This diagram also includes a finger comprised of both YSO candidates and contaminants, extending to high $[4.5]-[5.8]$ values ranging from $\sim$2 to $\sim$3.5, but with $[5.8]-[8.0]$ colors in a restricted range $1.5\lesssim [5.8] - [8.0] \lesssim 2.6$. Previously published YSO catalogs \citep[e.g.,][]{Gutermuth2009,Rebull2011} have included a few YSOs in this region of color space; however, the high number of sources identified when examining the entire inner Galactic midplane makes this feature appear much more pronounced. These stars have colors similar to the PAH nebulosity found in star-forming regions \citep{Povich13}; however, visual inspection of a sample of these sources suggests that the majority are {\it bona fide} point sources in all four IRAC bands. 

In the bottom two panels of \autoref{fig:mir_ccds}, the peaks of the YSO candidate distributions are redder than the peaks of the probable contaminant distributions for each IRAC color. Nevertheless, while the objects with reddest $[3.6]-[4.5]$ tend to be YSO candidates, the objects with most extreme red $[4.5]-[8.0]$ or $[5.8]-[8.0]$ colors are almost all classified as contaminants.

\autoref{fig:nir_mir_ccds} shows the $J-H$ colors, which are the most sensitive to extinction, versus the IRAC colors, which are the most sensitive to IR excess. In $J-H$, the peaks of the density distributions are slightly redder for the contaminants than for the YSO candidates, but in IRAC colors, the peaks are significantly redder for the YSO candidates than the contaminants. In both cases, the objects with most extreme red colors tend to be classified as contaminants. However, some of the reddest IRAC sources do not appear on these plots because they lack $J$-band magnitudes. 

On all these color-color plots, the blue ends of the distributions are artificially truncated by the selection rules imposed to ensure that the IR excesses are real. Thus, our catalogs will not be sensitive to certain classes of YSOs, including some YSOs with anemic disks or pre--main-sequence stars without disks. 

\vspace*{0.05in}
\subsection{Optical Color-Magnitude Diagram}

Less than half the YSO candidates are optically visible, for example Gaia DR2 detects $\sim$36,000 of them, which comprise $\sim$30\% of the entire sample. The candidates detected by Gaia tend not to be as red in the mid-IR as other candidates (e.g., $[3.6]-[4.5] \lesssim1$ and $[4.5]-[8.0]\lesssim1.2$). 

\autoref{fig:fig_Gaia_CMD} shows a Gaia color-magnitude diagram for the visible YSO candidates. Absolute $G$-band magnitudes, computed using Gaia parallaxes $\varpi$, are plotted against Gaia $G-RP$ colors. Only sources with signal-to-noise $\varpi/\sigma_\varpi > 3$ are included, meaning that the sample of 7686 sources is small compared with the total number of YSO candidates. Nevertheless, this sample is useful for evaluating whether the optically bright candidates have properties consistent with pre--main-sequence stars.

On the Gaia color-magnitude diagram, we show isochrones for young stars at several ages ranging from 1~Myr to 50~Myr from the \citet{Bressan2012} models.  We also indicate the effects of reddening, which would shift points down and to the right on this diagram. The wide Gaia bands mean that the effect of reddening depends on the spectrum of the object, so we show three approximate reddening vectors for three colors; more discussion of how this affects selection of pre--main-sequence stars can be found in \citet{Herczeg2019} and \citet{Kuhn2020}. Nearly all the candidates lie above the 50~Myr isochrone, and the majority also lie above the 1~Myr isochrone, which is consistent with most of these candidates being very young pre--main-sequence stars.

\subsection{24~Micron Photometry}

When photometry is available in the MIPS 24~$\mu$m band (or the W4 band at 22~$\mu$m), it can be useful for corroborating classifications based on IRAC. For example, the SED at $\sim$24~$\mu$m tends to be more steeply declining for AGB stars, where IR excess is produced in hot dusty winds, in contrast with YSOs' relatively cooler disks and envelopes. 

\autoref{fig:explanatory24micron} shows the candidate YSOs and contaminants in $J-K_s$ vs.\ $[4.5]-[24]$ colors. These colors may be useful for distinguishing between AGB stars and YSOs because the typical AGB star has a precipitous rise in the $JHK$ bands followed by a drop in the mid-IR. The figure shows that the YSO candidates tend to be in the middle of the $[4.5]-[24]$ distribution. The objects with $[4.5]-[24] \lesssim 2.4$ are almost all classified as probable contaminants; however, there is a red tail to the probable contaminant distribution, with a high percentage of objects redder than $[4.5]-[24] \gtrsim 7$ also being considered contaminants. An examination of the spatial distribution (not shown) of the probable contaminants in this red tail reveals that many of them are non-clustered higher latitude objects in the Galactic bulge. Of the contaminants with low $[4.5]-[24]$, the distribution of $J-K_s$ ranges from $\sim$0 to $\sim$9, extending redward of the YSOs. Such red $J-K_s$ colors combined with relatively blue $[4.5]-[24]$ colors would be consistent with our hypothesis that many of these probable contaminants are AGB stars.  

\subsection{SED Class}\label{sec:yso_class}

Spectral index in the infrared, defined as 
\begin{equation}
    \alpha = \frac{\mathrm{d}\log(\lambda f_\lambda)}{\mathrm{d}\log\lambda},
\end{equation}
is frequently used to assess the evolutionary stages of YSOs \citep[e.g.,][]{Lada1987,Andre1994,diskionary,Rebull2014}.
However, the value of $\alpha$ depends on what spectral range is used, with the largest available range typically being favored by most studies. Furthermore, the calculation of spectral index may also be affected by reddening. To estimate $\alpha$ values that are minimally affected by reddening, we use the wavelength range from 4.5~$\mu$m to 24~$\mu$m, since interstellar extinction in these bands is smaller than at shorter wavelengths and the reddening curve is flatter 
\citep{Indebetouw2005,McClure09,Xue2016}. For these bands, 
\begin{align}
    \alpha_{[4.5] - [24]} & \approx 0.55\,([4.5]-[24]) - 2.94\\
    \alpha_{[4.5] - W4} & \approx 0.58\,([4.5]-W4) - 2.92\\
    \alpha_{[4.5] - [8.0]} & \approx 1.64\,([4.5]-[8]) - 2.82.
\end{align}
Where available, we prefer the $\alpha$ estimate based $[4.5]-[24]$, followed by $[4.5] - W4$, and finally $[4.5]-[8.0]$. For YSOs suspected of having strong silicate absorption or PAH emission (Sections~\ref{sec:silicate}--\ref{sec:PAH}) we do not use the $[4.5]-[8.0]$ color to estimate YSO class because either feature could affect the 8.0~$\mu$m band. 

\autoref{fig:alpha} shows the distribution of spectral indices calculated for candidate YSOs. 
Based on these estimates there are 15,943 Class I ($\alpha > 0.3$), 23,810 flat spectrum ($0.3 \leq \alpha < -0.3$), 59,949 Class II ($-0.3\leq \alpha < -1.6$), and 5,352 Class III ($\alpha \leq 1.6$) YSOs, using the $\alpha$ boundaries from \citet{Greene1994}. In addition there are 12,392 candidate YSOs with uncertain class due to missing photometry. This classification scheme roughly reflects the YSO evolutionary sequence from deeply embedded sources with massive envelopes (Class~I and flat spectrum) to stars with disks (Class~II) and systems where the disk has mostly dispersed (Class~III). However, viewing geometry may also affect the assigned YSO class; for example, a YSO that would otherwise be considered Class~II may have a Class~I SED if viewed at high inclination \citep{Williams2011}. Finally, we clarify that even though some classification schemes regard Class~III sources as having no IR excess \citep[see][]{diskionary}, in our scheme Class~III implies weak, but detectable excess. 

\subsection{Possible Silicate Absorption}\label{sec:silicate}

Broad silicate dust absorption or emission features, centered at $\sim$9.7 and $\sim$18~$\mu$m, are frequently detected in the mid-IR spectra of YSOs \citep[e.g.,][]{Furlan2006,Furlan2008,Furlan2011,Oliveira2010}. The 9.7~$\mu$m feature overlaps the IRAC 8~$\mu$m band, so these features can affect YSO colors observed by IRAC. 

In the color-color diagram shown in \autoref{fig:explanatory} (left panel), a group of $\sim$2000 YSO candidates stand out due to their unusually blue $[5.8]-[8.0] < 0$ colors -- these objects are flagged in \autoref{tab:spicy}. Given the lack of a red color in $[5.8]-[8.0]$, the classification of these stars as YSO candidates was based mainly on their $[3.6]-[4.5]\gtrsim0.5$ and $[4.5]-[5.8]\gtrsim0.5$ colors, both of which tend to be redder than most of the other YSO candidates.  

\autoref{fig:SED} shows three example SEDs that we have fit with YSO models from \citet{Robitaille17} -- for each source the ten best-fitting convolved models are indicated by the gray lines. \citet{Robitaille17} include multiple configurations of disks and/or envelopes, so we used the simplest model forms capable of explaining the data: a star and disk model (sp-s-i\footnote{The designations correspond to models from \citet{Robitaille17}.}) for SPICY~75228; a star, disk, and envelope models with variable inner radius (spu-hmi) for SPICY~85135; and a star and disk model with variable inner radius (sp-h-i) for SPICY~99415. 
Although these fits are not all formally good given the reported photometric uncertainties, they illustrate the range of SED morphologies that could produced the colors that we observe.
Each case requires a strong silicate absorption feature at 9.7~$\mu$m to reproduce the lower 8.0~$\mu$m band emission. The best models also tend to also have nearly edge-on inclination to provide the high absorbing column density. 

Silicate absorption in YSO SEDs can come from the object itself or from foreground interstellar dust \citep{vanBreemen2011}.  We would expect a YSO with strong intrinsic silicate absorption to have a substantial disk or envelope that can produce the extinction, and \citet{Forbrich2010} find that YSOs with positive spectral indices are more likely to have strong silicate absorption.  \autoref{fig:explanatory} (right panel) shows our strong silicate absorption candidates on a plot of $[3.6]-[4.5]$ vs.\ $[4.5]-[24]$. The color $[4.5]-[24]$ is a good indicator for SED spectral index that is not affected by silicate absorption. Most of the objects with possible silicate absorption have $[4.5]-[24]>5$, higher than average for the YSOs, but $\sim$20 sources have $[4.5]-[24]$ colors bluer than this. In the interstellar medium, the relation between the optical depth of the 9.7~$\mu$m feature and optical extinction is approximately $\tau_{9.7}\sim A_V/20$ \citep{Roche1984,Chiar2007,Shao2018}. Although most stars in our sample would not have sufficiently high foreground extinction for the feature to become optically thick, this can be achieved along lines of sight that pass through dense molecular clouds or near the Galactic Center.  

\vspace*{0.2in}
\subsection{Possible PAH Emission}\label{sec:PAH}

Another salient feature in the $[4.5]-[5.8]$ vs.\ $[5.8]-[8.0]$ diagram is the finger-like structure at $[5.8]-[8.0]\approx1.6$. Most of the sources with these colors in the full IRAC catalogs were classified as probable contaminants, but a minority ($\sim$490 objects) were classified as YSO candidates. These colors match those expected for sources dominated by PAH emission bands. For an astronomical PAH emission spectrum, the ratio of flux in the 5.8~$\mu$m band to the 8.0~$\mu$m band ranges from 0.31 to 0.41 \citep[][and references therein]{Draine2007}, corresponding to $[5.8]-[8.0] = 1.6$--1.9. There is little PAH emission in the 4.5~$\mu$m band, leading to a red $[4.5] - [5.8]$ color. Candidates are flagged in \autoref{tab:spicy} for strong PAH emission if they meet the criteria $[4.5]-[5.8]>2$ and $[5.8]-[8.0] > 1$, which are based on the observed morphology of this feature in color space. 

Although IR nebulosity in star-forming regions is dominated by PAH emission, inspection of the flagged YSO candidates suggests that most are valid point sources in all 4 IRAC bands, not spurious detection of nebular knots. For example, $\sim$90\% of these sources have $M=2$ detections in both the 5.8~$\mu$m and 8.0~$\mu$m bands, indicating reliable detections. We examined the images of a subset of these objects by eye and found that even in cases with surrounding nebulosity, the sources themselves appeared to match the PSF. PAH emission may be intrinsic to massive YSOs with sufficiently high ultraviolet luminosities \citep[e.g.,][]{Whitney2013}.
Spitzer IRS spectroscopy of massive YSOs has shown PAH emission to be nearly ubiquitous and correlated with YSO luminosity \citep{Oliveira2013}.

In the MYStIX IR-excess catalog that we used for training, \citet{Povich13} aggressively filtered sources with PAH emission to avoid contamination by nebular PAH knots. To be included, they required sources to exhibit red $K_s-[4.5]$ colors (avoiding bands with PAH emission), as would be expected for a massive YSO. This requirement will be reflected in our classifications via the random forest classifier. In the SPICY catalog, YSO candidates with possible PAH emission have median $K_s - [4.5]=3.3$ compared with a median for the entire sample of 2.0.

\autoref{fig:explanatory} (right panel) shows the 24~$\mu$m emission for these objects. The YSO candidates with possible PAH emission have $[4.5]-[24]$ colors ranging from 5 to 10, much higher than the average for YSOs. This result is consistent with these objects being massive YSOs. 

\section{Environment}\label{sec:images}

\begin{figure*}
	\centering
	\includegraphics[width=0.9\textwidth]{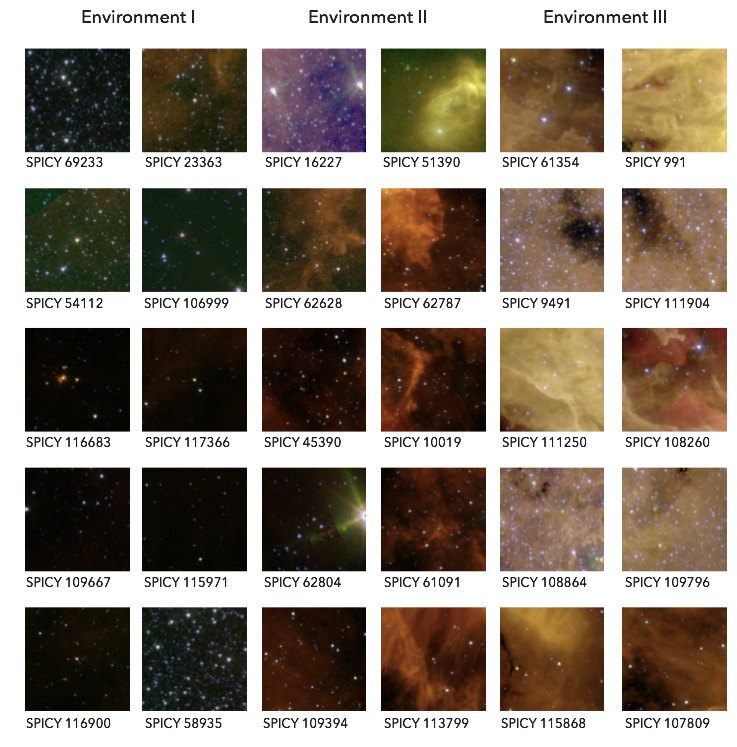}
    \caption{Color $3^\prime\times3^\prime$ cutouts    (IRAC 3.6~$\mu$m in blue; 5.8~$\mu$m in green; 8.0~$\mu$m in red) centered around a sample of YSO candidates from the SPICY Album. The first two columns show examples of stamps from the  Environemnt 1 class, corresponding to images with no or minimal nebulosity. The two middle columns correspond to examples of stamps from the Environment 2 class, which is a mixed class mostly containing regions at the transition between the Environments 1 and 3 or miss-classifications from those two extreme classes. The last two columns show examples of regions classified as Environment 3, that clearly correspond to cloud-like environments.}
  \label{fig:cutout}
\end{figure*}

Many dynamical processes sculpt the interstellar medium in star-forming regions and affect the spatial relationship between the clouds and young stars \citep{Shu1987,McKeeOstriker2007}.
Infrared nebulosity, however, can be considered a strong proxy for star formation, as newly-formed massive stars illuminate the primordial clouds in the star-forming complex. The Spitzer images reveal features ranging from IR dark clouds to bright PAH-dominated nebulosity, which can trace the photodissociation regions at the edges of clouds and bubbles \citep[e.g.,][]{Churchwell09,Pari2020}.  

To facilitate the study of the local environments around YSOs, we have created an album of $3'\times3'$ image cutouts\footnote{
To produce the album, we constructed an infrastructure to query the IPAC archive at \url{http://irsa.ipac.caltech.edu} that tracks the FITS transfers and that also tracks and verifies the local generation of the PNG stamps. This infrastructure makes use of a PostgreSQL database \citep{PostgreSQL} and is parallelized. However, we kept the number of parallel data transfers from IPAC low to avoid overloading their servers, enabling the extraction of all IRAC images and the construction of all the stamps in about three days.}
in all four Spitzer bands with additional false-color image files ready for visual inspection or generic image processing frameworks \citep[e.g.,][]{Yang2012}. The false-color images (see examples in \autoref{fig:cutout}) were created with a heuristic based on \citet{2004PASP..116..133L}, mapping the IRAC 3.6~$\mu$m to the blue channel, 5.8~$\mu$m to green and 8.0~$\mu$m to red. Here, we applied a hyperbolic arcsin transform to each IRAC band, and we selected the range of the color intervals from the mode of the distribution of the lower $2\times10^{-2}\%$ of the pixels for the minimum value, and the mode of the distribution of the upper $6\times10^{-5}\%$ pixels for the maximum. These values were chosen to optimize the visual experience while minimizing information loss and excluding extreme outlier pixels. The modes were estimated using the \citet{venter1967} estimator, as implemented by the {\sc modeest} package \citep{modeest2019}. 

The SPICY album comprises a total of 117,224 PNG stamps. A total of 222 YSOs candidates from the SPICY catalog miss their stamps due to numerical problems in the original FITS files and/or the lack of response from the IPAC archive in one or more bands at the time of the album creation. Its size is 251~GB, and all PNG and FITS files are publicly distributed and archived long-term at Zenodo
hosted at CERN facilities. 

\subsection{A Simple Characterization}

Below, we demonstrate an example application for these cutouts, using a simple unsupervised image clustering strategy to characterize environments in which the YSOs candidates are found. 

We avoid clustering in the pixel space because it is not invariant to image translations and rotations, which are properties that any proper content-based image clustering solution should have. Two candidate transforms that can introduce these properties via the power spectrum are wavelets \citep[as used in][for a similar application]{KM2019} and Fourier transforms \citep[e.g.,][]{Kauppinen,VANDERSCHAAF1996}; here, we adopt the latter. This is partially motivated because the Fourier power spectra is linked to the turbulent properties of the star formation medium \citep[e.g.,][]{2004ARA&A..42..211E}, revealing signatures of different physical phenomena.

We first compute the 2D Fourier power spectra of each cutout in each IRAC band. Then, we compute 1D radially medianized power spectra from each of the original 2D power spectra and concatenate these 1D power spectra to form a vector for each YSO candidate. Next, we organize the vectors of all environments into a single matrix and perform principal components analysis \citep[PCA;][]{Pearson01, Hotelling33}, from which we select the most relevant dimensions \citep[see also][for PCA variants]{Ishida2013,deSouza2014}, which acts as feature compression \citep[see e.g.,][]{Sasdelli2016}. Finally, we model the distribution using a multivariate Gaussian mixture model \citep[GMM;][]{1894RSPTA.185...71P, mclust,deSouza2017,Melchior2018} in the space defined by the first two principal components of the power spectra and the modes of the pixel values in each cutout, which we transform using an inverse $\sinh$ function. The distribution in this space is complex and requires many (25) Gaussian components, with model selection using the Bayesian information criterion \citep{schwarz1978}. 

Visual inspection shows that the GMM components tend to correspond to three types of environments: those that are nebulosity-free (or minimal nebulosity) environments, mixed environments, and cloud-like environments. These are labeled environments 1, 2, and 3 in \autoref{tab:spicy}. We also found outliers on the boundary of the distribution. Examination of the cutouts showed that the outliers correspond to severe image reconstruction errors and/or missing data in one or more bands and are located at the edges of the surveys.

A total of 66,539 stamps, or $\sim$57\%, of the valid stamps, were classified as cloud-like, while 32,790 stamps, or $\sim$28\%, were classified as nebulosity-free. The mixed class and the outliers correspond to 15,462 and 2,433 stamps, or  $\sim$13\% and $\sim$2\%, respectively. These numbers indicate that most YSO candidates are indeed in cloud-dominated environments, as would be expected for YSOs in star-forming regions. However, the number of candidates in environments presenting diffuse nebulosity or even no detectable nebulosity is not negligible. There is a slight, but statistically significant\footnote{The Kolmogorov-Smirnov test yields a $p$-value $<$0.01 for the null hypothesis that the $\alpha$ values are drawn from the same distribution for cloud-dominated versus nebulosity-free regions.} correlation between spectral slope $\alpha$ and environment class, with stars in nebulosity-free environments having preferentially more negative spectral indices (indicating later evolutionary stages).

The cloud-like environments are most prevalent in the inner regions of the Galaxy, between approximately $\ell=300^\circ$ and $50^\circ$ and $|b|\leq1^\circ$. Cloud-like environments are also associated with large star-forming complexes outside this coordinate range, including some of the star-forming regions in Cygnus~X. 

The mixed environment is most prevalent further from the Galactic Center (e.g., $\ell\leq310^\circ$ or $\ell\geq30^\circ$). Some stellar associations include both cloud-like and mixed classes (e.g., the Carina Nebula). 

The image cutouts with no or minimal nebulosity are found throughout the entirety of the survey, except within $\sim$1$^\circ$ of the Galactic Center. Stars in this environment are the most evenly distributed, but even among these stars several clusters can be seen. For example, the Sco~OB1 Association is made up of both cloud-like and nebulosity-free classes.    

This simple application can certainly be significantly improved by adopting tailored methodologies and signal representations, for instance, curvelet transforms \citep[e.g.][]{Candes00curvelets, 2003A&A...398..785S}, to characterize the signal power contained in filamentary structures, and customized clustering methods. Moreover, a proper physical characterization of the YSO environment requires consideration of effects of, for example, the distances to the objects, accounting for the distinct physical scales probed, differences of PSFs in different IRAC bands and their impacts on the power spectra, projection effects of the nebulous matter in the plane of the sky, etc. However, as we show here, even a simple analysis already reveals that, curiously, a significant fraction of the YSO candidates in the SPICY catalog do not seem to be lying in environments dominated by clouds. 

\section{Spatial Clustering}\label{sec:clustering}

Spatial aggregation is a well recognized property of YSOs \citep[e.g.,][]{vanDenBergh1964,Carpenter00,Allen2007} that  can be observed in the distribution of our YSO candidates (\autoref{fig:atlas}). However, the best clustering algorithm to use for stars, or even what is the most meaningful definition of star cluster, is not obvious \citep[see][]{KuhnFeigelson2017,Gouliermis18,Ascenso2018}. For example, different groups may have vastly different numbers and densities of stars, and the spatial distributions are complex and often fractal-like. Thus, different cluster analysis methods that yield different segmentations may be appropriate for different scientific applications \citep[e.g.,][]{Everitt2011}. 

\begin{figure}[t]
	\centering
	\includegraphics[width=0.47\textwidth]{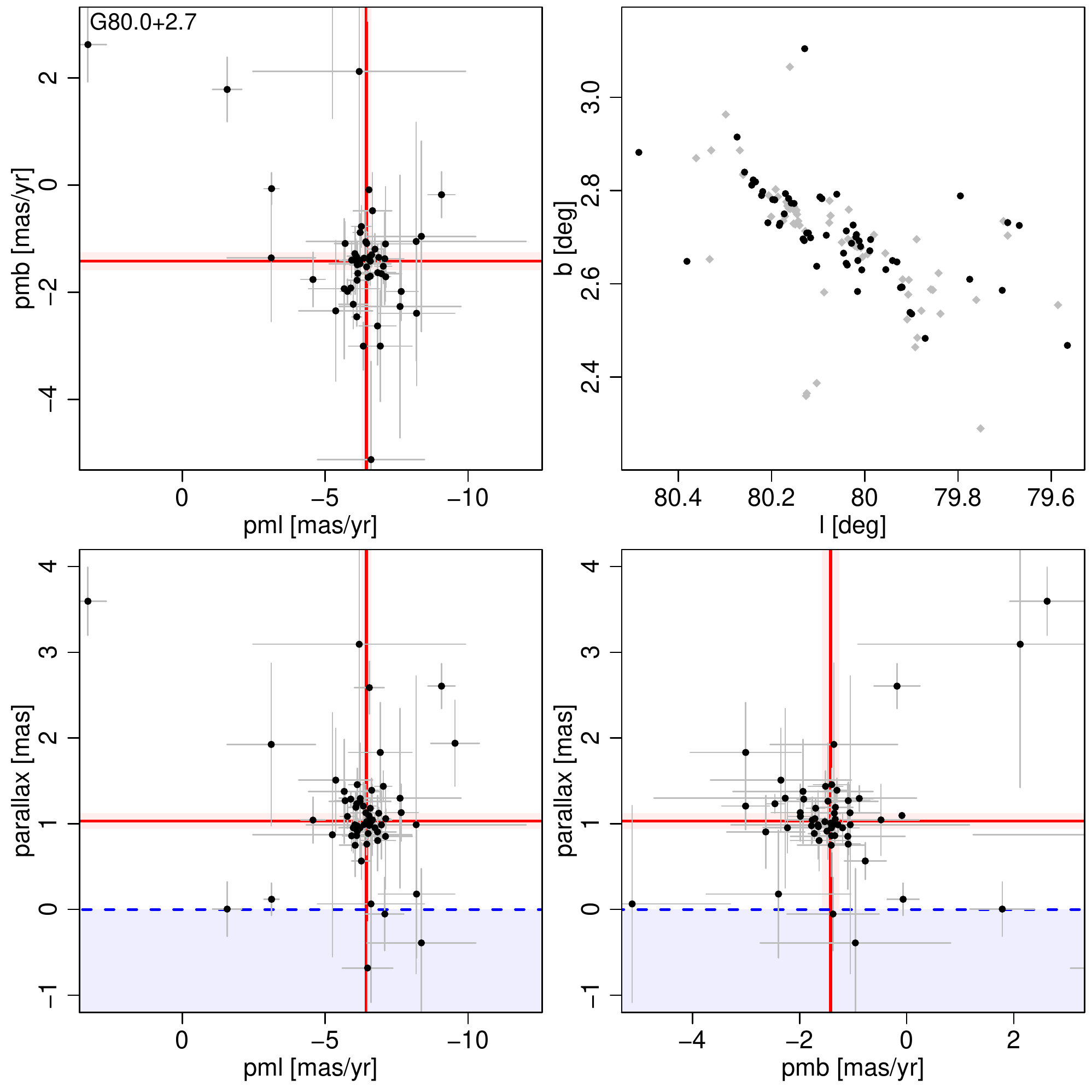}
    \caption{
    Scatter plots of astrometric properties for YSO groups; G80.0+2.7 (a group in the Cygnus~X field) is shown as an example. These plots include proper-motion vs.\ proper motion, parallax vs.\ proper motion, and $(\ell,b)$ positions. The estimated means from the hierarchical Bayesian model are shown by the red lines, and 2$\sigma$ formal uncertainties are illustrated by the pink shaded areas. Values excluded by our prior are shaded blue. Stars with Gaia DR2 5-parameter astrometry are black circles with gray 1$\sigma$ error bars, and stars with only position information are gray diamonds. \\
    (The complete figure set (406 images) is available in the online version.)
    }
  \label{fig:means}
\end{figure}

\begin{figure*}
	\centering
	\includegraphics[width=0.75\textwidth]{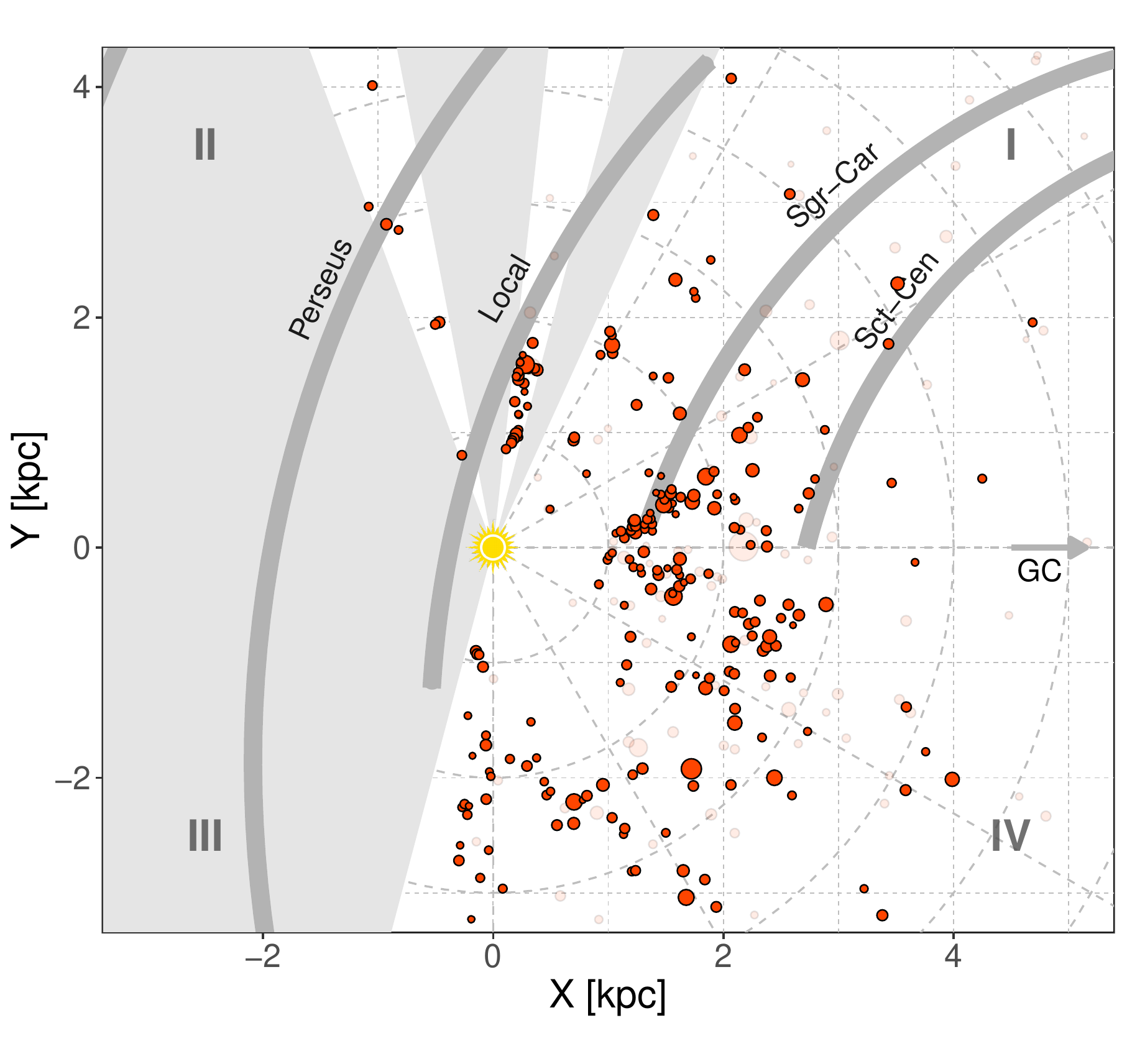}
    \caption{Spatial distribution of YSO groups in heliocentric Galactic XY coordinates. Groups with more reliable distances are depicted by the darker red circles, while those with more uncertain distances ($\varpi/\sigma_\varpi < 2$ or flagged) are lighter red. Circle sizes are proportional to the total numbers of members. The approximate centers of spiral arms from \citet{Reid2019} are indicated by the gray curves. The Sun's location is indicated by the yellow symbol. Wedges of the $XY$ plane not covered by the catalog are shaded gray. The Galactic quadrants and the direction of the Galactic Center are labeled. For conversion of Gaia DR2 parallaxes to distance in this figure, we use the $-0.0523$~mas zero-point offset estimated by \citet{Leung2019}}
  \label{fig:galactic_distribution}
\end{figure*}

We choose the algorithm ``Hierarchical Density-Based Spatial Clustering of Applications with Noise'' \citep[HDBSCAN;][]{Campello2013}, which has been successfully applied to Gaia DR2 data to detect hundreds of new open clusters \citep[e.g.,][]{Kounkel2019,2020A&A...635A..45C}. The HDBSCAN algorithm \citep{Campello2013} allows for groups of stars with different numbers, densities, and morphologies, it permits stars to not belong to any group, and it can be applied across the entire survey area in a uniform way providing a reasonable looking clustering solution. We apply this algorithm to each of the contiguous survey regions using a Python implementation\footnote{\url{https://hdbscan.readthedocs.io/en/latest/api.html}}. The main parameter we choose is the minimum number of stars in a group, which we set to $n=30$, and we run the algorithm using the ``excess of mass'' method for cutting the tree. The algorithm is run on the Galactic $\ell$ and $b$ coordinates for each contiguous segment of the IRAC survey area. The resulting groups are not necessarily gravitationally bound systems, but rather collections of YSO candidates that appear to be spatially aggregated. The two groups, labeled in \autoref{fig:COIN_cluster}, were selected using this algorithm. A list of these groups, along with their properties, are provided in \autoref{tab:groups}.

The method found 406 stellar groups, collectively including 58,084 ($=49$\%) of the YSO candidates. This suggests that roughly half the YSO candidates are spatially clustered while the other half are more widely distributed. The choice of $n$ does affect the solution, particularly whether groups are subdivided into smaller groups are unified into larger groups. However, the percentage of stars in groups stays relatively constant (i.e.\ 47--56\%) when $n$ is varied from 15 to 60. The median angular diameters of the groups increase from $\sim$0.2$^\circ$ to $\sim$0.7$^\circ$ over this range of $n$, with a value of $\sim$0.4$^\circ$ at $n=30$. We pick $n=30$ because the resulting solution appears to avoid chaining together unrelated stars over large areas of the sky, but the groups are large enough to include enough Gaia sources for their astrometric properties to be estimated. An examples of a complicated structure identified by us as a single group is the Carina Nebula complex, an association made up of multiple star clusters. We also note that HDBSCAN collects the over-density of candidate YSOs toward the Galactic center into a single group labeled ``G0.2-0.1''; these stars do not all come from the same star-forming regions but this region of the Galaxy is challenging for our algorithms.

\subsection{Galactic {\it XY} Distribution}\label{sec:xy}

To estimate the heliocentric distances ($d_{\odot}$)  to each of the YSO groups we employ a hierarchical Bayesian model \citep{Hilbe2017}  to account for the measurement errors in parallaxes and presence of outliers. The use of robust statistics is particularly suitable given the non-negligible presence of unknown contaminants in each group. Normality assumptions are sensitive to noise and outliers, which may result in a biased estimate of the mean distance. Replacing a Gaussian likelihood by a $t$-distribution is a relatively easy fix. The $t$-distribution has an extra $\nu$ parameter called degrees of freedom, which controls close the distribution resembles the normal distribution. Larger values $\nu > 30$ essentially recover the normal distribution, while smaller values result in a distribution with heavier tails. This extra flexibility enables it to adapt to the extra noise in the data, without introducing a bias in the underlying relationship. 

The model formulation for the robust estimate is given below, where we define a $t$-likelihood for the observed $\varpi$, and suitably vague priors on all the model parameters: uniform for $d_{\odot}$ over 25 kpc, and a gamma ($\Gamma$) prior (to ensure positivity) for $\nu$.

\begin{align}
\label{eq:model}
   &\rm \varpi_{i}\sim \rm {\mathscr{T}}(1/d_{\odot},\sigma_{\varpi_{i}}^2,\nu), \notag  \\
   &\nu \sim {\Gamma}(2,0.1) ,  \\
   &d_{\odot} \sim {\rm Uniform}(0,25), \notag \\ 
   & i = 1\dots n_\mathrm{Gaia} \notag
    \end{align}
The index $i$ runs over the members of each group $n_\mathrm{Gaia}$ with Gaia astrometric information. Although distance is constrained to be positive,  our likelihood model permits the parallax measurements for individual stars,  $\varpi_i$,  to be either positive or negative. We evaluate the model using a Gibbs sampler, for which we use the \textsc{jags}\footnote{http://cran.r-project.org/package=rjags} package \citep{plummer2017jags} within the \textsc{R} language. 
We initiate three Markov Chains by starting the Gibbs sampler at different initial values. Initial burn-in phases were set to 5,000 steps followed by 5,000 integration steps for each YSO group, which are sufficient to guarantee the convergence of each chain.  

Table~\ref{tab:groups} provides group parallaxes and proper motions estimated from the posterior medians. Uncertainties are estimated from the mean absolute deviation (MAD) of the posterior (scaled to approximate 1$\sigma$ uncertainties) and added in quadrature to the $\pm$0.04~mas and $\pm$0.07~mas~yr$^{-1}$ spatially correlated systematic errors on DR2 zero points \citep{Lindegren2018}. Out of 406 groups, 402 have some Gaia astrometry, giving at least a rough estimate of parallax and proper motion. Of these, most groups include at least  $n_\mathrm{Gaia}=10$ members having Gaia 5-parameter astrometric solutions, enabling estimates that are more precise than those based on individual stars.
 
For each group, we show scatter plots of stellar proper motions, parallaxes, and positions (\autoref{fig:means}), with the groups' mean parallaxes and proper motions indicated. In most cases, the stars form a single clump in $\mu_{\ell^\star}$--$\mu_b$--$\varpi$ space, suggesting that most of the group members are spatially and kinematically associated. In other cases (e.g., G77.8+1.0), multiple clumps are apparent, which may imply that distinct stellar groups with chance alignment have been merged by the HDBSCAN algorithm. In the example G77.8+1.0, the estimated properties correspond to the more distant but more numerous of the two groups. We visually inspected all groups in \autoref{fig:means} and flag those in Table~\ref{tab:groups} where problems could affect the interpretation of the Bayesian models, such as the suggestion of multimodality, groups that appear dominated by field stars, or a single data point with too much leverage. We also flag groups with $n_\mathrm{Gaia}<3$. 

The locations of these stellar groups in heliocentric Galactic $XY$ coordinates are plotted in \autoref{fig:galactic_distribution}. For this plot, we converted parallaxes to distance using an average $-0.0523$~mas Gaia DR2 zero-point correction estimated by \citet{Leung2019}.
Then $X = d_{\odot}\cos(b)\cos(\ell)$, and $Y = d_{\odot}\cos(b)\sin(\ell)$. The estimated centers of spiral arms from \citet{Reid2019} are also indicated. Fainter red points are groups with $\varpi/\sigma_\varpi<2$ or groups that have been flagged.  Few YSO groups are detected within 1~kpc, but the footprint of the GLIMPSE (and GLIMPSE extensions) surveys exclude many of the nearest star-forming regions, which are located more than several degrees above or below the Galactic midplane. The bulk of the YSO candidates for which we have accurate measurements have heliocentric distances that range from 1 to 3~kpc. There may be some bias in the distances we are sensitive to because this range resembles the range in distance of objects in our training set. Nevertheless, there are groups that appear to lie beyond $\sim$3~kpc, but Gaia-based distance become more uncertain at this range. 

\begin{figure*}
	\centering
	\includegraphics[width=0.75\textwidth]{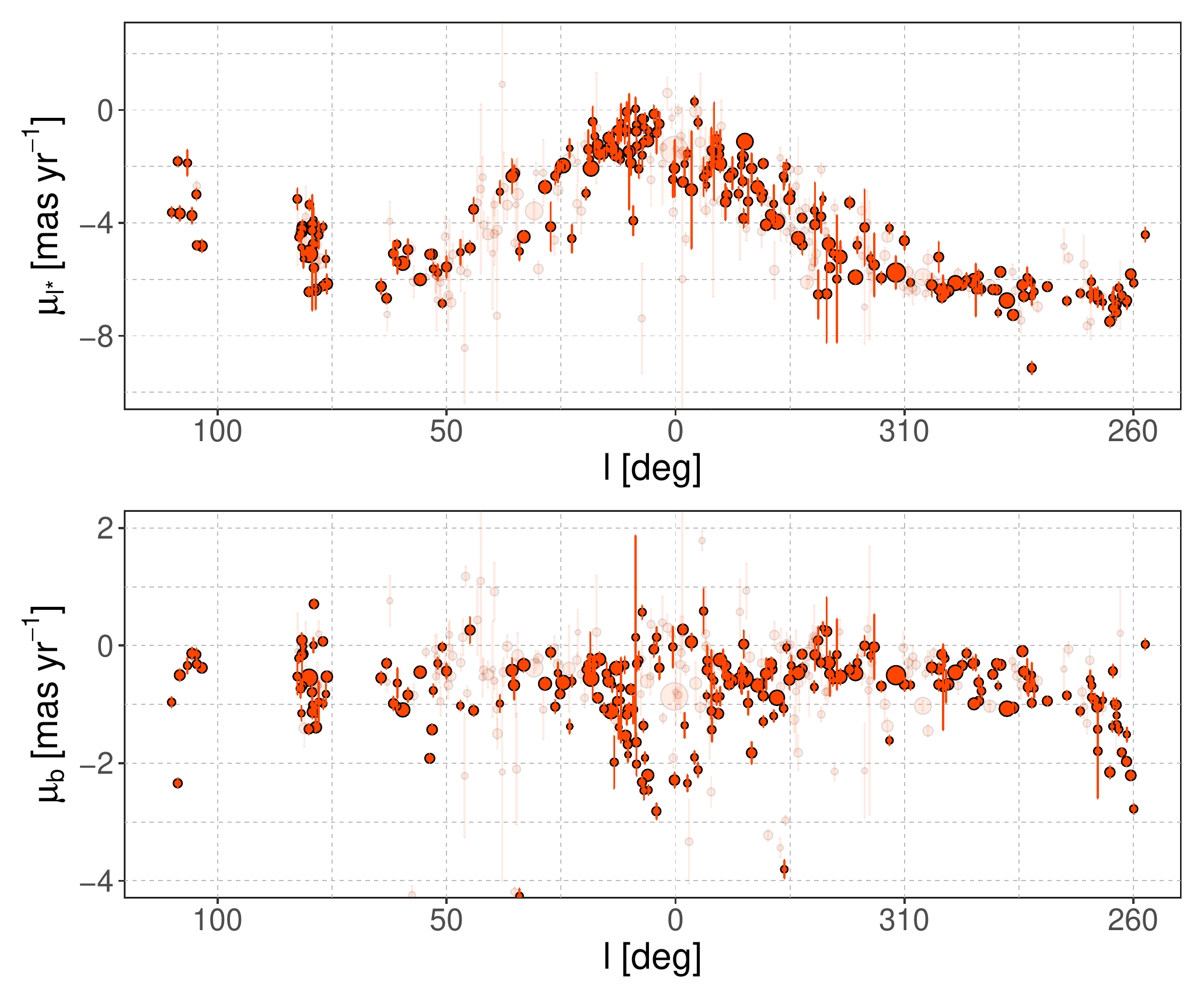}
    \caption{Position-velocity diagrams for the YSOs candidates: $\mu_{\ell^\star}$ vs.\ $\ell$ (top) and $\mu_b$ vs.\ $\ell$ (bottom). The distributions are the cumulative effects of Galactic rotation, Solar motion, and peculiar velocities of YSO associations. Size and shading of circles is the same as \autoref{fig:galactic_distribution}. Error bars combine the statistical uncertainties on group motions with the $\pm$0.07~mas~yr$^{-1}$ Gaia DR2 systematic uncertainty.}
  \label{fig:pmlb_l}
\end{figure*}

The YSO groups are not distributed smoothly within the Galaxy, but instead reveal Galactic structure. The survey areas intersect several spiral arms, and crucially provide information about Galactic structure at the boundary between Quadrants I and IV, where structures traced by $v_\mathrm{lsr}$ measurements of gas become degenerate. We discuss the relation of the stellar groups to the spiral arms below. 
\begin{description}
    \item[Local (Orion) Arm] In Quadrant~I and II, the local Arm intersects both the SMOG field and the Cygnus~X field. In SMOG, there is one group at the approximate distance of this arm. We find the YSO groups in Cygnus~X to be spread linearly, spanning a factor of $\sim$2 in distance ($\sim$1--2~kpc). This situation is consistent with looking down the length of the arm. Cygnus~X is thought to lie at an end of a long molecular filament \citep{Alves2020,Zucker2020} that contains multiple prominent star-forming regions (e.g., Orion, Taurus, the North America Nebula). The additional length of Cygnus~X may increase the total length of this structure by 50\%. A similar result is reported by \citet{Xu2016}. In Quadrant~IV, a chain of groups is located at $X\sim 0$ and extends from $\sim$1 to $\sim$3~kpc in distance toward the constellation Vela. The orientation of this chain suggests that these groups could connect with the Local Arm. 
    \item[Sagittarius-Carina Arm] Numerous star-forming regions can be found within 20$^\circ$ of the direction of the Galactic center, including some famous regions like the Trifid Nebula, the Lagoon Nebula, and NGC~6334. Some of these groups (including the aforementioned famous regions) form a coherent, chevron-shaped structure, that has a vertex pointing toward us at a distance of $\sim$1.0~kpc and edges that extend away with lengths of $\sim$1~kpc. This collection of star-forming regions is at the approximate distance of the Sagittarius-Carina Arm from \citet{Reid2019}, implying that the arm is an active site of star-formation activity. However, the angles of the edges of the chevron are inconsistent with the angle of the arm predicted by \citet{Reid2019}. The linear structure making up the edges of the chevron cannot be a result of the ``Fingers of God'' effect because it is not oriented along our line of sight. We find relatively little sign of YSO groups associated with the Sagittarius-Carina Arm at Galactic longitudes beyond $\ell>30^\circ$ in Quadrant~I or between $300<\ell<330^\circ$ in Quadrant~IV.
    \item[Scutum-Centaurus Arm] This arm is less clearly delineated by stellar groups than the others, possibly owing to large distance uncertainties at the distance of this arm. However, there is an increase in the density of groups near this arm in Quadrants~I and IV.  
    \item[Perseus Arm] This arm is intersected by the SMOG field, and three groups have distance estimates consistent with the center of this arm from \citet{Reid2019}. The large distance of this arm may decrease our sensitivity to YSOs associated with it. 
    \item[Inter-arm] There are multiple stellar groups that appear to be located between the spiral arms from \citet{Reid2019}. For example, within $\sim$10$^\circ$ of the Galactic Center, many groups appear located between the Sagittarius-Carina and Scutum-Centaurus arms. In Quadrant~I, several groups are located between the Local arm and the Sagittarius-Carina Arm.
\end{description}

\subsection{Galactic Rotation}

The procedure for calculation of mean proper motions for the groups is similar to the calculation of heliocentric distances, using a weakly Gaussian prior for $\mu_{\alpha^\star,0}$ and $\mu_{\delta,0}$ instead.

\autoref{fig:pmlb_l} displays the proper motions in Galactic longitude and latitude as function of $\ell$. The expected distribution for stars in circular Galactic orbits would be governed by Galactic rotation, parameterized by the \citet{Oort1927} constants, and effects from Solar motion, which are distance dependent \citep{Bovy2017}. On the $\mu_{\ell^\star}$ vs.\ $\ell$ diagram, to first order in the Galactic plane, this would be a sinusoid with period $180^\circ$ for Galactic rotation added to a sinusoid with period $360^\circ$ for Solar motion. This overall structure appears to dominate the $\mu_{\ell^\star}$ vs.\ $\ell$ plot for our YSO groups, but there are hints of deviations that we will discuss in more detail in a subsequent paper.   

On the $\mu_{b}$ vs.\ $\ell$ diagram, several structures can be seen. For example, between $\ell\sim5$--$25^\circ$, there is a diagonal chain of groups in position--proper-motion space. These groups correspond to one of the edges of the chevron-like structure detected in the $XY$ diagram. The dispersion in $\mu_b$ is slightly higher around $\ell\approx75^\circ$ (Cygnus~X) and around $\ell\approx260^\circ$ (Vela) -- both of which correspond to extended structures along the line of sight. 

\vspace*{0.2in}
\subsection{Spatial distribution of YSOs by Class}\label{sec:spatial_class}

\begin{figure*}
	\centering
		\includegraphics[width=0.48\textwidth]{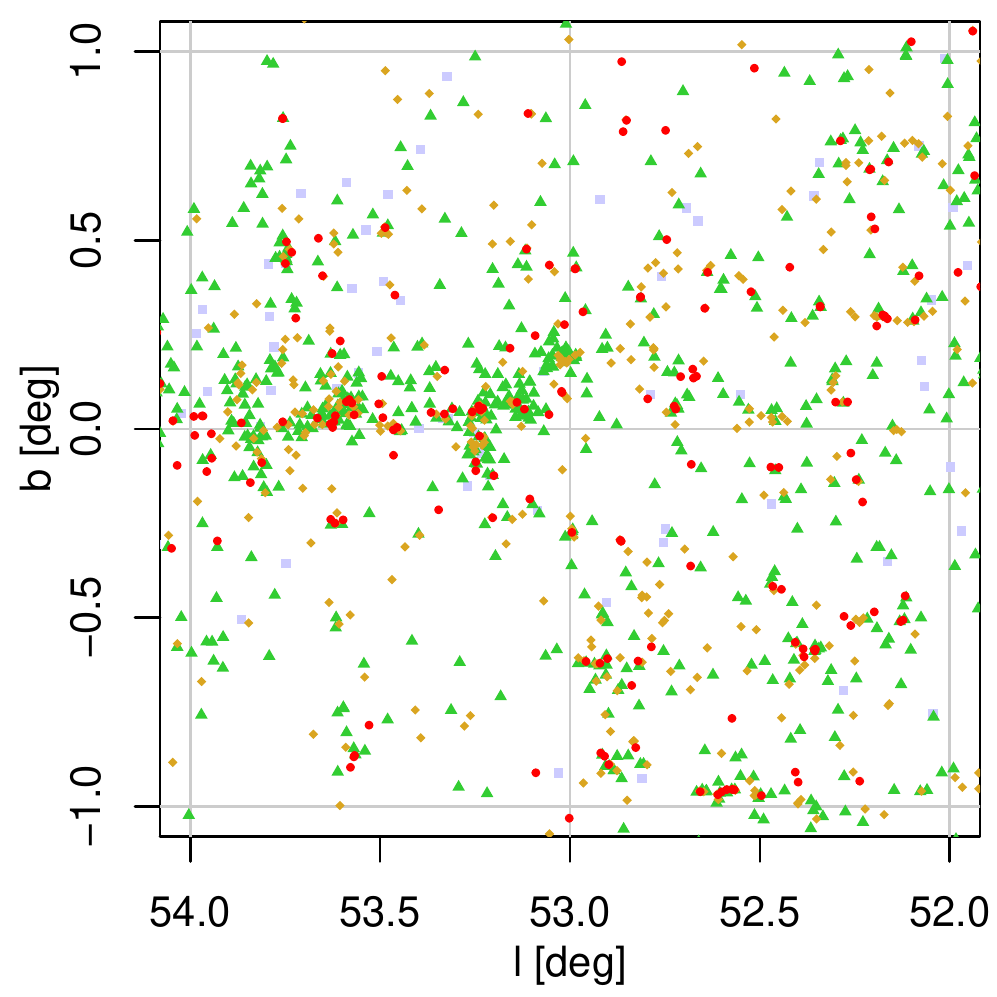}
	\includegraphics[width=0.48\textwidth]{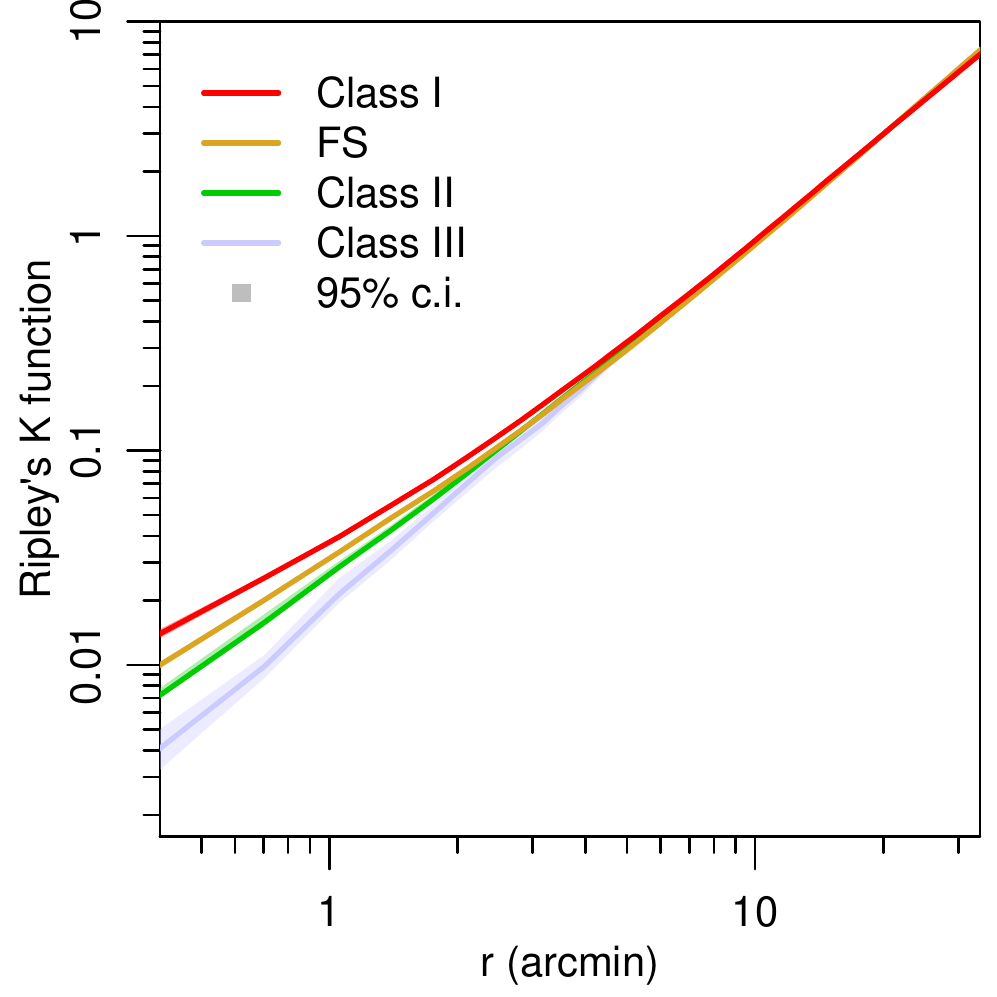}
    \caption{Diagnostics of spatial clustering for stars of different classes. The same color-coding is used to represent each class in both panels.  Left: The spatial distribution of YSO candidates in a $\sim$2$^\circ\times$2$^\circ$ sample area in Sagitta.  Right: Ripley's reduced second moment $K(r)$ function, with 95\% confidence intervals calculated for all stars. The flatter slope for the Class I and flat-SED YSOs at small angular separations implies these sources are more strongly clustered than the Class~II and Class~III objects. All classes exhibit some spatial clustering, but the earlier evolutionary classes tend to be more clumped. Nevertheless, examples of isolated YSOs of all classes can also be found.}
  \label{fig:class_map}
\end{figure*}

\begin{figure*}
	\centering
	\includegraphics[width=0.85\textwidth]{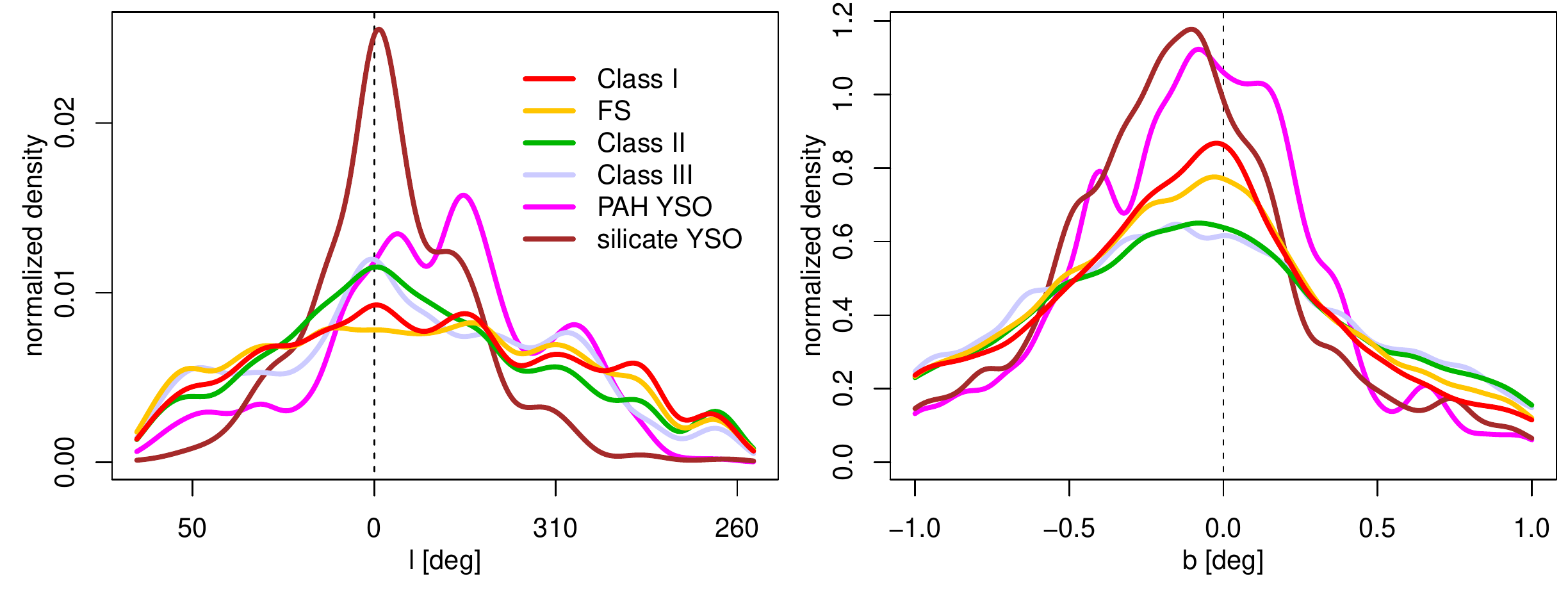}
    \caption{The distributions of $\ell$ and $b$ for stars of different YSO classes and stars with probable PAH emission or silicate absorption. Densities are calculated with a 5$^\circ$ kernel in $\ell$ and a 0.05$^\circ$ kernel in $b$. Only GLIMPSE~I, II, 3D, and Vela-Carina data are included in this plot. In Galactic longitude, the distributions of Class~I/FS/II/III stars are similar, but in Galactic latitude, the younger YSO classes are more strongly peaked toward the midplane. Sources with silicate absorption are strongly peaked near the Galactic center, while the peak in PAH source density is offset by $\sim$30$^\circ$ from the center.}
   \label{fig:gal_distribution}
\end{figure*}

YSO candidates of all SED classes are spatially clustered, but the classes corresponding to earlier evolutionary stages tend to be more strongly clustered, as can be seen in the section of the Galactic midplane in \autoref{fig:class_map} (left panel). In this region near the young cluster NGC~6823, Class~I sources mostly lie within the densest groups, but the other classes are comparatively more distributed. The $K$ function \citep{Ripley1976} can be used to quantitatively compare the relative strength of clustering for these populations. We used the \textsc{spatstat} package \citep{Baddeley2017} to estimate $K$ as a function of angular separation $r$ for Class~I, flat spectrum, Class~II, and Class~III objects with 95\% confidence intervals estimate using the bootstrap method of \citet{Loh_2008}. On this log-log plot (\autoref{fig:class_map}, right panel), at angular separations of several arcminutes, the slope for Class~I YSOs is significantly flatter than for flat spectrum YSOs, which is also significantly flatter than for Class~II and III YSOs, implying that the earlier stages are more clustered. This finding agrees with numerous other examinations of the spatial distribution of sources by YSO class \citep[e.g.,][]{Sung2009,Samal2010,Buckner2020}. We note that even some candidate Class~I YSOs appear isolated, for example $\sim$100 of these objects ($<$1\% of the Class~I YSOs) are separated from their nearest neighbors in our catalog by more than 10$^\prime$.  

\autoref{fig:gal_distribution} shows the smoothed distributions of sources of various classes in both Galactic longitude and latitude. In longitude, the normalized distributions of YSOs of all SED classes are similar, whereas, in latitude, the distributions of earlier classes (e.g., Class~I and flat spectrum) are more strongly concentrated near the midplane than the later classes (e.g., Class~II and III). This may be a result of the dispersal of YSOs, if stars are born in regions nearest the midplane, and then drift away. For example, a YSO traveling at a tangential velocity of $\sim$2~km~s$^{-1}$ at a distance of $\sim$2~kpc could travel 0.25$^\circ$ from its point of origin in $\sim$5~Myr. This is enough to flatten the distribution of $b$ shown in the figure but not the distribution of $\ell$.

The sources with strong silicate absorption are more concentrated toward the Galactic Center than other YSOs. This could be an effect of the higher interstellar dust column densities in this direction. The YSOs with strong PAH emission also appear to be preferentially concentrated toward the inner Galaxy, but the peak of the distribution appears to be in star-forming regions around $\ell\sim330^\circ$. 

\section{Optical Variability}\label{sec:variability}

\begin{figure}
	\centering
	\includegraphics[width=0.45\textwidth]{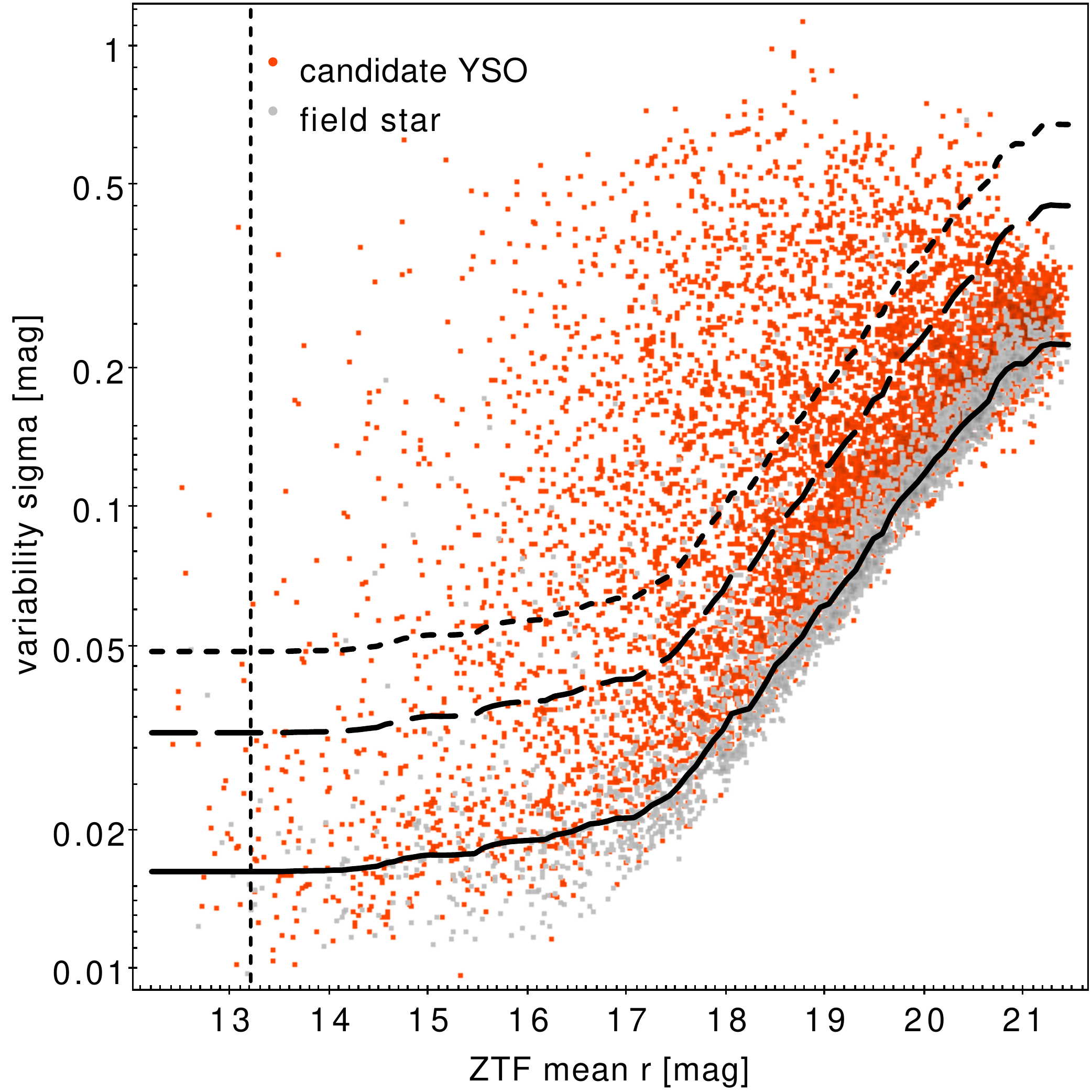}
    \caption{The standard deviation of the ZTF light curve's variability in the $r$-band ($\sigma_\mathrm{var}$) vs.\ mean $r$ for 7,585 YSO candidates (red) and $\sim$2,400 randomly selected field stars (gray) from the same areas of the sky. The solid, black line is the median $\sigma_\mathrm{var}$ as a function of magnitude for the field stars, and the lines above are 2 times (dashed) and 3 times (dotted) this level. We label stars above the dotted line as having ``strongly variability,'' between the dashed and dotted lines ``moderately variability,'' and below the dashed line ``low or insignificant variability.''}
  \label{fig:ztf_var}
\end{figure}

\begin{figure*}
	\centering
 	\includegraphics[width=0.95\textwidth]{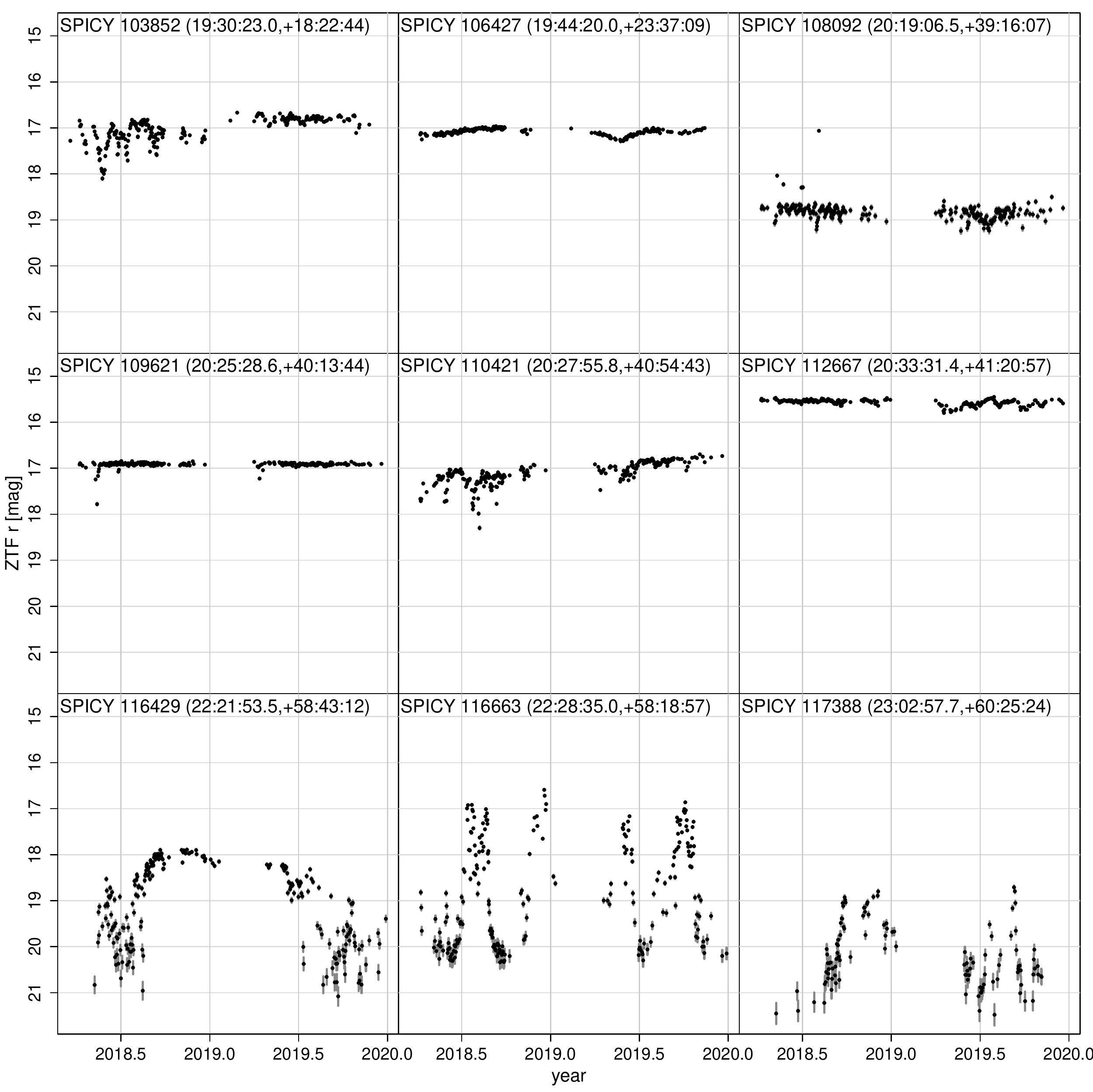}
    \caption{Sample ZTF light curves for several YSO candidates showing strong variability. These light curves exhibit diverse behaviors, including dipping, slow variation in brightness, outbursts, and periodicity.\\
    (The ZTF light curves for 7,585 SPICY sources are provided as data behind the figure.)}
  \label{fig:ztf_lc}
\end{figure*}

Optical variability, with amplitudes ranging from several tenths of a magnitude to outbursts of multiple magnitudes, is associated with YSOs \citep[e.g.,][]{Joy1945,Herbig1954,Cody2018} and has even been used as a criterion for identifying previously unrecognized YSOs \citep[e.g.,][]{ContrerasPena2017}. Thus, strong optical variability from YSO candidates in our Spitzer selected sample can be regarded as corroborating evidence for the youth of these objects. To investigate which sources show optical variability, we use photometric measurements from the Zwicky Transient Facility \citep[ZTF;][]{Bellm2019}, which is sensitive to a variety of variability phenomena from YSOs with its cadence of approximately 1 observation per night \citep{Graham2019}, including dips due to occultation from circumstellar dust, variations in accretion rate, magnetic flares, and rotational modulation due to large star spots. 

We cross-match our YSO candidates to the ZTF DR3 \citep{Masci2019} catalog using a match radius of 1$^{\prime\prime}$, and use ZTF sources with at least 10 measurements in the $r$ band between April 2018 and June 2019, excluding observations from the high cadence deep-drilling program; the median number of observations is $\sim$130. This yields 7,585 YSO candidates with usable ZTF light curves. This represents a relatively small fraction of our entire catalog because many of the Spitzer sources are not detected in the optical and ZTF is only available for the Northern Hemisphere. Nevertheless, in absolute numbers of sources this sample is moderately large and useful for statistical analysis. The sample is skewed toward Class II and III YSOs, but over a thousand sources categorized as Class~I or flat SED YSOs are included as well. 

To characterize variables, we calculate the $r$-band light curve's standard deviation $\sigma_\mathrm{var}$, the mean magnitude $\bar{r}$, and skewness of the distribution. 
\autoref{fig:ztf_var} shows $\sigma_\mathrm{var}$ vs.\ $\bar{r}$ for our YSO candidates (red points) as well as $\sim$2400 randomly selected field stars (gray points) from the same region of the sky. For the field stars, we have estimated the running median $\sigma_\mathrm{var}$ as a function of $\bar{r}$, shown as the solid black line. 88\% of the YSO candidates have $\sigma_\mathrm{var}$ values greater than this line, indicating that the YSO candidates have higher variability on average than the field stars. Following \citet{Fang2020}, we use the median $\sigma_\mathrm{var}$ for field stars to delineate variability thresholds for the YSO candidates. Objects where $\sigma_\mathrm{var}$ is more than 3 times the median value for field stars are considered to show high variability (above the dotted line), objects between 2--3 times the median value are considered to show moderate variability (between the dashed and dotted lines), and objects $<$2 times the median value have low or insignificant variability (below the dashed line). Using these definitions we find 1695 with high variability, 914 with moderate variability, and 4976 with low or insignificant variability. 

Many interesting variables have asymmetric variability, as indicated by skewness. For example, ``dipper'' YSOs, where fading is caused by circumstellar material passing in front of the star, tend to have positive skewness \citep[e.g.,][]{Cody2018}. Of the highly variable stars in our sample, 76\% have positive skewness, with a mean skewness of 0.3.

We find a slight tendency for strong or moderate YSO candidates to be more spatially clustered than weak or non-variable YSO candidates. For example, 56\% of stars with strong variability are members of HDBSCAN groups, while 51\% of stars with moderate variability are members, and 47\% of stars with weak or no variability are members.\footnote{Fisher's exact test of contingency tables \citep{Fisher1922} indicates that, even though the effect size is small, these differences are statistically significant at the $p<0.01$ level.} If non-variable candidates have a higher probability of being non-YSO contaminants, this could influence the observed trend, since contaminants are not expected to be clustered. However, even among the YSO candidates with high variability, 44\% are not members of HDBSCAN groups, providing further evidence suggesting that many of the relatively isolated candidates may still be legitimate YSOs.  

\autoref{fig:ztf_lc} shows a sample of light curves from sources showing high variability. Many of these stars exhibit dipping features with sharp bottoms, a morphological feature typically associated with extinction from dust (possibly in the circumstellar disk) briefly passing in front of the stars. The timescales of the dips seen in these ZTF light curves can range from several days to multiple months. Some light curves show long timescale trends, like the gradual brightening seen SPICY~110421. Other stars' light curves exhibit outbursts, with SPICY~116663 shown as an example with a particularly large $>$3~mag amplitude. These features are similar to the categories of YSO variability identified by \citet{Cody2018}, albeit many of the structures identified in their K2 study occur on a shorter timescale than we are sensitive to with the cadence of ZTF. Some of our candidates YSOs also exhibit periodic behavior, which is thought to be associated with the rotation periods of the stars due either to star spots or  material orbiting at the co-rotation radius \citep{Herbst1994,Stauffer2017}. SPICY~108092 is an example of one such star, which includes periodic rotation along with dips, bursts, and long time-scale changes.

The ZTF YSO light curves exhibit considerable diversity in their morphologies. Given that the objects in the SPICY catalog were selected in a uniform way independent of their variability, this dataset may be useful as a training set for future efforts to develop a classifier of YSOs based on optical variability. 

\section{Comparison to Other YSO Catalogs}\label{sec:comparison}

\begin{figure*}
	\centering
 	\includegraphics[width=0.95\textwidth]{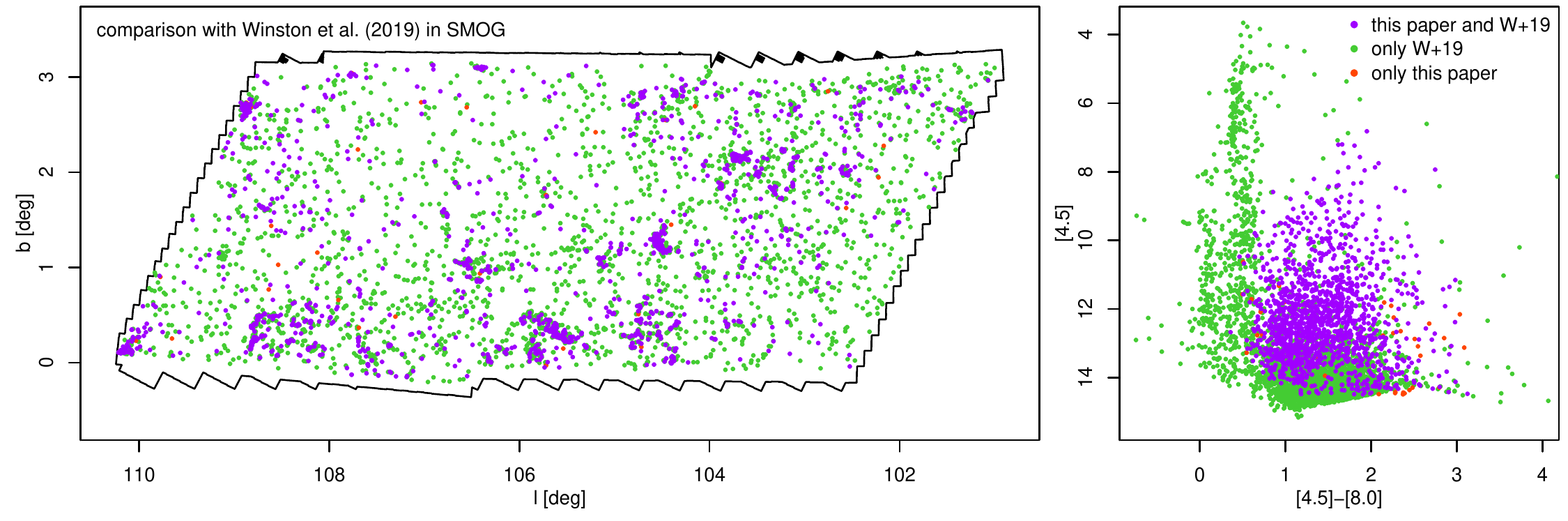}
    \caption{Comparison between our catalog and \citet{Winston19} in the SMOG field, where Spitzer observations are deeper than the main GLIMPSE survey. YSOs in both catalogs are color-coded purple, objects only in  \citet{Winston19} are green, and a low number of objects only in our catalog are red. The left panel shows the spatial distribution and the right panel shows a color-magnitude diagram. These diagrams indicate that our criteria are more selective than \citet{Winston19}, and the sources we omit tend to either have bluer IR colors or be fainter. Consistency between the catalogs is greatest among the spatially clustered sources, but our catalog does not include many of the non-clustered objects found by \citet{Winston19}. However, formal assessment of the accuracy of either catalog would require additional information from follow-up spectroscopic observations.}
  \label{fig:smog}
\end{figure*}

\begin{figure}
	\centering
 	\includegraphics[width=0.475\textwidth]{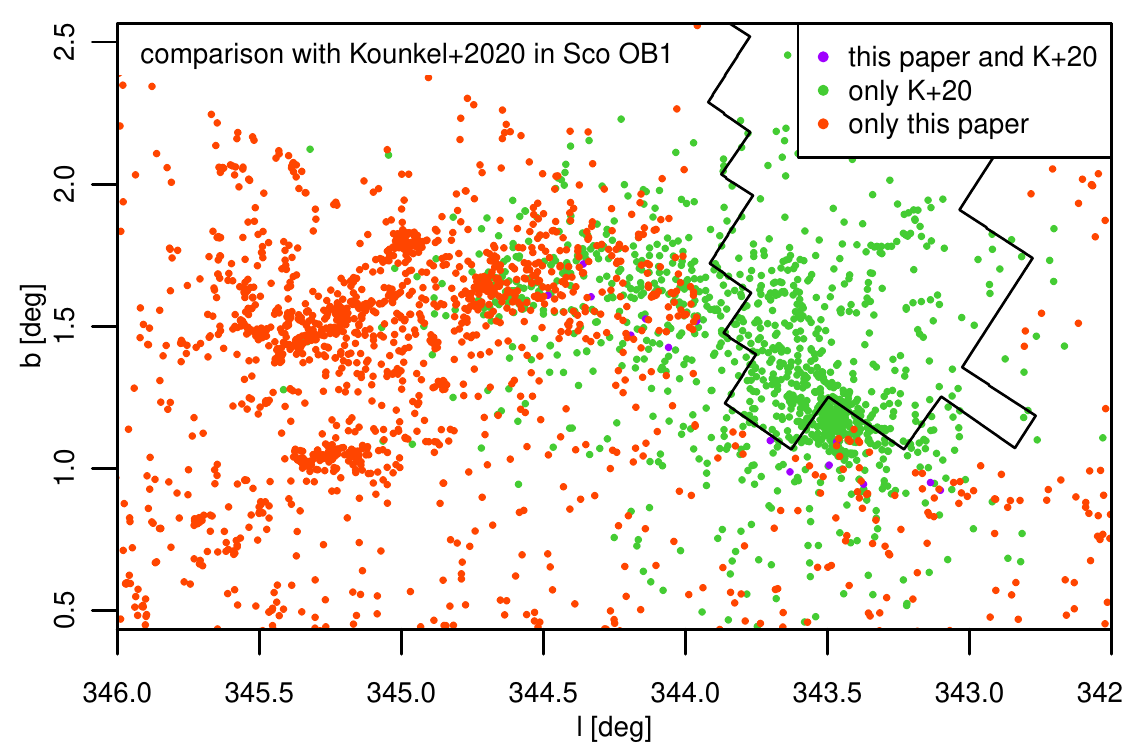}
    \caption{Comparison between our catalog and \citet{Kounkel2020} ($\mathrm{age}<10$~Myr) in a field centered around the Sco~OB1 association. There are few individual stars in common (purple points) between our infrared-excess selected YSOs (red points) and their astrometrically selected sample (green points). Nevertheless, these catalogs appear complementary because they trace different components of the same stellar association. The boundary of the GLIMPSE field (black lines) excludes a section of the sky in the upper right of this figure from our survey. }
  \label{fig:scoob1}
\end{figure}

Both we and \citet{Robitaille08} search GLIMPSE catalogs for YSOs, but our catalog extends the search to much fainter magnitudes. We have less stringent source quality criteria, we do not impose an {\it ab initio} $[4.5]-[8.0]\geq 1$ color cut, and, most significantly, we include sources fainter than the flux limits imposed in their catalog. Their 10~mJy limit in the 8.0~$\mu$m band ($[8.0]<9.52$~mag) would discount 73\% of our YSOs. In the overlapping regions, the GLIMPSE I and II survey areas, we identify $>$4 times more YSO candidates than the red sources from \citet{Robitaille08} catalog. 

Unlike our catalog, most of the sources from \citet{Robitaille08} are bright enough to have been detected by observations at 24~$\mu$m. For these objects, they use a simple heuristic set of color criteria to separate YSO and AGB candidates: sources with $[4.5]\le7.8$ or $[8.0]-[24] < 2.5$ are considered likely AGB stars, while sources with $[4.5]>7.8$ and $[8.0]-[24] \ge 2.5$ are likely YSOs. They acknowledge that a division like this is likely to produce erroneous classifications in either direction. Out of 16,670 ``red sources'' from \citet{Robitaille08} that they labeled either YSO (9,387) or AGB (7,283), 13,290 (80\%) were re-identified as candidate YSOs by our analysis, including 8,637 (92\%) of those labeled YSO and 4,653 (64\%) of those labeled AGB. Assuming the \citet{Robitaille08} classifications are accurate, and extrapolating to fainter objects, this could suggest that up to $\sim$35\% of our YSO candidates are misclassified.  However, of the 4,653 sources classified as YSOs by us but as AGB by \citet{Robitaille08}, 33\% are members of the spatial clusters from \autoref{sec:clustering}, suggesting that some of the objects they label AGB stars could be YSOs. 

As discussed in the introduction, a variety of strategies have been used to identify YSOs from IRAC colors. The SMOG field, which was published by \citet{Winston19} using a modified version of the \citet{Gutermuth2009} color selection rules, provides an excellent testbed for such a comparison. Comparison between our SMOG candidates (1524 objects) and theirs (4648 objects) shows that our selection methodology is more restrictive; 97\% of our YSO candidates were also classified as YSOs by \citet{Winston19}, while only 32\% of their YSO candidates were classified as YSOs by us.  \autoref{fig:smog} (right) shows that objects from their list that not included by us tend to be either objects with bluer colors or objects fainter than most of our sample. The difference in magnitude distributions -- ours peaking at $[4.5]=13$~mag and theirs peaking at $[4.5]=14.5$~mag -- may be a limitation related to our training set which was dominated by objects from the shallower GLIMPSE survey areas (\autoref{sec:data}). In spatial distribution (\autoref{fig:smog}, left) our candidate YSOs tend to be more spatially clustered than those from \citet{Winston19}.

The ``Star Formation in the Outer Galaxy'' \citep[SFOG;][]{Winston2020} YSO catalog was recently produced for the GLIMPSE~360 fields, observed during Spitzer's warm mission, meaning that Spitzer's 5.8 and 8.0~$\mu$m bands were unavailable. In Galactic coverage, this catalog is largely complementary to ours, but overlaps in the regions of SMOG and part of Cygnus~X. The catalogs also overlap in Vela in Galactic longitude, but cover different ranges of Galactic latitude. 

WISE covers similar wavelengths as Spitzer, but provides photometry for the whole sky. All-sky searches for YSOs in WISE data include \citet{Marton2016} and \citet{Marton19}, with the latter using cross-matches with Gaia. Below, we compare our catalog to the list of $\sim$130,000 candidate Class~I--II objects from \citet{Marton2016}; this paper also lists $>$600,000 candidate Class~III sources, but we do not include these in our comparison because Class~III sources are a minority in our catalog but make up the majority of the candidates from \citet{Marton2016}. Within the footprint of our catalog, \citet{Marton2016} identify $\sim$75,000 Class I--II WISE sources, whereas we identify $\sim$110,000 Class I--II IRAC sources. It is unsurprising that Spitzer can identify more YSOs in the Galactic midplane due to IRAC's higher spatial resolution and WISE's greater susceptibility to detector saturation from bright nebulosity. Using a $2^{\prime\prime}$ match radius, there are only $\sim$5000 sources in common between our catalog and theirs; visual inspection of the spatial distributions of the unmatched candidates reveal that we include more clustered YSO candidates (often more difficult to observe with WISE), while they include more spatially distributed candidates.

An effort to identify intermediate-mass young stars (e.g., Herbig Ae/Be stars) via machine learning was made by \citet{Vioque20}, who identify 8,470 candidates using public optical and infrared catalogs. Their list, focused on the higher end of the initial mass function, includes many fewer stars than our catalog; however, within the spatial overlap area, most of their candidates were re-selected by us. 

Another relevant catalog is provided by \citet{Kounkel2020} who identify groups of co-moving stars, including clusters, associations, moving groups, and stellar streams, in Gaia DR2 using a search radius of 3~kpc. Although these systems are not necessarily young, their catalog does include $\sim$35,000 members of groups with ages $<$10~Myr, many of which are located near the Galactic midplane. Only $\sim$300 objects are in common between our catalog and theirs, but this low fraction appears to be related to different selection biases, in particular their stringent Gaia quality cuts, which are only met by 4\% of our YSO candidates. The objects in common are mostly assigned to groups with ages from \citet{Kounkel2020} between 4--10~Myr; a handful of objects with older ages may result from either errors in their age estimates or contaminants in our catalog. Visual examination suggests that the catalogs reveal complementary aspects of stellar associations, with \citet{Kounkel2020} mostly selecting diskless members and us the disk/envelope-bearing members. In \autoref{fig:scoob1}, we show Sco~OB1 as an example where the combination of both lists provides a more complete picture of the association. 

\section{Conclusions}
\label{sec:conclusions}

We present a catalog of 117,446 candidate YSOs ($\sim$90,000 of which are new identifications) in the Galactic midplane from the GLIMPSE survey \citep{Benjamin03,Churchwell09} and extensions of this survey observed during Spitzer's cryogenic mission. We classify objects obtained from the GLIMPSE I, II, 3D, Vela-Carina \citep{Majewski07,Zasowski2009}, Cygnus~X \citep{Beerer10}, and SMOG \citep{Winston19} IRAC catalogs, using ancillary data from the near-IR 2MASS \citep{Skrutskie06}, UKIDSS \citep{Lawrence07}, and VVV \citep{Minniti10} surveys in the most comprehensive search for YSOs in the inner Galactic midplane to date. This catalog is largely restricted to the inner Galaxy and, thus, is complementary to YSO searches in the outer Galaxy \citep[e.g.,][]{Winston2020}. 

Classification of candidates was entirely based on near-IR and IRAC photometry, and our random forest diagnostics confirm that the IRAC bands were the most important for classification. The IRAC catalogs contain many sources not detected by MIPS or WISE because IRAC, particularly when processed by the GLIMPSE pipeline, is more sensitive in regions of the Galaxy with high crowding and nebulosity. By focusing on IRAC, we are able to identify tens-of-thousands of new YSO candidates that we would not have been able to if we required additional bands. Depending on the science application, future studies that use our YSO list may wish to augment our catalog with YSOs selected using other wavelengths. 

The spatial distribution of the candidates, as projected on the sky, is highly structured with cluster-like and filament-like patterns, but also includes a substantial non-clustered population. We have not used spatial information as an input to the classifier because the extent of spatial clustering of YSOs is still an open question and we wish to minimize the influence of selection effects on the observed spatial distributions. From the HDBSCAN algorithm, we identify $\sim$400 groups of YSOs and estimate their distances and proper motions from the mean astrometry of members detected by Gaia DR2. 

The YSOs we identify in the Galactic midplane are mostly at distances $\gtrsim$1~kpc. Some YSO groups appear associated with the Orion, Sagittarius-Carina, and the Scutum-Centaurus arms of the Galaxy, but do not appear to closely trace the estimated arm centers found by other methods \citep[e.g.,][]{Reid2019}. Near the boundary between Galactic Quadrants I and IV a large collection of YSO groups are located at the approximate distance of the Sagittarius-Carina Arm. However, these groups are not aligned parallel to the arm but, instead, form a chevron-like shape.

From the portion of our catalog visible to the ZTF survey, our YSO candidates tend to be more variable than field stars in the same region of the Galaxy. Nearly half the stars measured have statistically significant ZTF variability. Visual examination of the sources with the highest variability amplitudes suggests that most of them have light curve morphologies that resemble those expected for YSOs, with large dipping or bursting features. This dataset provides a useful testbed for future work on statistical classification of YSO light curves.

Although the properties of our sample, including optical/infrared photometry, spatial clustering, and variability, are consistent with most of the candidates being YSOs, the level of contamination is difficult to constrain without follow up observation. Although most objects are optically faint, the large total number of YSO candidates means that there are plenty of objects bright enough to follow up with optical spectroscopy. For example, $\sim$66,000 YSO candidates have $G<19$~mag, the faint limit for future large spectroscopic surveys such as WEAVE \citep{2014SPIE.9147E..0LD} or 4MOST \citep{2019Msngr.175....3D}, and more than 85\% of them are newly proposed in this paper. Furthermore, in the IR, $\sim$2000 YSO candidates have $H<11.5$~mag, bright  enough for an instrument like APOGEE \citep{Blanton2017,Cottle2018}, approximately half of which are newly proposed in this paper. Our candidates have been selected with nearly uniform methodology, so they should provide a useful statistical sample for further studies.  

\appendix

\section{GLIMPSE Flags}\label{sec:GLIMPSE_flags}
We visually examined a sample of IRAC images from crowded, nebulous regions of the Galaxy to investigate what GLIMPSE flags, including the ``close source flag'' and the number of detections in each band, imply about the reliability of source detection. The close neighbors for YSO candidates can usually be seen in the 3.6 micron image, but rarely in the 8.0 micron images, possibly because they are fainter at this wavelength, since most neighbors lack IR excess. In such cases, it appears that the GLIMPSE pipeline has correctly identified both sources. Nearly all sources with two or more detections in a band look like a {\it bona fide} point sources in that band's images; however, some of the sources detected only once look like point sources while others do not. Thus, having fewer than 2 detections is an indicator that a source could be less reliable. Nevertheless, we do not filter out sources based on either the ``close source flag'' or the number of detections in each band because any such cuts would remove numerous objects that appear to be good YSOs. These GLIMPSE flags are included in \autoref{tab:spicy} for any users of our YSO catalog who would like to make alternate choices for their own scientific applications. 

\section{Red Non-YSOs in the IRAC Catalogs}\label{sec:contaminants}

\subsection{Evolved Stars}

Certain evolved stars, including dusty red giants, AGB stars, post-AGB stars, and red super giants (RSGs), have red IR SEDs due to dusty stellar winds \citep[e.g.,][]{Marengo1997,Marengo1999,Groenewegen2012,Chun2015,Suh2020}, making such objects a significant category of contaminant in infrared YSO catalogs \citep[e.g.,][]{Robitaille08,Povich13}. \citet{Reiter2015} present a sample of AGB stars (including O-rich, S-rich, and C-rich stars) and RSG stars with $JHK$ and IRAC photometry. In the near-IR, the $J-K$ colors of this sample ($J-K\gtrsim 0.9$) are consistent with the group of probable contaminates on the $J$ vs.\ $J-K_s$ diagram that are brighter and redder than the typical YSOs (\autoref{fig:ir_cmds}, left panel). In IRAC color space, the distribution of the AGB+RGS sources partially overlap the distribution of YSO candidates.
However, most of them have IRAC colors (e.g., $[3.6]-[4.5] \lesssim 0.5$, $[4.5]-[8.0]  \lesssim 1$) bluer than the typical colors of YSOs, but similar in color to the bright sources classified as probable contaminants (e.g., \autoref{fig:ir_cmds}, right panel).

\subsection{Extragalactic Sources}

Extragalactic sources, including active and star-forming galaxies, can have mid-IR colors that mimic the IR excesses of YSOs \citep[e.g.,][]{Stern2005,Jarrett2011}. These sources can contribute a significant number of possible contaminants in some YSO searches \citep[e.g.,][]{Harvey2007,Gutermuth2008}. However, in the GLIMPSE survey, extragalactic contamination is expected to be lower, owing to shallower IRAC observations and high extinction near the Galactic midplane \citep[e.g.,][]{Kang2009}. \citet{Jarrett2011} provide a sample of IRAC sources in fields dominated by extragalactic sources. The galaxies in these fields tend to have $[3.6]-[4.5]\approx0$--1.25 and $[4.5]-[8.0]\approx1$--4. This distribution more closely resembles the distribution of probable contaminants in our sample than it does our YSO candidates (\autoref{fig:mir_ccds}, lower right). Their extragalactic sources mostly have $[3.6]\gtrsim14$~mag, which is fainter than most of our candidate YSOs.

\subsection{Labeled Non-YSOs in the Training Set}\label{sec:field}

For the random forest classifier, the sample of non-YSOs in our training set is equal in importance to the sample of YSOs.  As with the labeled YSOs, the labeled field objects are obtained from the set of NIR+IRAC sources that could not be fit by a reddened stellar photosphere (\autoref{sec:sed_reddened_photosphere}). Thus, they also represent objects with red IR colors.

We generate our list of ``non-YSOs'' using both sources within the boundaries of the MYStIX star-forming regions that were classified as non-YSOs and field stars in regions of the Galaxy where there is no star-formation activity. For the first category, we include all IRAC sources that lie within the MYStIX fields that show no indication of youth -- requiring both rejection as a YSO based on its SED and non detection of X-ray emission. 

We also include stars from rectangular regions of the sky in areas where there is no evidence for star-formation or the presence of YSOs. These regions, listed in Table~\ref{tab:rectangles}, have been chosen to sample stars along different Galactic latitudes and longitudes and along lines of sight with different levels of extinction. Given the high numbers of stars within these fields, a random subsample of these stars are included in the training data for the classifier.  Altogether, there are 14,019 labeled field objects for the 2MASS+IRAC sample, 5,320 for UKIDSS+IRAC, and 4,047 for VVV+IRAC. 

\section{Training Sets and Imputed Colors}\label{sec:training_ccds}
 
\begin{figure*}
	\centering
	\includegraphics[width=0.42\textwidth]{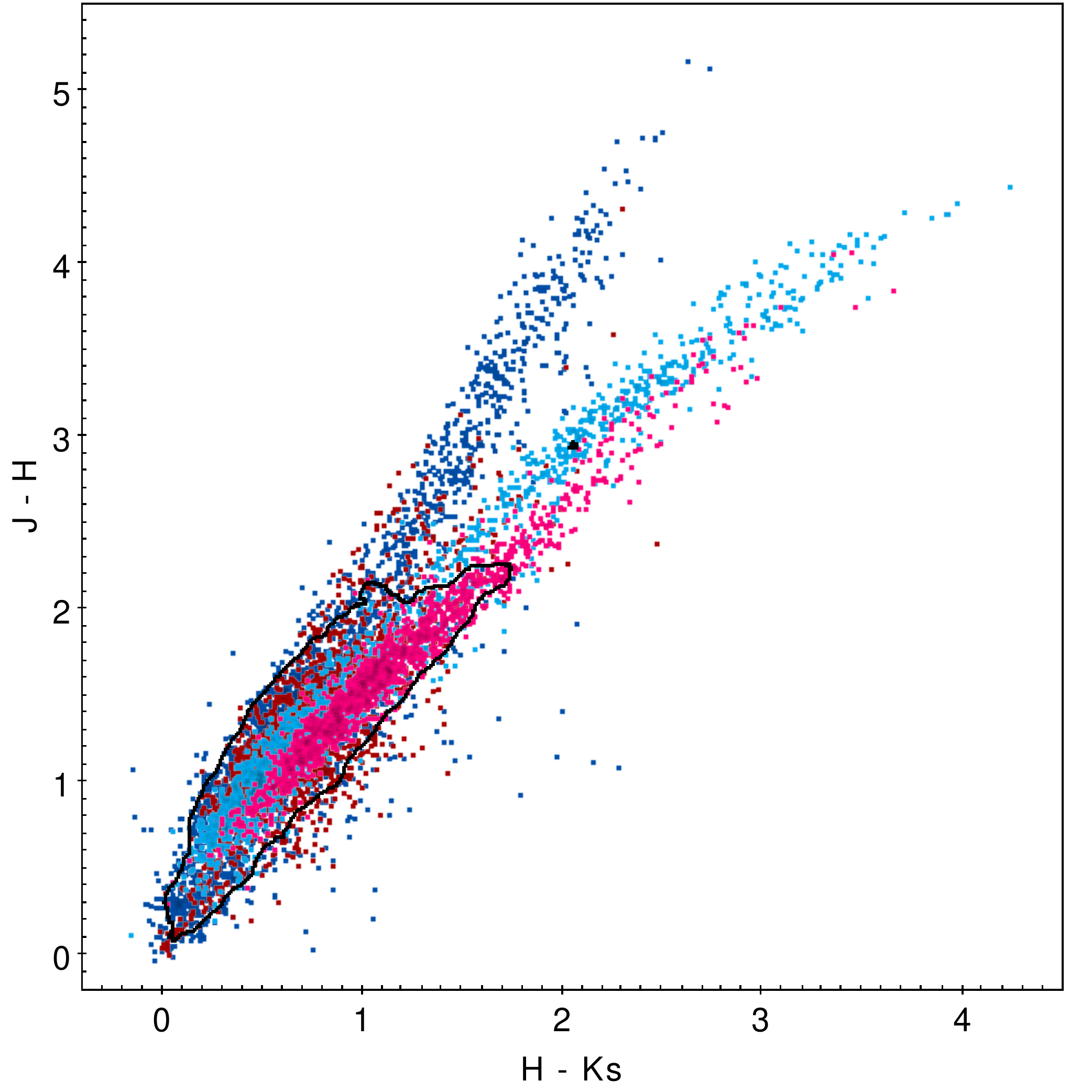}
	\includegraphics[width=0.42\textwidth]{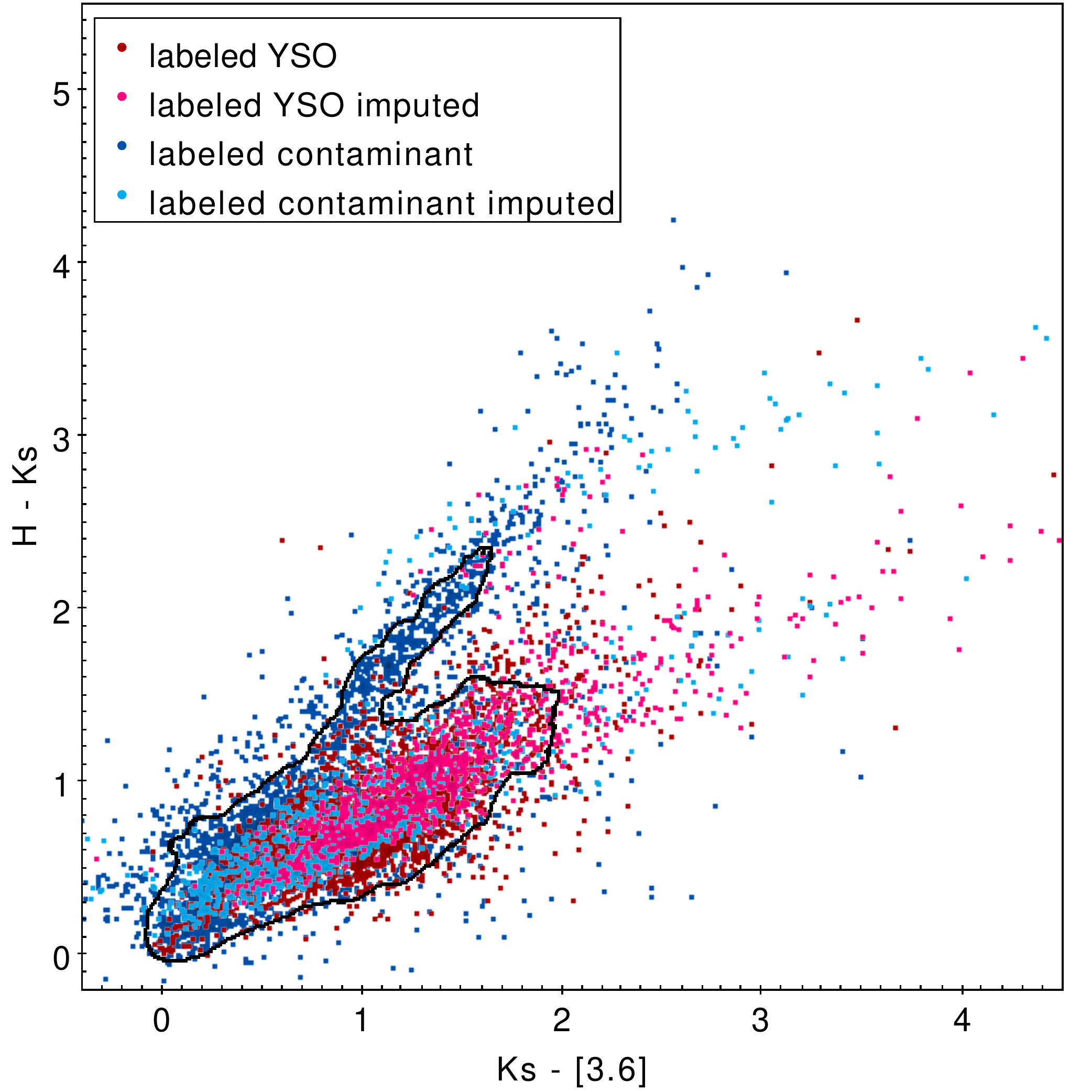}
	\includegraphics[width=0.42\textwidth]{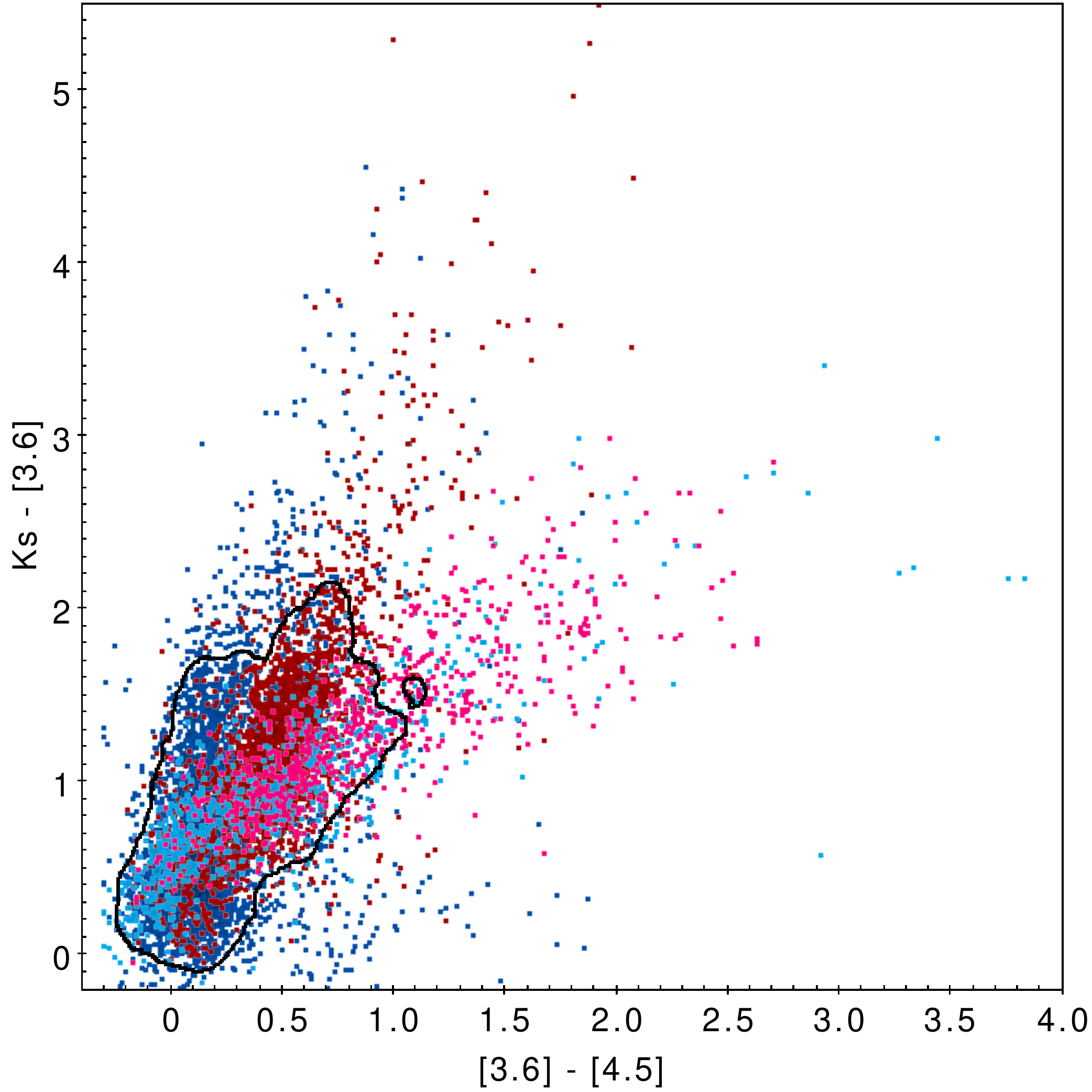}
	\includegraphics[width=0.42\textwidth]{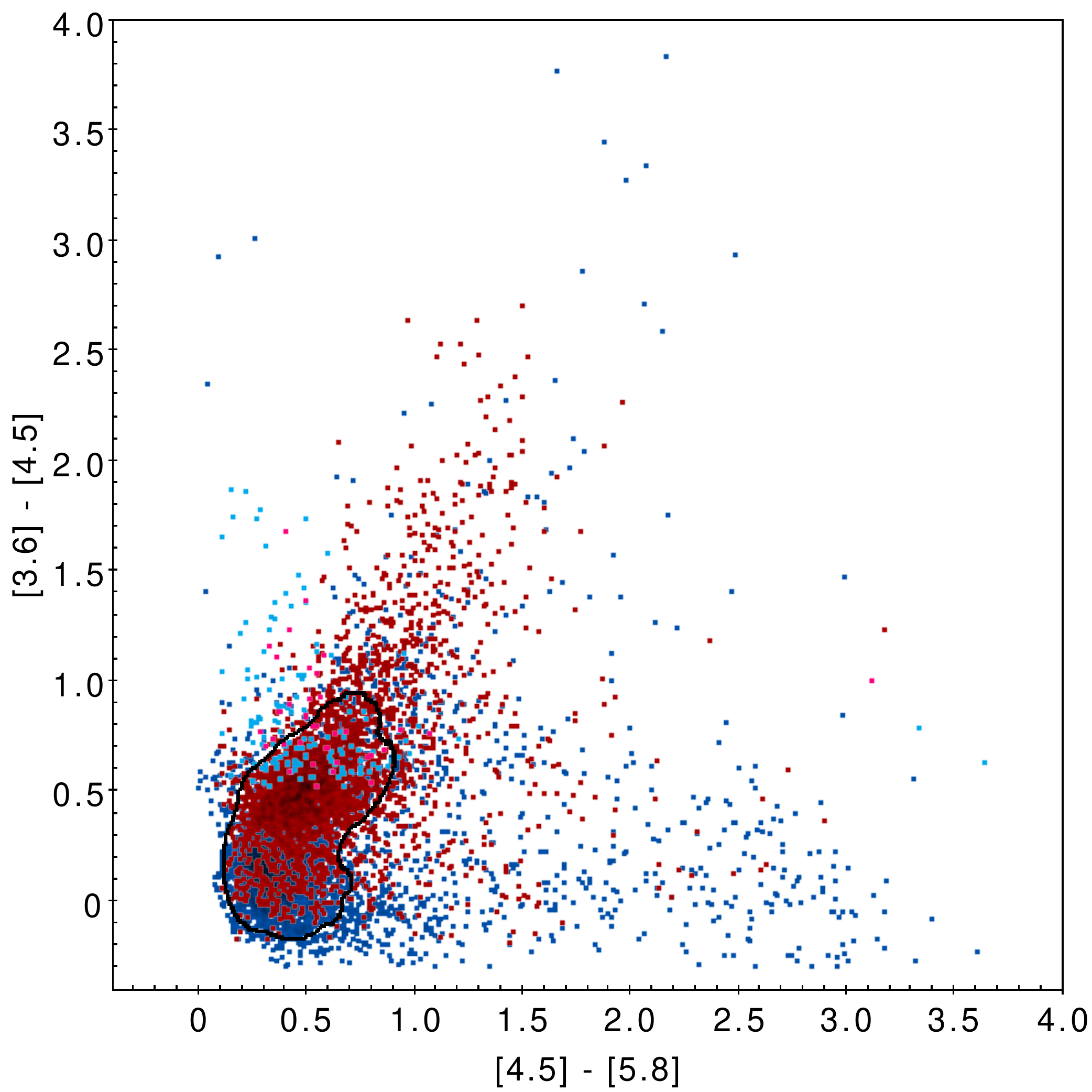}
	\includegraphics[width=0.42\textwidth]{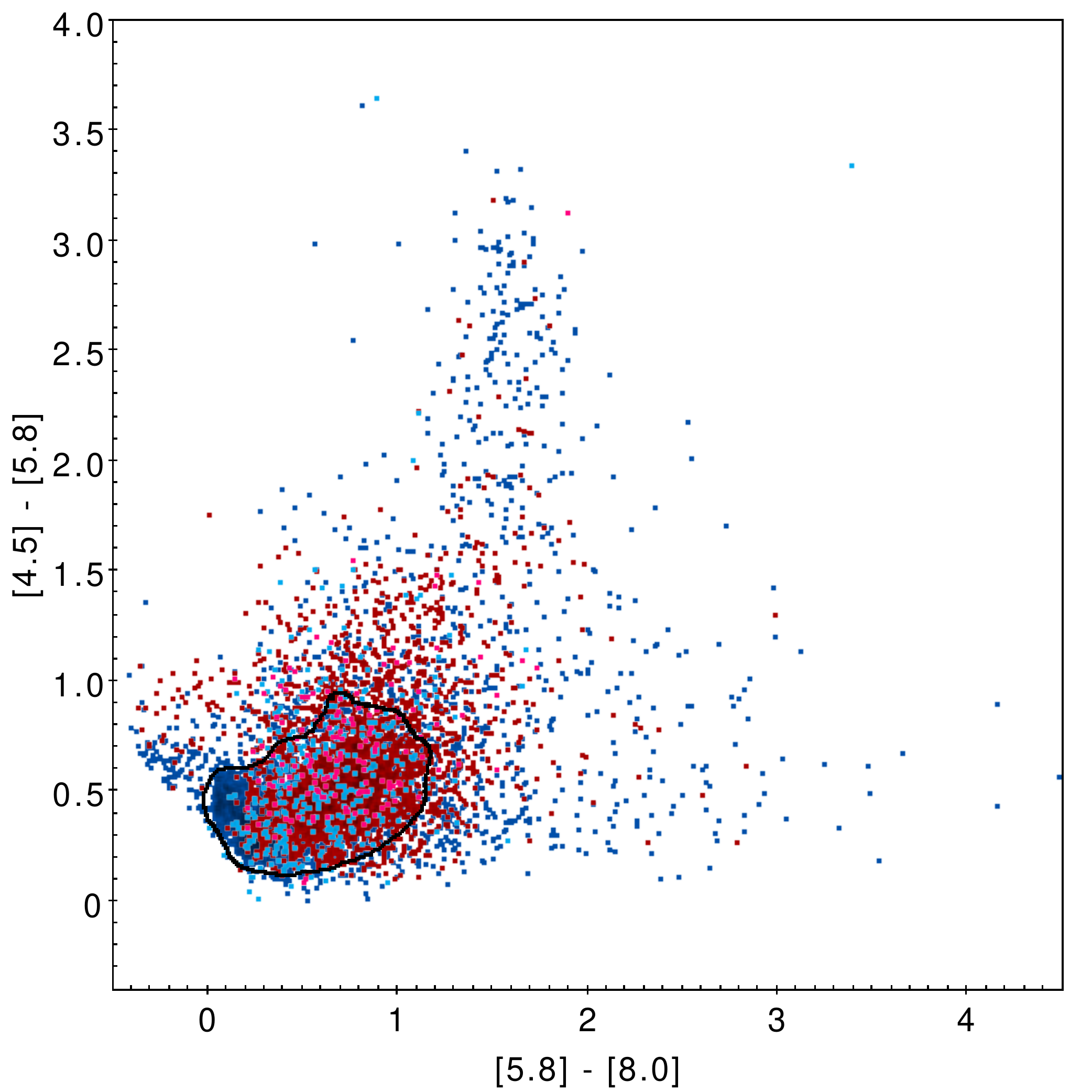}
	\includegraphics[width=0.42\textwidth]{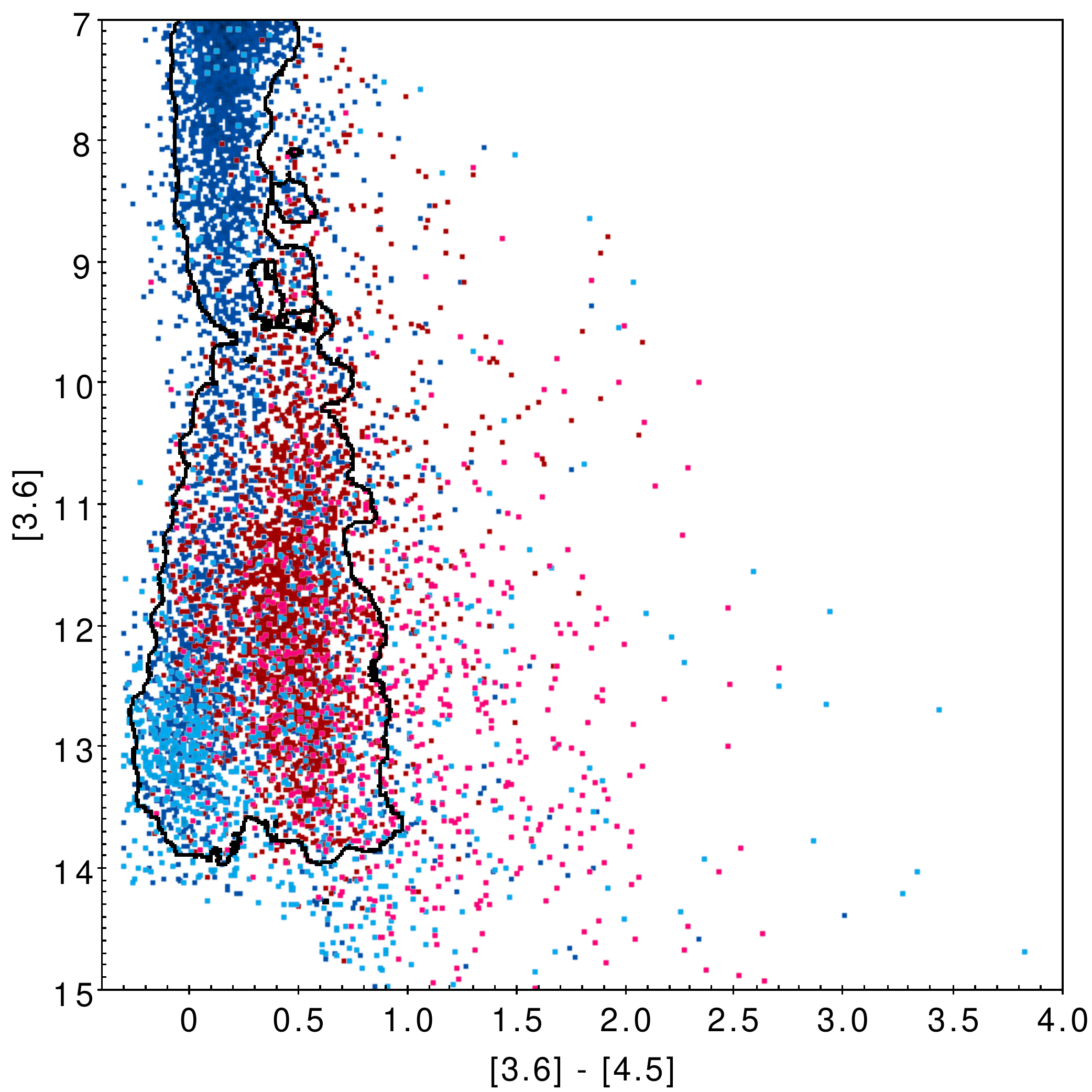}
    \caption{Scatter plots of labeled training data after copula imputation. Objects labeled ``YSO'' are indicated by reddish points and objects labeled ``field'' are indicated by bluish points. On each panel, the darker points have measurements of both colors, while points where one or both colors are imputed are marked with a lighter color as indicated by the legend.}
  \label{fig:training}
\end{figure*}
 
\autoref{fig:training} shows color-color/magnitude diagrams for the training set, containing both labeled YSOs (reddish points) and non-YSOs (bluish points) and observed (dark points) and imputed (light points) colors. Here, we show only the 2MASS+IRAC training data, but the general morphologies of the distributions are similar for target dataset, as well as for the training/target UKIDSS+IRAC and VVV+IRAC datasets. For each NIR+IRAC combination, an identical copula was used for every data point regardless of label (YSO or non-YSO) or whether the data point belongs to the training or target set. Thus, any differences that emerge in the distributions of imputed data must emerge from the data itself. 

In general the distribution of the imputed data lies within the distributions traced out by the observed data, with the $J-H$ vs.\ $H-K_s$ diagram being the main exception. In the $JHK_s$ diagram, many sources are missing the $J-H$ color (presumably due to high extinction), and follow a locus with a slightly flatter slope than the objects for which both $J-H$ and $H-K_s$ have been measured. This behavior also appears when using UKIDSS or VVV $JHK$ photometry, and it may suggest that there is an intrinsic difference in distributions for sources with and without measured $J-H$. The sources without measured $J-H$ also tend to have higher than average $H-K_s$ values, possibly influencing how the distribution of the imputed sources. Nevertheless, the analysis of the random forest classifier suggests that $J-H$ color is one of the less important features for producing a classification.
 
None of the other diagrams reveal such large deviations between observed and imputed data, but differences are visible in the distributions of YSOs and non-YSOs. On some diagrams there are regions with no imputed data (e.g., the lower left of the $[3.6]-[4.5]$ vs.\ $[4.5]-[5.8]$ diagram) because any missing data would have meant that the source would not be under consideration (see \autoref{sec:sed_reddened_photosphere}). 

In the target set, there is a low number of outliers in regions of color space that are not well populated by sources from the training set, either by sources labeled ``YSO'' or ``non-YSO,'' meaning that our classifier would not be able to generate reliable classifications for these objects. These are either rare objects that arise due to the large size ($\sim5\times 10^7$ sources) of the full IRAC photometric catalogs
or are sources unlikely to have infrared excess but that we failed to remove with our procedures in \autoref{sec:sed_reddened_photosphere}. Given that we have no basis for classifying these objects with a RF, we are cautious and exclude them from our lists of candidate YSOs. We use the following criteria to define these outliers: $[3.6]-[4.5] < -0.3$, $[4.5]-[5.8]< 0$, $[4.5]-[8.0]<0$, $[3.6]-[5.8]<0$, $[3.6]-[5.8]>6$, $[4.5]-[5.8]<0$, or $[3.6]-[8.0]<0$. 
 
\acknowledgements
We thank Robert Benjamin for useful discussions about GLIMPSE and spiral structure of the Galaxy, and Philip Lucas and Leigh Smith for assistance with the UKIDSS and VVV catalogs. This work is a result of the $\rm 6^{th}$ COIN Residence Program (CRP\#6; \url{https://cosmostatistics-initiative.org/residence-programs/crp6}) held in Chamonix, France in August 2019. COIN is financially supported by CNRS as part of its MOMENTUM programme over the 2018--2020 period. This work is based on observations made with the Spitzer Space Telescope, which is operated by the Jet Propulsion Laboratory, California Institute of Technology under a contract with NASA. 
This work has also made use of data from the European Space Agency mission Gaia, processed by the Gaia Data Processing and Analysis Consortium. Funding for the DPAC has been provided by national institutions, in particular the institutions participating in the Gaia Multilateral Agreement. This work is also based in part on observations obtained with the Samuel Oschin 48-inch Telescope at the Palomar Observatory as part of the Zwicky Transient Facility project. ZTF is supported by the National Science Foundation under Grant No.\ AST-1440341 and a collaboration including Caltech, IPAC, the Weizmann Institute for Science, the Oskar Klein Center at Stockholm University, the University of Maryland, the University of Washington, Deutsches Elektronen-Synchrotron and Humboldt University, Los Alamos National Laboratories, the TANGO Consortium of Taiwan, the University of Wisconsin at Milwaukee, and Lawrence Berkeley National Laboratories. This research has made use of the NASA/IPAC Infrared Science Archive, which is funded by the National Aeronautics and Space Administration and operated by the California Institute of Technology. AKM acknowledges the support from the Portuguese Funda\c c\~ao para a Ci\^encia e a Tecnologia (FCT) through grants SFRH/BPD/74697/2010, PTDC/FIS-AST/31546/2017 and from the Portuguese Strategic Programme UID/FIS/00099/2013 for CENTRA. M.A.K. acknowledges support from the Chandra grant GO9-20002X for analysis of data related to the Trifid Nebula.

\facility{2MASS, Gaia, Spitzer (IRAC, MIPS), UKIRT, VISTA/VIRCAM, WISE, ZTF, IRSA} 

\software{
          caret \citep{caret},
          mclust \citep{mclust},
          hdbscan \citep{McInnes2017},
          modeest \citep{modeest2019},
          mclust \citep{mclust},
          PostgreSQL \citep{PostgreSQL},
          Python,
          R \citep{rcore19}, 
          rjags \citep{plummer2017jags,plummer2019},
          SAOImage DS9 \citep{2003ASPC..295..489J},
          sbgcop \citep{sbgcop},
          TOPCAT \& STILTS \citep{Taylor2005}
          }

\clearpage
\bibliography{main.bbl}

\begin{thebibliography}{}
\expandafter\ifx\csname natexlab\endcsname\relax\def\natexlab#1{#1}\fi
\providecommand{\url}[1]{\href{#1}{#1}}
\providecommand{\dodoi}[1]{doi:~\href{http://doi.org/#1}{\nolinkurl{#1}}}
\providecommand{\doeprint}[1]{\href{http://ascl.net/#1}{\nolinkurl{http://ascl.net/#1}}}
\providecommand{\doarXiv}[1]{\href{https://arxiv.org/abs/#1}{\nolinkurl{https://arxiv.org/abs/#1}}}

\bibitem[{{Allen} {et~al.}(2007){Allen}, {Megeath}, {Gutermuth}, {Myers},
  {Wolk}, {Adams}, {Muzerolle}, {Young}, \& {Pipher}}]{Allen2007}
{Allen}, L., {Megeath}, S.~T., {Gutermuth}, R., {et~al.} 2007, in Protostars
  and Planets V, ed. B.~{Reipurth}, D.~{Jewitt}, \& K.~{Keil} (Tucson:
  University of Arizona Press), 361.
\newblock \doarXiv{astro-ph/0603096}

\bibitem[{{Allen} {et~al.}(2004){Allen}, {Calvet}, {D'Alessio}, {Merin},
  {Hartmann}, {Megeath}, {Gutermuth}, {Muzerolle}, {Pipher}, {Myers}, \&
  {Fazio}}]{Allen2004}
{Allen}, L.~E., {Calvet}, N., {D'Alessio}, P., {et~al.} 2004, \apjs, 154, 363,
  \dodoi{10.1086/422715}

\bibitem[{{Alves} {et~al.}(2020){Alves}, {Zucker}, {Goodman}, {Speagle},
  {Meingast}, {Robitaille}, {Finkbeiner}, {Schlafly}, \& {Green}}]{Alves2020}
{Alves}, J., {Zucker}, C., {Goodman}, A.~A., {et~al.} 2020, \nat, 578, 237,
  \dodoi{10.1038/s41586-019-1874-z}

\bibitem[{{Anderson} {et~al.}(2014){Anderson}, {Bania}, {Balser}, {Cunningham},
  {Wenger}, {Johnstone}, \& {Armentrout}}]{Anderson2014}
{Anderson}, L.~D., {Bania}, T.~M., {Balser}, D.~S., {et~al.} 2014, \apjs, 212,
  1, \dodoi{10.1088/0067-0049/212/1/1}

\bibitem[{{Andre} \& {Montmerle}(1994)}]{Andre1994}
{Andre}, P., \& {Montmerle}, T. 1994, \apj, 420, 837, \dodoi{10.1086/173608}

\bibitem[{{Andreani} {et~al.}(2018){Andreani}, {Boselli}, {Ciesla}, {Vio},
  {Cortese}, {Buat}, \& {Miyamoto}}]{Andreani2018}
{Andreani}, P., {Boselli}, A., {Ciesla}, L., {et~al.} 2018, \aap, 617, A33,
  \dodoi{10.1051/0004-6361/201832873}

\bibitem[{Ascenso(2018)}]{Ascenso2018}
Ascenso, J. 2018, in The Birth of Star Clusters, ed. S.~Stahler (Cham: Springer
  International Publishing), 1--37, \dodoi{10.1007/978-3-319-22801-3_1}

\bibitem[{Baddeley(2017)}]{Baddeley2017}
Baddeley, A. 2017, Spatial Statistics, 22, 261 ,
  \dodoi{https://doi.org/10.1016/j.spasta.2017.03.001}

\bibitem[{{Beerer} {et~al.}(2010){Beerer}, {Koenig}, {Hora}, {Gutermuth},
  {Bontemps}, {Megeath}, {Schneider}, {Motte}, {Carey}, {Simon}, {Keto},
  {Smith}, {Allen}, {Fazio}, {Kraemer}, {Price}, {Mizuno}, {Adams},
  {Hern{\'a}ndez}, \& {Lucas}}]{Beerer10}
{Beerer}, I.~M., {Koenig}, X.~P., {Hora}, J.~L., {et~al.} 2010, \apj, 720, 679,
  \dodoi{10.1088/0004-637X/720/1/679}

\bibitem[{{Bellm} {et~al.}(2019){Bellm}, {Kulkarni}, {Graham}, {Dekany},
  {Smith}, {Riddle}, {Masci}, {Helou}, {Prince}, {Adams}, {Barbarino},
  {Barlow}, {Bauer}, {Beck}, {Belicki}, {Biswas}, {Blagorodnova}, {Bodewits},
  {Bolin}, {Brinnel}, {Brooke}, {Bue}, {Bulla}, {Burruss}, {Cenko}, {Chang},
  {Connolly}, {Coughlin}, {Cromer}, {Cunningham}, {De}, {Delacroix}, {Desai},
  {Duev}, {Eadie}, {Farnham}, {Feeney}, {Feindt}, {Flynn}, {Franckowiak},
  {Frederick}, {Fremling}, {Gal-Yam}, {Gezari}, {Giomi}, {Goldstein},
  {Golkhou}, {Goobar}, {Groom}, {Hacopians}, {Hale}, {Henning}, {Ho}, {Hover},
  {Howell}, {Hung}, {Huppenkothen}, {Imel}, {Ip}, {Ivezi{\'c}}, {Jackson},
  {Jones}, {Juric}, {Kasliwal}, {Kaspi}, {Kaye}, {Kelley}, {Kowalski},
  {Kramer}, {Kupfer}, {Landry}, {Laher}, {Lee}, {Lin}, {Lin}, {Lunnan},
  {Giomi}, {Mahabal}, {Mao}, {Miller}, {Monkewitz}, {Murphy}, {Ngeow},
  {Nordin}, {Nugent}, {Ofek}, {Patterson}, {Penprase}, {Porter}, {Rauch},
  {Rebbapragada}, {Reiley}, {Rigault}, {Rodriguez}, {van Roestel}, {Rusholme},
  {van Santen}, {Schulze}, {Shupe}, {Singer}, {Soumagnac}, {Stein}, {Surace},
  {Sollerman}, {Szkody}, {Taddia}, {Terek}, {Van Sistine}, {van Velzen},
  {Vestrand}, {Walters}, {Ward}, {Ye}, {Yu}, {Yan}, \& {Zolkower}}]{Bellm2019}
{Bellm}, E.~C., {Kulkarni}, S.~R., {Graham}, M.~J., {et~al.} 2019, \pasp, 131,
  018002, \dodoi{10.1088/1538-3873/aaecbe}

\bibitem[{{Benjamin} {et~al.}(2003){Benjamin}, {Churchwell}, {Babler}, {Bania},
  {Clemens}, {Cohen}, {Dickey}, {Indebetouw}, {Jackson}, {Kobulnicky},
  {Lazarian}, {Marston}, {Mathis}, {Meade}, {Seager}, {Stolovy}, {Watson},
  {Whitney}, {Wolff}, \& {Wolfire}}]{Benjamin03}
{Benjamin}, R.~A., {Churchwell}, E., {Babler}, B.~L., {et~al.} 2003, \pasp,
  115, 953, \dodoi{10.1086/376696}

\bibitem[{{Blanton} {et~al.}(2017){Blanton}, {Bershady}, {Abolfathi},
  {Albareti}, {Allende Prieto}, {Almeida}, {Alonso-Garc{\'\i}a}, {Anders},
  {Anderson}, {Andrews}, {Aquino-Ort{\'\i}z}, {Arag{\'o}n-Salamanca},
  {Argudo-Fern{\'a}ndez}, {Armengaud}, {Aubourg}, {Avila-Reese}, {Badenes},
  {Bailey}, {Barger}, {Barrera-Ballesteros}, {Bartosz}, {Bates}, {Baumgarten},
  {Bautista}, {Beaton}, {Beers}, {Belfiore}, {Bender}, {Berlind}, {Bernardi},
  {Beutler}, {Bird}, {Bizyaev}, {Blanc}, {Blomqvist}, {Bolton}, {Boquien},
  {Borissova}, {van den Bosch}, {Bovy}, {Brandt}, {Brinkmann}, {Brownstein},
  {Bundy}, {Burgasser}, {Burtin}, {Busca}, {Cappellari}, {Delgado Carigi},
  {Carlberg}, {Carnero Rosell}, {Carrera}, {Chanover}, {Cherinka}, {Cheung},
  {G{\'o}mez Maqueo Chew}, {Chiappini}, {Choi}, {Chojnowski}, {Chuang},
  {Chung}, {Cirolini}, {Clerc}, {Cohen}, {Comparat}, {da Costa}, {Cousinou},
  {Covey}, {Crane}, {Croft}, {Cruz-Gonzalez}, {Garrido Cuadra}, {Cunha},
  {Damke}, {Darling}, {Davies}, {Dawson}, {de la Macorra}, {Dell'Agli}, {De
  Lee}, {Delubac}, {Di Mille}, {Diamond-Stanic}, {Cano-D{\'\i}az}, {Donor},
  {Downes}, {Drory}, {du Mas des Bourboux}, {Duckworth}, {Dwelly}, {Dyer},
  {Ebelke}, {Eigenbrot}, {Eisenstein}, {Emsellem}, {Eracleous}, {Escoffier},
  {Evans}, {Fan}, {Fern{\'a}ndez-Alvar}, {Fernandez-Trincado}, {Feuillet},
  {Finoguenov}, {Fleming}, {Font-Ribera}, {Fredrickson}, {Freischlad},
  {Frinchaboy}, {Fuentes}, {Galbany}, {Garcia-Dias},
  {Garc{\'\i}a-Hern{\'a}ndez}, {Gaulme}, {Geisler}, {Gelfand},
  {Gil-Mar{\'\i}n}, {Gillespie}, {Goddard}, {Gonzalez-Perez}, {Grabowski},
  {Green}, {Grier}, {Gunn}, {Guo}, {Guy}, {Hagen}, {Hahn}, {Hall}, {Harding},
  {Hasselquist}, {Hawley}, {Hearty}, {Gonzalez Hern{\'a}ndez}, {Ho}, {Hogg},
  {Holley-Bockelmann}, {Holtzman}, {Holzer}, {Huehnerhoff}, {Hutchinson},
  {Hwang}, {Ibarra-Medel}, {da Silva Ilha}, {Ivans}, {Ivory}, {Jackson},
  {Jensen}, {Johnson}, {Jones}, {J{\"o}nsson}, {Jullo}, {Kamble}, {Kinemuchi},
  {Kirkby}, {Kitaura}, {Klaene}, {Knapp}, {Kneib}, {Kollmeier}, {Lacerna},
  {Lane}, {Lang}, {Law}, {Lazarz}, {Lee}, {Le Goff}, {Liang}, {Li}, {Li},
  {Lian}, {Lima}, {Lin}, {Lin}, {Bertran de Lis}, {Liu}, {de Icaza Lizaola},
  {Long}, {Lucatello}, {Lundgren}, {MacDonald}, {Deconto Machado}, {MacLeod},
  {Mahadevan}, {Geimba Maia}, {Maiolino}, {Majewski}, {Malanushenko},
  {Malanushenko}, {Manchado}, {Mao}, {Maraston}, {Marques-Chaves}, {Masseron},
  {Masters}, {McBride}, {McDermid}, {McGrath}, {McGreer}, {Medina Pe{\~n}a},
  {Melendez}, {Merloni}, {Merrifield}, {Meszaros}, {Meza}, {Minchev},
  {Minniti}, {Miyaji}, {More}, {Mulchaey}, {M{\"u}ller-S{\'a}nchez}, {Muna},
  {Munoz}, {Myers}, {Nair}, {Nandra}, {Correa do Nascimento}, {Negrete},
  {Ness}, {Newman}, {Nichol}, {Nidever}, {Nitschelm}, {Ntelis}, {O'Connell},
  {Oelkers}, {Oravetz}, {Oravetz}, {Pace}, {Padilla}, {Palanque-Delabrouille},
  {Alonso Palicio}, {Pan}, {Parejko}, {Parikh}, {P{\^a}ris}, {Park}, {Patten},
  {Peirani}, {Pellejero-Ibanez}, {Penny}, {Percival}, {Perez-Fournon},
  {Petitjean}, {Pieri}, {Pinsonneault}, {Pisani}, {Poleski}, {Prada},
  {Prakash}, {Queiroz}, {Raddick}, {Raichoor}, {Barboza Rembold}, {Richstein},
  {Riffel}, {Riffel}, {Rix}, {Robin}, {Rockosi}, {Rodr{\'\i}guez-Torres},
  {Roman-Lopes}, {Rom{\'a}n-Z{\'u}{\~n}iga}, {Rosado}, {Ross}, {Rossi}, {Ruan},
  {Ruggeri}, {Rykoff}, {Salazar-Albornoz}, {Salvato}, {S{\'a}nchez}, {Aguado},
  {S{\'a}nchez-Gallego}, {Santana}, {Santiago}, {Sayres}, {Schiavon}, {da Silva
  Schimoia}, {Schlafly}, {Schlegel}, {Schneider}, {Schultheis}, {Schuster},
  {Schwope}, {Seo}, {Shao}, {Shen}, {Shetrone}, {Shull}, {Simon}, {Skinner},
  {Skrutskie}, {Slosar}, {Smith}, {Sobeck}, {Sobreira}, {Somers}, {Souto},
  {Stark}, {Stassun}, {Stauffer}, {Steinmetz}, {Storchi-Bergmann},
  {Streblyanska}, {Stringfellow}, {Su{\'a}rez}, {Sun}, {Suzuki}, {Szigeti},
  {Taghizadeh-Popp}, {Tang}, {Tao}, {Tayar}, {Tembe}, {Teske}, {Thakar},
  {Thomas}, {Thompson}, {Tinker}, {Tissera}, {Tojeiro}, {Hernandez Toledo}, {de
  la Torre}, {Tremonti}, {Troup}, {Valenzuela}, {Martinez Valpuesta},
  {Vargas-Gonz{\'a}lez}, {Vargas-Maga{\~n}a}, {Vazquez}, {Villanova}, {Vivek},
  {Vogt}, {Wake}, {Walterbos}, {Wang}, {Weaver}, {Weijmans}, {Weinberg},
  {Westfall}, {Whelan}, {Wild}, {Wilson}, {Wood-Vasey}, {Wylezalek}, {Xiao},
  {Yan}, {Yang}, {Ybarra}, {Y{\`e}che}, {Zakamska}, {Zamora}, {Zarrouk},
  {Zasowski}, {Zhang}, {Zhao}, {Zheng}, {Zheng}, {Zhou}, {Zhou}, {Zhu},
  {Zoccali}, \& {Zou}}]{Blanton2017}
{Blanton}, M.~R., {Bershady}, M.~A., {Abolfathi}, B., {et~al.} 2017, \aj, 154,
  28, \dodoi{10.3847/1538-3881/aa7567}

\bibitem[{{Bovy}(2017)}]{Bovy2017}
{Bovy}, J. 2017, \mnras, 468, L63, \dodoi{10.1093/mnrasl/slx027}

\bibitem[{Breiman(2001)}]{breiman2001random}
Breiman, L. 2001, Machine Learning, 45, 5, \dodoi{10.1023/A:1010933404324}

\bibitem[{{Bressan} {et~al.}(2012){Bressan}, {Marigo}, {Girardi}, {Salasnich},
  {Dal Cero}, {Rubele}, \& {Nanni}}]{Bressan2012}
{Bressan}, A., {Marigo}, P., {Girardi}, L., {et~al.} 2012, \mnras, 427, 127,
  \dodoi{10.1111/j.1365-2966.2012.21948.x}

\bibitem[{{Bressert} {et~al.}(2010){Bressert}, {Bastian}, {Gutermuth},
  {Megeath}, {Allen}, {Evans}, {Rebull}, {Hatchell}, {Johnstone}, {Bourke},
  {Cieza}, {Harvey}, {Merin}, {Ray}, \& {Tothill}}]{Bressert2010}
{Bressert}, E., {Bastian}, N., {Gutermuth}, R., {et~al.} 2010, \mnras, 409,
  L54, \dodoi{10.1111/j.1745-3933.2010.00946.x}

\bibitem[{{Buckner} {et~al.}(2020){Buckner}, {Khorrami}, {Gonz{\'a}lez},
  {Lumsden}, {Moraux}, {Oudmaijer}, {Clark}, {Joncour}, {Blanco}, {de la
  Calle}, {Hacar}, {Herrera-Fernandez}, {Motte}, {Salgado}, \&
  {Valero-Mart{\'\i}n}}]{Buckner2020}
{Buckner}, A. S.~M., {Khorrami}, Z., {Gonz{\'a}lez}, M., {et~al.} 2020, \aap,
  636, A80, \dodoi{10.1051/0004-6361/201936935}

\bibitem[{{Bufano} {et~al.}(2018){Bufano}, {Leto}, {Carey}, {Umana}, {Buemi},
  {Ingallinera}, {Bulpitt}, {Cavallaro}, {Riggi}, {Trigilio}, \&
  {Molinari}}]{Bufano2018}
{Bufano}, F., {Leto}, P., {Carey}, D., {et~al.} 2018, \mnras, 473, 3671,
  \dodoi{10.1093/mnras/stx2560}

\bibitem[{Campello {et~al.}(2013)Campello, Moulavi, \& Sander}]{Campello2013}
Campello, R.~J., Moulavi, D., \& Sander, J. 2013, in Pacific-Asia conference on
  knowledge discovery and data mining, Springer, 160--172

\bibitem[{Cand\`es \& Donoho(2000)}]{Candes00curvelets}
Cand\`es, E.~J., \& Donoho, D.~L. 2000, Curvelets -- A Surprisingly Effective
  Nonadaptive Representation For Objects with Edges (TN, Nashville: Vanderbilt
  Univ. Press), 1--10

\bibitem[{{Cardelli} {et~al.}(1989){Cardelli}, {Clayton}, \&
  {Mathis}}]{Cardelli1989}
{Cardelli}, J.~A., {Clayton}, G.~C., \& {Mathis}, J.~S. 1989, \apj, 345, 245,
  \dodoi{10.1086/167900}

\bibitem[{{Carey} {et~al.}(2009){Carey}, {Noriega-Crespo}, {Mizuno}, {Shenoy},
  {Paladini}, {Kraemer}, {Price}, {Flagey}, {Ryan}, {Ingalls}, {Kuchar},
  {Pinheiro Gon{\c{c}}alves}, {Indebetouw}, {Billot}, {Marleau}, {Padgett},
  {Rebull}, {Bressert}, {Ali}, {Molinari}, {Martin}, {Berriman}, {Boulanger},
  {Latter}, {Miville-Deschenes}, {Shipman}, \& {Testi}}]{Carey2009}
{Carey}, S.~J., {Noriega-Crespo}, A., {Mizuno}, D.~R., {et~al.} 2009, \pasp,
  121, 76, \dodoi{10.1086/596581}

\bibitem[{{Carpenter}(2000{\natexlab{a}})}]{Carpenter2000}
{Carpenter}, J.~M. 2000{\natexlab{a}}, \aj, 120, 3139, \dodoi{10.1086/316845}

\bibitem[{{Carpenter}(2000{\natexlab{b}})}]{Carpenter00}
---. 2000{\natexlab{b}}, \aj, 120, 3139, \dodoi{10.1086/316845}

\bibitem[{{Castelli} \& {Kurucz}(2003)}]{CastelliKurucz2004}
{Castelli}, F., \& {Kurucz}, R.~L. 2003, in IAU Symposium, Vol. 210, Modelling
  of Stellar Atmospheres, ed. N.~{Piskunov}, W.~W. {Weiss}, \& D.~F. {Gray},
  A20.
\newblock \doarXiv{astro-ph/0405087}

\bibitem[{{Castro-Ginard} {et~al.}(2020){Castro-Ginard}, {Jordi}, {Luri},
  {{\'A}lvarez Cid-Fuentes}, {Casamiquela}, {Anders}, {Cantat-Gaudin},
  {Mongui{\'o}}, {Balaguer-N{\'u}{\~n}ez}, {Sol{\`a}}, \&
  {Badia}}]{2020A&A...635A..45C}
{Castro-Ginard}, A., {Jordi}, C., {Luri}, X., {et~al.} 2020, \aap, 635, A45,
  \dodoi{10.1051/0004-6361/201937386}

\bibitem[{{Chiar} {et~al.}(2007){Chiar}, {Ennico}, {Pendleton}, {Boogert},
  {Greene}, {Knez}, {Lada}, {Roellig}, {Tielens}, {Werner}, \&
  {Whittet}}]{Chiar2007}
{Chiar}, J.~E., {Ennico}, K., {Pendleton}, Y.~J., {et~al.} 2007, \apjl, 666,
  L73, \dodoi{10.1086/521789}

\bibitem[{{Chun} {et~al.}(2015){Chun}, {Jung}, {Kang}, {Kim}, \&
  {Sohn}}]{Chun2015}
{Chun}, S.-H., {Jung}, M., {Kang}, M., {Kim}, J.-W., \& {Sohn}, Y.-J. 2015,
  \aap, 578, A51, \dodoi{10.1051/0004-6361/201525849}

\bibitem[{{Churchwell} {et~al.}(2006){Churchwell}, {Povich}, {Allen}, {Taylor},
  {Meade}, {Babler}, {Indebetouw}, {Watson}, {Whitney}, {Wolfire}, {Bania},
  {Benjamin}, {Clemens}, {Cohen}, {Cyganowski}, {Jackson}, {Kobulnicky},
  {Mathis}, {Mercer}, {Stolovy}, {Uzpen}, {Watson}, \&
  {Wolff}}]{Churchwell2006}
{Churchwell}, E., {Povich}, M.~S., {Allen}, D., {et~al.} 2006, \apj, 649, 759,
  \dodoi{10.1086/507015}

\bibitem[{{Churchwell} {et~al.}(2007){Churchwell}, {Watson}, {Povich},
  {Taylor}, {Babler}, {Meade}, {Benjamin}, {Indebetouw}, \&
  {Whitney}}]{Churchwell2007}
{Churchwell}, E., {Watson}, D.~F., {Povich}, M.~S., {et~al.} 2007, \apj, 670,
  428, \dodoi{10.1086/521646}

\bibitem[{{Churchwell} {et~al.}(2009){Churchwell}, {Babler}, {Meade},
  {Whitney}, {Benjamin}, {Indebetouw}, {Cyganowski}, {Robitaille}, {Povich},
  {Watson}, \& {Bracker}}]{Churchwell09}
{Churchwell}, E., {Babler}, B.~L., {Meade}, M.~R., {et~al.} 2009, \pasp, 121,
  213, \dodoi{10.1086/597811}

\bibitem[{{Cody} \& {Hillenbrand}(2018)}]{Cody2018}
{Cody}, A.~M., \& {Hillenbrand}, L.~A. 2018, \aj, 156, 71,
  \dodoi{10.3847/1538-3881/aacead}

\bibitem[{{Contreras Pe{\~n}a} {et~al.}(2017){Contreras Pe{\~n}a}, {Lucas},
  {Minniti}, {Kurtev}, {Stimson}, {Navarro Molina}, {Borissova}, {Kumar},
  {Thompson}, {Gledhill}, {Terzi}, {Froebrich}, \& {Caratti o
  Garatti}}]{ContrerasPena2017}
{Contreras Pe{\~n}a}, C., {Lucas}, P.~W., {Minniti}, D., {et~al.} 2017, \mnras,
  465, 3011, \dodoi{10.1093/mnras/stw2801}

\bibitem[{{Cottle} {et~al.}(2018){Cottle}, {Covey}, {Su{\'a}rez},
  {Rom{\'a}n-Z{\'u}{\~n}iga}, {Schlafly}, {Downes}, {Ybarra}, {Hernandez},
  {Stassun}, {Stringfellow}, {Getman}, {Feigelson}, {Borissova}, {Kim},
  {Roman-Lopes}, {Da Rio}, {De Lee}, {Frinchaboy}, {Kounkel}, {Majewski},
  {Mennickent}, {Nidever}, {Nitschelm}, {Pan}, {Shetrone}, {Zasowski},
  {Chambers}, {Magnier}, \& {Valenti}}]{Cottle2018}
{Cottle}, J.~N., {Covey}, K.~R., {Su{\'a}rez}, G., {et~al.} 2018, \apjs, 236,
  27, \dodoi{10.3847/1538-4365/aabada}

\bibitem[{{Dalton} {et~al.}(2014){Dalton}, {Trager}, {Abrams}, {Bonifacio},
  {L{\'o}pez Aguerri}, {Middleton}, {Benn}, {Dee}, {Say{\`e}de}, {Lewis},
  {Pragt}, {Pico}, {Walton}, {Rey}, {Allende Prieto}, {Pe{\~n}ate}, {Lhome},
  {Ag{\'o}cs}, {Alonso}, {Terrett}, {Brock}, {Gilbert}, {Ridings}, {Guinouard},
  {Verheijen}, {Tosh}, {Rogers}, {Steele}, {Stuik}, {Tromp}, {Jasko}, {Kragt},
  {Lesman}, {Mottram}, {Bates}, {Gribbin}, {Fernand o Rodriguez}, {Delgado},
  {Martin}, {Cano}, {Navarro}, {Irwin}, {Lewis}, {Gonzalez Solares},
  {O'Mahony}, {Bianco}, {Zurita}, {ter Horst}, {Molinari}, {Lodi}, {Guerra},
  {Vallenari}, \& {Baruffolo}}]{2014SPIE.9147E..0LD}
{Dalton}, G., {Trager}, S., {Abrams}, D.~C., {et~al.} 2014, in Society of
  Photo-Optical Instrumentation Engineers (SPIE) Conference Series, Vol. 9147,
  Ground-based and Airborne Instrumentation for Astronomy V, ed. S.~K.
  {Ramsay}, I.~S. {McLean}, \& H.~{Takami}, 91470L, \dodoi{10.1117/12.2055132}

\bibitem[{{de Jong} {et~al.}(2019){de Jong}, {Agertz}, {Berbel}, {Aird},
  {Alexander}, {Amarsi}, {Anders}, {Andrae}, {Ansarinejad}, {Ansorge},
  {Antilogus}, {Anwand -Heerwart}, {Arentsen}, {Arnadottir}, {Asplund},
  {Auger}, {Azais}, {Baade}, {Baker}, {Baker}, {Balbinot}, {Baldry}, {Banerji},
  {Barden}, {Barklem}, {Barth{\'e}l{\'e}my-Mazot}, {Battistini}, {Bauer},
  {Bell}, {Bellido-Tirado}, {Bellstedt}, {Belokurov}, {Bensby}, {Bergemann},
  {Bestenlehner}, {Bielby}, {Bilicki}, {Blake}, {Bland-Hawthorn}, {Boeche},
  {Boland}, {Boller}, {Bongard}, {Bongiorno}, {Bonifacio}, {Boudon}, {Brooks},
  {Brown}, {Brown}, {Br{\"u}ggen}, {Brynnel}, {Brzeski}, {Buchert},
  {Buschkamp}, {Caffau}, {Caillier}, {Carrick}, {Casagrande}, {Case}, {Casey},
  {Cesarini}, {Cescutti}, {Chapuis}, {Chiappini}, {Childress}, {Christlieb},
  {Church}, {Cioni}, {Cluver}, {Colless}, {Collett}, {Comparat}, {Cooper},
  {Couch}, {Courbin}, {Croom}, {Croton}, {Daguis{\'e}}, {Dalton}, {Davies},
  {Davis}, {de Laverny}, {Deason}, {Dionies}, {Disseau}, {Doel}, {D{\"o}scher},
  {Driver}, {Dwelly}, {Eckert}, {Edge}, {Edvardsson}, {Youssoufi}, {Elhaddad},
  {Enke}, {Erfanianfar}, {Farrell}, {Fechner}, {Feiz}, {Feltzing}, {Ferreras},
  {Feuerstein}, {Feuillet}, {Finoguenov}, {Ford}, {Fotopoulou}, {Fouesneau},
  {Frenk}, {Frey}, {Gaessler}, {Geier}, {Fusillo}, {Gerhard}, {Giannantonio},
  {Giannone}, {Gibson}, {Gillingham}, {Gonz{\'a}lez-Fern{\'a}ndez},
  {Gonzalez-Solares}, {Gottloeber}, {Gould}, {Grebel}, {Gueguen}, {Guiglion},
  {Haehnelt}, {Hahn}, {Hansen}, {Hartman}, {Hauptner}, {Hawkins}, {Haynes},
  {Haynes}, {Heiter}, {Helmi}, {Aguayo}, {Hewett}, {Hinton}, {Hobbs}, {Hoenig},
  {Hofman}, {Hook}, {Hopgood}, {Hopkins}, {Hourihane}, {Howes}, {Howlett},
  {Huet}, {Irwin}, {Iwert}, {Jablonka}, {Jahn}, {Jahnke}, {Jarno}, {Jin},
  {Jofre}, {Johl}, {Jones}, {J{\"o}nsson}, {Jordan}, {Karovicova}, {Khalatyan},
  {Kelz}, {Kennicutt}, {King}, {Kitaura}, {Klar}, {Klauser}, {Kneib}, {Koch},
  {Koposov}, {Kordopatis}, {Korn}, {Kosmalski}, {Kotak}, {Kovalev}, {Kreckel},
  {Kripak}, {Krumpe}, {Kuijken}, {Kunder}, {Kushniruk}, {Lam}, {Lamer},
  {Laurent}, {Lawrence}, {Lehmitz}, {Lemasle}, {Lewis}, {Li}, {Lidman}, {Lind},
  {Liske}, {Lizon}, {Loveday}, {Ludwig}, {McDermid}, {Maguire}, {Mainieri},
  {Mali}, {Mandel}, {Mandel}, {Mannering}, {Martell}, {Martinez Delgado},
  {Matijevic}, {McGregor}, {McMahon}, {McMillan}, {Mena}, {Merloni}, {Meyer},
  {Michel}, {Micheva}, {Migniau}, {Minchev}, {Monari}, {Muller}, {Murphy},
  {Muthukrishna}, {Nandra}, {Navarro}, {Ness}, {Nichani}, {Nichol}, {Nicklas},
  {Niederhofer}, {Norberg}, {Obreschkow}, {Oliver}, {Owers}, {Pai},
  {Pankratow}, {Parkinson}, {Paschke}, {Paterson}, {Pecontal}, {Parry},
  {Phillips}, {Pillepich}, {Pinard}, {Pirard}, {Piskunov}, {Plank},
  {Pl{\"u}schke}, {Pons}, {Popesso}, {Power}, {Pragt}, {Pramskiy}, {Pryer},
  {Quattri}, {Queiroz}, {Quirrenbach}, {Rahurkar}, {Raichoor}, {Ramstedt},
  {Rau}, {Recio-Blanco}, {Reiss}, {Renaud}, {Revaz}, {Rhode}, {Richard},
  {Richter}, {Rix}, {Robotham}, {Roelfsema}, {Romaniello}, {Rosario},
  {Rothmaier}, {Roukema}, {Ruchti}, {Rupprecht}, {Rybizki}, {Ryde}, {Saar},
  {Sadler}, {Sahl{\'e}n}, {Salvato}, {Sassolas}, {Saunders}, {Saviauk},
  {Sbordone}, {Schmidt}, {Schnurr}, {Scholz}, {Schwope}, {Seifert}, {Shanks},
  {Sheinis}, {Sivov}, {Sk{\'u}lad{\'o}ttir}, {Smartt}, {Smedley}, {Smith},
  {Smith}, {Sorce}, {Spitler}, {Starkenburg}, {Steinmetz}, {Stilz}, {Storm},
  {Sullivan}, {Sutherland}, {Swann}, {Tamone}, {Taylor}, {Teillon}, {Tempel},
  {ter Horst}, {Thi}, {Tolstoy}, {Trager}, {Traven}, {Tremblay}, {Tresse},
  {Valentini}, {van de Weygaert}, {van den Ancker}, {Veljanoski}, {Venkatesan},
  {Wagner}, {Wagner}, {Walcher}, {Waller}, {Walton}, {Wang}, {Winkler},
  {Wisotzki}, {Worley}, {Worseck}, {Xiang}, {Xu}, {Yong}, {Zhao}, {Zheng},
  {Zscheyge}, \& {Zucker}}]{2019Msngr.175....3D}
{de Jong}, R.~S., {Agertz}, O., {Berbel}, A.~A., {et~al.} 2019, The Messenger,
  175, 3, \dodoi{10.18727/0722-6691/5117}

\bibitem[{{de Souza} {et~al.}(2014){de Souza}, {Maio}, {Biffi}, \&
  {Ciardi}}]{deSouza2014}
{de Souza}, R.~S., {Maio}, U., {Biffi}, V., \& {Ciardi}, B. 2014, \mnras, 440,
  240, \dodoi{10.1093/mnras/stu274}

\bibitem[{{de Souza} {et~al.}(2017){de Souza}, {Dantas}, {Costa-Duarte},
  {Feigelson}, {Killedar}, {Lablanche}, {Vilalta}, {Krone-Martins}, {Beck}, \&
  {Gieseke}}]{deSouza2017}
{de Souza}, R.~S., {Dantas}, M.~L.~L., {Costa-Duarte}, M.~V., {et~al.} 2017,
  \mnras, 472, 2808, \dodoi{10.1093/mnras/stx2156}

\bibitem[{{Dewangan} \& {Ojha}(2013)}]{Dewangan2013}
{Dewangan}, L.~K., \& {Ojha}, D.~K. 2013, \mnras, 429, 1386,
  \dodoi{10.1093/mnras/sts430}

\bibitem[{{Draine} \& {Li}(2007)}]{Draine2007}
{Draine}, B.~T., \& {Li}, A. 2007, \apj, 657, 810, \dodoi{10.1086/511055}

\bibitem[{{Ducourant} {et~al.}(2017){Ducourant}, {Teixeira}, {Krone-Martins},
  {Bontemps}, {Despois}, {Galli}, {Bouy}, {Le Campion}, {Rapaport}, \&
  {Cuillandre}}]{2017A&A...597A..90D}
{Ducourant}, C., {Teixeira}, R., {Krone-Martins}, A., {et~al.} 2017, \aap, 597,
  A90, \dodoi{10.1051/0004-6361/201527574}

\bibitem[{{Elmegreen} \& {Scalo}(2004)}]{2004ARA&A..42..211E}
{Elmegreen}, B.~G., \& {Scalo}, J. 2004, \araa, 42, 211,
  \dodoi{10.1146/annurev.astro.41.011802.094859}

\bibitem[{{Evans} {et~al.}(2018){Evans}, {Riello}, {De Angeli}, {Carrasco},
  {Montegriffo}, {Fabricius}, {Jordi}, {Palaversa}, {Diener}, {Busso},
  {Cacciari}, {van Leeuwen}, {Burgess}, {Davidson}, {Harrison}, {Hodgkin},
  {Pancino}, {Richards}, {Altavilla}, {Balaguer-N{\'u}{\~n}ez}, {Barstow},
  {Bellazzini}, {Brown}, {Castellani}, {Cocozza}, {De Luise}, {Delgado},
  {Ducourant}, {Galleti}, {Gilmore}, {Giuffrida}, {Holl}, {Kewley}, {Koposov},
  {Marinoni}, {Marrese}, {Osborne}, {Piersimoni}, {Portell}, {Pulone},
  {Ragaini}, {Sanna}, {Terrett}, {Walton}, {Wevers}, \&
  {Wyrzykowski}}]{2018A&A...616A...4E}
{Evans}, D.~W., {Riello}, M., {De Angeli}, F., {et~al.} 2018, \aap, 616, A4,
  \dodoi{10.1051/0004-6361/201832756}

\bibitem[{{Evans} {et~al.}(2009{\natexlab{a}}){Evans}, {Dunham},
  {J{\o}rgensen}, {Enoch}, {Mer{\'\i}n}, {van Dishoeck}, {Alcal{\'a}}, {Myers},
  {Stapelfeldt}, {Huard}, {Allen}, {Harvey}, {van Kempen}, {Blake}, {Koerner},
  {Mundy}, {Padgett}, \& {Sargent}}]{Evans2009}
{Evans}, Neal~J., I., {Dunham}, M.~M., {J{\o}rgensen}, J.~K., {et~al.}
  2009{\natexlab{a}}, \apjs, 181, 321, \dodoi{10.1088/0067-0049/181/2/321}

\bibitem[{{Evans} {et~al.}(2009{\natexlab{b}}){Evans}, {Calvet}, {Cieza},
  {Forbrich}, {Hillenbrand}, {Lada}, {Mer{\'\i}n}, {Strom}, \&
  {Watson}}]{diskionary}
{Evans}, N., {Calvet}, N., {Cieza}, L., {et~al.} 2009{\natexlab{b}}, arXiv
  e-prints, arXiv:0901.1691.
\newblock \doarXiv{0901.1691}

\bibitem[{Everitt {et~al.}(2001)Everitt, Landau, Leese, \& Stahl}]{Everitt2011}
Everitt, B.~S., Landau, S., Leese, M., \& Stahl, D. 2001, Cluster Analysis, A
  Hodder Arnold Publication (Wiley).
\newblock \url{https://books.google.com.br/books?id=htZzDGlCnQYC}

\bibitem[{{Fang} {et~al.}(2020){Fang}, {Hillenbrand}, {Kim}, {Findeisen},
  {Herczeg}, {Carpenter}, {Rebull}, \& {Wang}}]{Fang2020}
{Fang}, M., {Hillenbrand}, L.~A., {Kim}, J.~S., {et~al.} 2020, arXiv e-prints,
  arXiv:2009.11995.
\newblock \doarXiv{2009.11995}

\bibitem[{{Fazio} {et~al.}(2004){Fazio}, {Hora}, {Allen}, {Ashby}, {Barmby},
  {Deutsch}, {Huang}, {Kleiner}, {Marengo}, {Megeath}, {Melnick}, {Pahre},
  {Patten}, {Polizotti}, {Smith}, {Taylor}, {Wang}, {Willner}, {Hoffmann},
  {Pipher}, {Forrest}, {McMurty}, {McCreight}, {McKelvey}, {McMurray}, {Koch},
  {Moseley}, {Arendt}, {Mentzell}, {Marx}, {Losch}, {Mayman}, {Eichhorn},
  {Krebs}, {Jhabvala}, {Gezari}, {Fixsen}, {Flores}, {Shakoorzadeh}, {Jungo},
  {Hakun}, {Workman}, {Karpati}, {Kichak}, {Whitley}, {Mann}, {Tollestrup},
  {Eisenhardt}, {Stern}, {Gorjian}, {Bhattacharya}, {Carey}, {Nelson},
  {Glaccum}, {Lacy}, {Lowrance}, {Laine}, {Reach}, {Stauffer}, {Surace},
  {Wilson}, {Wright}, {Hoffman}, {Domingo}, \& {Cohen}}]{Fazio2004}
{Fazio}, G.~G., {Hora}, J.~L., {Allen}, L.~E., {et~al.} 2004, \apjs, 154, 10,
  \dodoi{10.1086/422843}

\bibitem[{Feigelson(2018)}]{Feigelson18}
Feigelson, E.~D. 2018, in The Birth of Star Clusters, ed. S.~Stahler (Cham:
  Springer International Publishing), 119--141,
  \dodoi{10.1007/978-3-319-22801-3_5}

\bibitem[{{Feigelson} {et~al.}(2013){Feigelson}, {Townsley}, {Broos}, {Busk},
  {Getman}, {King}, {Kuhn}, {Naylor}, {Povich}, {Baddeley}, {Bate},
  {Indebetouw}, {Luhman}, {McCaughrean}, {Pittard}, {Pudritz}, {Sills}, {Song},
  \& {Wadsley}}]{Feigelson13}
{Feigelson}, E.~D., {Townsley}, L.~K., {Broos}, P.~S., {et~al.} 2013, \apjs,
  209, 26, \dodoi{10.1088/0067-0049/209/2/26}

\bibitem[{{Fisher}(1922)}]{Fisher1922}
{Fisher}, R.~A. 1922, Journal of the Royal Statistical Society, 85, 87.
\newblock \url{http://www.jstor.org/stable/2340521}

\bibitem[{{Forbrich} {et~al.}(2010){Forbrich}, {Tappe}, {Robitaille}, {Muench},
  {Teixeira}, {Lada}, {Stolte}, \& {Lada}}]{Forbrich2010}
{Forbrich}, J., {Tappe}, A., {Robitaille}, T., {et~al.} 2010, \apj, 716, 1453,
  \dodoi{10.1088/0004-637X/716/2/1453}

\bibitem[{{Furlan} {et~al.}(2006){Furlan}, {Hartmann}, {Calvet}, {D'Alessio},
  {Franco-Hern{\'a}ndez}, {Forrest}, {Watson}, {Uchida}, {Sargent}, {Green},
  {Keller}, \& {Herter}}]{Furlan2006}
{Furlan}, E., {Hartmann}, L., {Calvet}, N., {et~al.} 2006, \apjs, 165, 568,
  \dodoi{10.1086/505468}

\bibitem[{{Furlan} {et~al.}(2008){Furlan}, {McClure}, {Calvet}, {Hartmann},
  {D'Alessio}, {Forrest}, {Watson}, {Uchida}, {Sargent}, {Green}, \&
  {Herter}}]{Furlan2008}
{Furlan}, E., {McClure}, M., {Calvet}, N., {et~al.} 2008, \apjs, 176, 184,
  \dodoi{10.1086/527301}

\bibitem[{{Furlan} {et~al.}(2011){Furlan}, {Luhman}, {Espaillat}, {D'Alessio},
  {Adame}, {Manoj}, {Kim}, {Watson}, {Forrest}, {McClure}, {Calvet}, {Sargent},
  {Green}, \& {Fischer}}]{Furlan2011}
{Furlan}, E., {Luhman}, K.~L., {Espaillat}, C., {et~al.} 2011, \apjs, 195, 3,
  \dodoi{10.1088/0067-0049/195/1/3}

\bibitem[{{Gaia Collaboration} {et~al.}(2016){Gaia Collaboration}, {Prusti},
  {de Bruijne}, {Brown}, {Vallenari}, {Babusiaux}, {Bailer-Jones}, {Bastian},
  {Biermann}, {Evans}, \& et~al.}]{2016A&A...595A...1G}
{Gaia Collaboration}, {Prusti}, T., {de Bruijne}, J.~H.~J., {et~al.} 2016,
  \aap, 595, A1, \dodoi{10.1051/0004-6361/201629272}

\bibitem[{{Gaia Collaboration} {et~al.}(2018){Gaia Collaboration}, {Brown},
  {Vallenari}, {Prusti}, {de Bruijne}, {Babusiaux}, {Bailer-Jones}, {Biermann},
  {Evans}, {Eyer}, \& et~al.}]{2018A&A...616A...1G}
{Gaia Collaboration}, {Brown}, A.~G.~A., {Vallenari}, A., {et~al.} 2018, \aap,
  616, A1, \dodoi{10.1051/0004-6361/201833051}

\bibitem[{{Gieles} {et~al.}(2012){Gieles}, {Moeckel}, \& {Clarke}}]{Gieles2012}
{Gieles}, M., {Moeckel}, N., \& {Clarke}, C.~J. 2012, \mnras, 426, L11,
  \dodoi{10.1111/j.1745-3933.2012.01312.x}

\bibitem[{{Gouliermis}(2018)}]{Gouliermis18}
{Gouliermis}, D.~A. 2018, \pasp, 130, 072001, \dodoi{10.1088/1538-3873/aac1fd}

\bibitem[{{Graham} {et~al.}(2019){Graham}, {Kulkarni}, {Bellm}, {Adams},
  {Barbarino}, {Blagorodnova}, {Bodewits}, {Bolin}, {Brady}, {Cenko}, {Chang},
  {Coughlin}, {De}, {Eadie}, {Farnham}, {Feindt}, {Franckowiak}, {Fremling},
  {Gezari}, {Ghosh}, {Goldstein}, {Golkhou}, {Goobar}, {Ho}, {Huppenkothen},
  {Ivezi{\'c}}, {Jones}, {Juric}, {Kaplan}, {Kasliwal}, {Kelley}, {Kupfer},
  {Lee}, {Lin}, {Lunnan}, {Mahabal}, {Miller}, {Ngeow}, {Nugent}, {Ofek},
  {Prince}, {Rauch}, {van Roestel}, {Schulze}, {Singer}, {Sollerman}, {Taddia},
  {Yan}, {Ye}, {Yu}, {Barlow}, {Bauer}, {Beck}, {Belicki}, {Biswas}, {Brinnel},
  {Brooke}, {Bue}, {Bulla}, {Burruss}, {Connolly}, {Cromer}, {Cunningham},
  {Dekany}, {Delacroix}, {Desai}, {Duev}, {Feeney}, {Flynn}, {Frederick},
  {Gal-Yam}, {Giomi}, {Groom}, {Hacopians}, {Hale}, {Helou}, {Henning},
  {Hover}, {Hillenbrand}, {Howell}, {Hung}, {Imel}, {Ip}, {Jackson}, {Kaspi},
  {Kaye}, {Kowalski}, {Kramer}, {Kuhn}, {Landry}, {Laher}, {Mao}, {Masci},
  {Monkewitz}, {Murphy}, {Nordin}, {Patterson}, {Penprase}, {Porter},
  {Rebbapragada}, {Reiley}, {Riddle}, {Rigault}, {Rodriguez}, {Rusholme}, {van
  Santen}, {Shupe}, {Smith}, {Soumagnac}, {Stein}, {Surace}, {Szkody}, {Terek},
  {Van Sistine}, {van Velzen}, {Vestrand}, {Walters}, {Ward}, {Zhang}, \&
  {Zolkower}}]{Graham2019}
{Graham}, M.~J., {Kulkarni}, S.~R., {Bellm}, E.~C., {et~al.} 2019, \pasp, 131,
  078001, \dodoi{10.1088/1538-3873/ab006c}

\bibitem[{{Greene} {et~al.}(1994){Greene}, {Wilking}, {Andre}, {Young}, \&
  {Lada}}]{Greene1994}
{Greene}, T.~P., {Wilking}, B.~A., {Andre}, P., {Young}, E.~T., \& {Lada},
  C.~J. 1994, \apj, 434, 614, \dodoi{10.1086/174763}

\bibitem[{Greenwell(2017)}]{Brandon2017}
Greenwell, B.~M. 2017, {The R Journal}, 9, 421, \dodoi{10.32614/RJ-2017-016}

\bibitem[{{Groenewegen}(2012)}]{Groenewegen2012}
{Groenewegen}, M.~A.~T. 2012, \aap, 540, A32,
  \dodoi{10.1051/0004-6361/201118287}

\bibitem[{{Gutermuth} \& {Heyer}(2015)}]{Gutermuth2015}
{Gutermuth}, R.~A., \& {Heyer}, M. 2015, \aj, 149, 64,
  \dodoi{10.1088/0004-6256/149/2/64}

\bibitem[{{Gutermuth} {et~al.}(2009){Gutermuth}, {Megeath}, {Myers}, {Allen},
  {Pipher}, \& {Fazio}}]{Gutermuth2009}
{Gutermuth}, R.~A., {Megeath}, S.~T., {Myers}, P.~C., {et~al.} 2009, \apjs,
  184, 18, \dodoi{10.1088/0067-0049/184/1/18}

\bibitem[{{Gutermuth} {et~al.}(2008){Gutermuth}, {Myers}, {Megeath}, {Allen},
  {Pipher}, {Muzerolle}, {Porras}, {Winston}, \& {Fazio}}]{Gutermuth2008}
{Gutermuth}, R.~A., {Myers}, P.~C., {Megeath}, S.~T., {et~al.} 2008, \apj, 674,
  336, \dodoi{10.1086/524722}

\bibitem[{{Hartmann} {et~al.}(2005){Hartmann}, {Megeath}, {Allen}, {Luhman},
  {Calvet}, {D'Alessio}, {Franco-Hernandez}, \& {Fazio}}]{Hartmann2005}
{Hartmann}, L., {Megeath}, S.~T., {Allen}, L., {et~al.} 2005, \apj, 629, 881,
  \dodoi{10.1086/431472}

\bibitem[{{Harvey} {et~al.}(2007){Harvey}, {Mer{\'\i}n}, {Huard}, {Rebull},
  {Chapman}, {Evans}, \& {Myers}}]{Harvey2007}
{Harvey}, P., {Mer{\'\i}n}, B., {Huard}, T.~L., {et~al.} 2007, \apj, 663, 1149,
  \dodoi{10.1086/518646}

\bibitem[{{Herbig}(1954)}]{Herbig1954}
{Herbig}, G.~H. 1954, \apj, 119, 483, \dodoi{10.1086/145854}

\bibitem[{{Herbst} {et~al.}(1994){Herbst}, {Herbst}, {Grossman}, \&
  {Weinstein}}]{Herbst1994}
{Herbst}, W., {Herbst}, D.~K., {Grossman}, E.~J., \& {Weinstein}, D. 1994, \aj,
  108, 1906, \dodoi{10.1086/117204}

\bibitem[{{Herczeg} {et~al.}(2019){Herczeg}, {Kuhn}, {Zhou}, {Hatchell},
  {Manara}, {Johnstone}, {Dunham}, {Bhardwaj}, {Jose}, \& {Yuan}}]{Herczeg2019}
{Herczeg}, G.~J., {Kuhn}, M.~A., {Zhou}, X., {et~al.} 2019, \apj, 878, 111,
  \dodoi{10.3847/1538-4357/ab1d67}

\bibitem[{{Hilbe} {et~al.}(2017){Hilbe}, {de Souza}, \& {Ishida}}]{Hilbe2017}
{Hilbe}, J.~M., {de Souza}, R.~S., \& {Ishida}, E. E.~O. 2017, {Bayesian Models
  for Astrophysical Data Using R, JAGS, Python, and Stan} (Cambridge University
  Press), \dodoi{10.1017/CBO9781316459515}

\bibitem[{Ho(1995)}]{ho1995}
Ho, T.~K. 1995, in Proceedings of the Third International Conference on
  Document Analysis and Recognition (Volume 1) - Volume 1, ICDAR ’95 (USA:
  IEEE Computer Society), 278

\bibitem[{{Hodgkin} {et~al.}(2009){Hodgkin}, {Irwin}, {Hewett}, \&
  {Warren}}]{Hodgkin2009}
{Hodgkin}, S.~T., {Irwin}, M.~J., {Hewett}, P.~C., \& {Warren}, S.~J. 2009,
  \mnras, 394, 675, \dodoi{10.1111/j.1365-2966.2008.14387.x}

\bibitem[{{Hodgkin} {et~al.}(2013){Hodgkin}, {Wyrzykowski}, {Blagorodnova}, \&
  {Koposov}}]{Hodgkin2013}
{Hodgkin}, S.~T., {Wyrzykowski}, L., {Blagorodnova}, N., \& {Koposov}, S. 2013,
  Philosophical Transactions of the Royal Society of London Series A, 371,
  20120239, \dodoi{10.1098/rsta.2012.0239}

\bibitem[{Hoff(2018)}]{sbgcop}
Hoff, P. 2018, sbgcop: Semiparametric Bayesian Gaussian Copula Estimation and
  Imputation.
\newblock \url{https://CRAN.R-project.org/package=sbgcop}

\bibitem[{{Hoff}(2007)}]{hoff2007}
{Hoff}, P.~D. 2007, Ann. Appl. Stat., 1, 265, \dodoi{10.1214/07-AOAS107}

\bibitem[{Honaker {et~al.}(2011)Honaker, King, \& Blackwell}]{Amelia2011}
Honaker, J., King, G., \& Blackwell, M. 2011, Journal of Statistical Software,
  45, 1

\bibitem[{Hotelling(1933)}]{Hotelling33}
Hotelling, H. 1933, Journal of Educational Psychology, 24, 417

\bibitem[{{Indebetouw} {et~al.}(2005){Indebetouw}, {Mathis}, {Babler}, {Meade},
  {Watson}, {Whitney}, {Wolff}, {Wolfire}, {Cohen}, {Bania}, {Benjamin},
  {Clemens}, {Dickey}, {Jackson}, {Kobulnicky}, {Marston}, {Mercer},
  {Stauffer}, {Stolovy}, \& {Churchwell}}]{Indebetouw2005}
{Indebetouw}, R., {Mathis}, J.~S., {Babler}, B.~L., {et~al.} 2005, \apj, 619,
  931, \dodoi{10.1086/426679}

\bibitem[{{Ishida} \& {de Souza}(2013)}]{Ishida2013}
{Ishida}, E.~E.~O., \& {de Souza}, R.~S. 2013, \mnras, 430, 509,
  \dodoi{10.1093/mnras/sts650}

\bibitem[{Jaeger {et~al.}(2020)Jaeger, Tierney, \& Simon}]{jaeger2020impute}
Jaeger, B.~C., Tierney, N.~J., \& Simon, N.~R. 2020, When to Impute? Imputation
  before and during cross-validation.
\newblock \doarXiv{2010.00718}

\bibitem[{{Jarrett} {et~al.}(2011){Jarrett}, {Cohen}, {Masci}, {Wright},
  {Stern}, {Benford}, {Blain}, {Carey}, {Cutri}, {Eisenhardt}, {Lonsdale},
  {Mainzer}, {Marsh}, {Padgett}, {Petty}, {Ressler}, {Skrutskie}, {Stanford},
  {Surace}, {Tsai}, {Wheelock}, \& {Yan}}]{Jarrett2011}
{Jarrett}, T.~H., {Cohen}, M., {Masci}, F., {et~al.} 2011, \apj, 735, 112,
  \dodoi{10.1088/0004-637X/735/2/112}

\bibitem[{{Jayasinghe} {et~al.}(2019){Jayasinghe}, {Dixon}, {Povich}, {Binder},
  {Velasco}, {Lepore}, {Xu}, {Offner}, {Kobulnicky}, {Anderson}, {Kendrew}, \&
  {Simpson}}]{Jayasinghe2019}
{Jayasinghe}, T., {Dixon}, D., {Povich}, M.~S., {et~al.} 2019, \mnras, 488,
  1141, \dodoi{10.1093/mnras/stz1738}

\bibitem[{{Joy}(1945)}]{Joy1945}
{Joy}, A.~H. 1945, \apj, 102, 168, \dodoi{10.1086/144749}

\bibitem[{{Joye} \& {Mandel}(2003)}]{2003ASPC..295..489J}
{Joye}, W.~A., \& {Mandel}, E. 2003, in Astronomical Society of the Pacific
  Conference Series, Vol. 295, Astronomical Data Analysis Software and Systems
  XII, ed. H.~E. {Payne}, R.~I. {Jedrzejewski}, \& R.~N. {Hook}, 489

\bibitem[{{Kang} {et~al.}(2009){Kang}, {Bieging}, {Povich}, \&
  {Lee}}]{Kang2009}
{Kang}, M., {Bieging}, J.~H., {Povich}, M.~S., \& {Lee}, Y. 2009, \apj, 706,
  83, \dodoi{10.1088/0004-637X/706/1/83}

\bibitem[{{Kauppinen} {et~al.}(1995){Kauppinen}, {Seppanen}, \&
  {Pietikainen}}]{Kauppinen}
{Kauppinen}, H., {Seppanen}, T., \& {Pietikainen}, M. 1995, IEEE Transactions
  on Pattern Analysis and Machine Intelligence, 17, 201,
  \dodoi{10.1109/34.368168}

\bibitem[{{Kobulnicky} {et~al.}(2013){Kobulnicky}, {Babler}, {Alexand er},
  {Meade}, {Whitney}, \& {Churchwell}}]{Kobulnicky2013}
{Kobulnicky}, H.~A., {Babler}, B.~L., {Alexand er}, M.~J., {et~al.} 2013,
  \apjs, 207, 9, \dodoi{10.1088/0067-0049/207/1/9}

\bibitem[{{Koenig} \& {Leisawitz}(2014)}]{Koenig14}
{Koenig}, X.~P., \& {Leisawitz}, D.~T. 2014, \apj, 791, 131,
  \dodoi{10.1088/0004-637X/791/2/131}

\bibitem[{{Kounkel} \& {Covey}(2019)}]{Kounkel2019}
{Kounkel}, M., \& {Covey}, K. 2019, \aj, 158, 122,
  \dodoi{10.3847/1538-3881/ab339a}

\bibitem[{{Kounkel} {et~al.}(2020){Kounkel}, {Covey}, \&
  {Stassun}}]{Kounkel2020}
{Kounkel}, M., {Covey}, K., \& {Stassun}, K.~G. 2020, arXiv e-prints,
  arXiv:2004.07261.
\newblock \doarXiv{2004.07261}

\bibitem[{{Krone-Martins} {et~al.}(2019){Krone-Martins}, {Graham}, {Stern},
  {Djorgovski}, {Delchambre}, {Ducourant}, {Teixeira}, {Drake}, {Scarano},
  {Surdej}, {Galluccio}, {Jalan}, {Wertz}, {Kl{\"u}ter}, {Mignard},
  {Spindola-Duarte}, {Dobie}, {Slezak}, {Sluse}, {Murphy}, {Boehm},
  {Nierenberg}, {Bastian}, {Wambsganss}, \& {LeCampion}}]{KM2019}
{Krone-Martins}, A., {Graham}, M.~J., {Stern}, D., {et~al.} 2019,
  arXiv:1912.08977

\bibitem[{{Kuhn}(2015)}]{caret}
{Kuhn}, M. 2015, {caret: Classification and Regression Training}.
\newblock \doeprint{1505.003}

\bibitem[{{Kuhn} \& {Feigelson}(2019)}]{KuhnFeigelson2017}
{Kuhn}, M.~A., \& {Feigelson}, E.~D. 2019, in Handbook of Mixture Analysis, ed.
  S.~Fruhwirth-Schnatter, G.~Celeux, \& C.~Robert (New York: Chapman and
  Hall/CRC), 463--489, \dodoi{10.1201/9780429055911}

\bibitem[{{Kuhn} {et~al.}(2015){Kuhn}, {Getman}, \& {Feigelson}}]{Kuhn2015}
{Kuhn}, M.~A., {Getman}, K.~V., \& {Feigelson}, E.~D. 2015, \apj, 802, 60,
  \dodoi{10.1088/0004-637X/802/1/60}

\bibitem[{Kuhn {et~al.}(2020)Kuhn, Hillenbrand, Carpenter, \&
  Menendez}]{Kuhn2020}
Kuhn, M.~A., Hillenbrand, L.~A., Carpenter, J.~M., \& Menendez, A. R.~A. 2020,
  \apj, 899, 128, \dodoi{10.3847/1538-4357/aba19a}

\bibitem[{{Lada}(1987)}]{Lada1987}
{Lada}, C.~J. 1987, in IAU Symposium, Vol. 115, Star Forming Regions, ed.
  M.~{Peimbert} \& J.~{Jugaku}, 1

\bibitem[{{Lawrence} {et~al.}(2007){Lawrence}, {Warren}, {Almaini}, {Edge},
  {Hambly}, {Jameson}, {Lucas}, {Casali}, {Adamson}, {Dye}, {Emerson},
  {Foucaud}, {Hewett}, {Hirst}, {Hodgkin}, {Irwin}, {Lodieu}, {McMahon},
  {Simpson}, {Smail}, {Mortlock}, \& {Folger}}]{Lawrence07}
{Lawrence}, A., {Warren}, S.~J., {Almaini}, O., {et~al.} 2007, \mnras, 379,
  1599, \dodoi{10.1111/j.1365-2966.2007.12040.x}

\bibitem[{{Leung} \& {Bovy}(2019)}]{Leung2019}
{Leung}, H.~W., \& {Bovy}, J. 2019, \mnras, 489, 2079,
  \dodoi{10.1093/mnras/stz2245}

\bibitem[{{Lin} {et~al.}(2016){Lin}, {Kilbinger}, \& {Pires}}]{Lin2016}
{Lin}, C.-A., {Kilbinger}, M., \& {Pires}, S. 2016, \aap, 593, A88,
  \dodoi{10.1051/0004-6361/201628565}

\bibitem[{{Lindegren} {et~al.}(2018){Lindegren}, {Hern{\'a}ndez}, {Bombrun},
  {Klioner}, {Bastian}, {Ramos-Lerate}, {de Torres}, {Steidelm{\"u}ller},
  {Stephenson}, {Hobbs}, {Lammers}, {Biermann}, {Geyer}, {Hilger}, {Michalik},
  {Stampa}, {McMillan}, {Casta{\~n}eda}, {Clotet}, {Comoretto}, {Davidson},
  {Fabricius}, {Gracia}, {Hambly}, {Hutton}, {Mora}, {Portell}, {van Leeuwen},
  {Abbas}, {Abreu}, {Altmann}, {Andrei}, {Anglada}, {Balaguer-N{\'u}{\~n}ez},
  {Barache}, {Becciani}, {Bertone}, {Bianchi}, {Bouquillon}, {Bourda},
  {Br{\"u}semeister}, {Bucciarelli}, {Busonero}, {Buzzi}, {Cancelliere},
  {Carlucci}, {Charlot}, {Cheek}, {Crosta}, {Crowley}, {de Bruijne}, {de
  Felice}, {Drimmel}, {Esquej}, {Fienga}, {Fraile}, {Gai}, {Garralda},
  {Gonz{\'a}lez-Vidal}, {Guerra}, {Hauser}, {Hofmann}, {Holl}, {Jordan},
  {Lattanzi}, {Lenhardt}, {Liao}, {Licata}, {Lister}, {L{\"o}ffler},
  {Marchant}, {Martin-Fleitas}, {Messineo}, {Mignard}, {Morbidelli}, {Poggio},
  {Riva}, {Rowell}, {Salguero}, {Sarasso}, {Sciacca}, {Siddiqui}, {Smart},
  {Spagna}, {Steele}, {Taris}, {Torra}, {van Elteren}, {van Reeven}, \&
  {Vecchiato}}]{Lindegren2018}
{Lindegren}, L., {Hern{\'a}ndez}, J., {Bombrun}, A., {et~al.} 2018, \aap, 616,
  A2, \dodoi{10.1051/0004-6361/201832727}

\bibitem[{Loh(2008)}]{Loh_2008}
Loh, J.~M. 2008, The Astrophysical Journal, 681, 726, \dodoi{10.1086/588631}

\bibitem[{{Lucas} {et~al.}(2008){Lucas}, {Hoare}, {Longmore}, {Schr{\"o}der},
  {Davis}, {Adamson}, {Bandyopadhyay}, {de Grijs}, {Smith}, {Gosling},
  {Mitchison}, {G{\'a}sp{\'a}r}, {Coe}, {Tamura}, {Parker}, {Irwin}, {Hambly},
  {Bryant}, {Collins}, {Cross}, {Evans}, {Gonzalez-Solares}, {Hodgkin},
  {Lewis}, {Read}, {Riello}, {Sutorius}, {Lawrence}, {Drew}, {Dye}, \&
  {Thompson}}]{Lucas08}
{Lucas}, P.~W., {Hoare}, M.~G., {Longmore}, A., {et~al.} 2008, \mnras, 391,
  136, \dodoi{10.1111/j.1365-2966.2008.13924.x}

\bibitem[{{Luhman}(2018)}]{Luhman2018_Taurus}
{Luhman}, K.~L. 2018, \aj, 156, 271, \dodoi{10.3847/1538-3881/aae831}

\bibitem[{{Lumsden} {et~al.}(2013){Lumsden}, {Hoare}, {Urquhart}, {Oudmaijer},
  {Davies}, {Mottram}, {Cooper}, \& {Moore}}]{Lumsden2013}
{Lumsden}, S.~L., {Hoare}, M.~G., {Urquhart}, J.~S., {et~al.} 2013, \apjs, 208,
  11, \dodoi{10.1088/0067-0049/208/1/11}

\bibitem[{{Lupton} {et~al.}(2004){Lupton}, {Blanton}, {Fekete}, {Hogg},
  {O'Mullane}, {Szalay}, \& {Wherry}}]{2004PASP..116..133L}
{Lupton}, R., {Blanton}, M.~R., {Fekete}, G., {et~al.} 2004, \pasp, 116, 133,
  \dodoi{10.1086/382245}

\bibitem[{{Majewski} {et~al.}(2007){Majewski}, {Babler}, {Churchwell},
  {Indebetouw}, {Meade}, {Nidever}, {Patterson}, {Rocha-Pinto}, {Skrutskie}, \&
  {Watson}}]{Majewski07}
{Majewski}, S., {Babler}, B., {Churchwell}, E., {et~al.} 2007, {Galactic
  Structure and Star Formation in Vela-Carina}, Spitzer Proposal

\bibitem[{{Mallick} {et~al.}(2015){Mallick}, {Ojha}, {Tamura}, {Linz}, {Samal},
  \& {Ghosh}}]{Mallick2015}
{Mallick}, K.~K., {Ojha}, D.~K., {Tamura}, M., {et~al.} 2015, \mnras, 447,
  2307, \dodoi{10.1093/mnras/stu2584}

\bibitem[{{Marengo} {et~al.}(1999){Marengo}, {Busso}, {Silvestro}, {Persi}, \&
  {Lagage}}]{Marengo1999}
{Marengo}, M., {Busso}, M., {Silvestro}, G., {Persi}, P., \& {Lagage}, P.~O.
  1999, \aap, 348, 501

\bibitem[{{Marengo} {et~al.}(1997){Marengo}, {Canil}, {Silvestro}, {Origlia},
  {Busso}, \& {Persi}}]{Marengo1997}
{Marengo}, M., {Canil}, G., {Silvestro}, G., {et~al.} 1997, \aap, 322, 924.
\newblock \doarXiv{astro-ph/9607129}

\bibitem[{{Marigo} {et~al.}(2013){Marigo}, {Bressan}, {Nanni}, {Girardi}, \&
  {Pumo}}]{Marigo2013}
{Marigo}, P., {Bressan}, A., {Nanni}, A., {Girardi}, L., \& {Pumo}, M.~L. 2013,
  \mnras, 434, 488, \dodoi{10.1093/mnras/stt1034}

\bibitem[{{Marton} {et~al.}(2016){Marton}, {T{\'o}th}, {Paladini}, {Kun},
  {Zahorecz}, {McGehee}, \& {Kiss}}]{Marton2016}
{Marton}, G., {T{\'o}th}, L.~V., {Paladini}, R., {et~al.} 2016, \mnras, 458,
  3479, \dodoi{10.1093/mnras/stw398}

\bibitem[{{Marton} {et~al.}(2019){Marton}, {{\'A}brah{\'a}m}, {Szegedi-Elek},
  {Varga}, {Kun}, {K{\'o}sp{\'a}l}, {Varga-Vereb{\'e}lyi}, {Hodgkin},
  {Szabados}, {Beck}, \& {Kiss}}]{Marton19}
{Marton}, G., {{\'A}brah{\'a}m}, P., {Szegedi-Elek}, E., {et~al.} 2019, \mnras,
  487, 2522, \dodoi{10.1093/mnras/stz1301}

\bibitem[{{Masci} {et~al.}(2019){Masci}, {Laher}, {Rusholme}, {Shupe}, {Groom},
  {Surace}, {Jackson}, {Monkewitz}, {Beck}, {Flynn}, {Terek}, {Landry},
  {Hacopians}, {Desai}, {Howell}, {Brooke}, {Imel}, {Wachter}, {Ye}, {Lin},
  {Cenko}, {Cunningham}, {Rebbapragada}, {Bue}, {Miller}, {Mahabal}, {Bellm},
  {Patterson}, {Juri{\'c}}, {Golkhou}, {Ofek}, {Walters}, {Graham}, {Kasliwal},
  {Dekany}, {Kupfer}, {Burdge}, {Cannella}, {Barlow}, {Van Sistine}, {Giomi},
  {Fremling}, {Blagorodnova}, {Levitan}, {Riddle}, {Smith}, {Helou}, {Prince},
  \& {Kulkarni}}]{Masci2019}
{Masci}, F.~J., {Laher}, R.~R., {Rusholme}, B., {et~al.} 2019, \pasp, 131,
  018003, \dodoi{10.1088/1538-3873/aae8ac}

\bibitem[{{McClure}(2009)}]{McClure09}
{McClure}, M. 2009, \apjl, 693, L81, \dodoi{10.1088/0004-637X/693/2/L81}

\bibitem[{{McInnes} {et~al.}(2017){McInnes}, {Healy}, \&
  {Astels}}]{McInnes2017}
{McInnes}, L., {Healy}, J., \& {Astels}, S. 2017, The Journal of Open Source
  Software, 2, 205, \dodoi{10.21105/joss.00205}

\bibitem[{McKee \& Ostriker(2007)}]{McKeeOstriker2007}
McKee, C.~F., \& Ostriker, E.~C. 2007, Annual Review of Astronomy and
  Astrophysics, 45, 565, \dodoi{10.1146/annurev.astro.45.051806.110602}

\bibitem[{{Melchior} \& {Goulding}(2018)}]{Melchior2018}
{Melchior}, P., \& {Goulding}, A.~D. 2018, Astronomy and Computing, 25, 183,
  \dodoi{10.1016/j.ascom.2018.09.013}

\bibitem[{{Melton}(2020)}]{Melton2020}
{Melton}, E. 2020, \aj, 159, 200, \dodoi{10.3847/1538-3881/ab72ac}

\bibitem[{{Minniti} {et~al.}(2010){Minniti}, {Lucas}, {Emerson}, {Saito},
  {Hempel}, {Pietrukowicz}, {Ahumada}, {Alonso}, {Alonso-Garcia}, {Arias},
  {Bandyopadhyay}, {Barb{\'a}}, {Barbuy}, {Bedin}, {Bica}, {Borissova},
  {Bronfman}, {Carraro}, {Catelan}, {Clari{\'a}}, {Cross}, {de Grijs},
  {D{\'e}k{\'a}ny}, {Drew}, {Fari{\~n}a}, {Feinstein}, {Fern{\'a}ndez
  Laj{\'u}s}, {Gamen}, {Geisler}, {Gieren}, {Goldman}, {Gonzalez}, {Gunthardt},
  {Gurovich}, {Hambly}, {Irwin}, {Ivanov}, {Jord{\'a}n}, {Kerins}, {Kinemuchi},
  {Kurtev}, {L{\'o}pez-Corredoira}, {Maccarone}, {Masetti}, {Merlo},
  {Messineo}, {Mirabel}, {Monaco}, {Morelli}, {Padilla}, {Palma}, {Parisi},
  {Pignata}, {Rejkuba}, {Roman-Lopes}, {Sale}, {Schreiber}, {Schr{\"o}der},
  {Smith}, {}, {Soto}, {Tamura}, {Tappert}, {Thompson}, {Toledo}, {Zoccali}, \&
  {Pietrzynski}}]{Minniti10}
{Minniti}, D., {Lucas}, P.~W., {Emerson}, J.~P., {et~al.} 2010, \na, 15, 433,
  \dodoi{10.1016/j.newast.2009.12.002}

\bibitem[{{Morales} \& {Robitaille}(2017)}]{Morales2017}
{Morales}, E. F.~E., \& {Robitaille}, T.~P. 2017, \aap, 598, A136,
  \dodoi{10.1051/0004-6361/201628450}

\bibitem[{Nelsen(2010)}]{Nelsen10}
Nelsen, R.~B. 2010, An Introduction to Copulas (Springer Publishing Company,
  Incorporated)

\bibitem[{{O'Donnell}(1994)}]{ODonnell1994}
{O'Donnell}, J.~E. 1994, \apj, 422, 158, \dodoi{10.1086/173713}

\bibitem[{{Oliveira} {et~al.}(2010){Oliveira}, {Pontoppidan}, {Mer{\'\i}n},
  {van Dishoeck}, {Lahuis}, {Geers}, {J{\o}rgensen}, {Olofsson}, {Augereau}, \&
  {Brown}}]{Oliveira2010}
{Oliveira}, I., {Pontoppidan}, K.~M., {Mer{\'\i}n}, B., {et~al.} 2010, \apj,
  714, 778, \dodoi{10.1088/0004-637X/714/1/778}

\bibitem[{{Oliveira} {et~al.}(2013){Oliveira}, {van Loon}, {Sloan},
  {Sewi{\l}o}, {Kraemer}, {Wood}, {Indebetouw}, {Filipovi{\'c}}, {Crawford},
  {Wong}, {Hora}, {Meixner}, {Robitaille}, {Shiao}, \& {Simon}}]{Oliveira2013}
{Oliveira}, J.~M., {van Loon}, J.~T., {Sloan}, G.~C., {et~al.} 2013, \mnras,
  428, 3001, \dodoi{10.1093/mnras/sts250}

\bibitem[{{Oort}(1927)}]{Oort1927}
{Oort}, J.~H. 1927, \bain, 3, 275

\bibitem[{{Pari} \& {Hora}(2020)}]{Pari2020}
{Pari}, J., \& {Hora}, J.~L. 2020, \pasp, 132, 054301,
  \dodoi{10.1088/1538-3873/ab7b39}

\bibitem[{{Pearson}(1894)}]{1894RSPTA.185...71P}
{Pearson}, K. 1894, Philosophical Transactions of the Royal Society of London
  Series A, 185, 71, \dodoi{10.1098/rsta.1894.0003}

\bibitem[{Pearson(1901)}]{Pearson01}
Pearson, K. 1901, The London, Edinburgh, and Dublin Philosophical Magazine and
  Journal of Science, 2, 559, \dodoi{10.1080/14786440109462720}

\bibitem[{{Pfalzner} {et~al.}(2012){Pfalzner}, {Kaczmarek}, \&
  {Olczak}}]{Pfalzner2012}
{Pfalzner}, S., {Kaczmarek}, T., \& {Olczak}, C. 2012, \aap, 545, A122,
  \dodoi{10.1051/0004-6361/201219881}

\bibitem[{Plummer(2017)}]{plummer2017jags}
Plummer, M. 2017, Retrieved from
  sourceforge.net/projects/mcmc-jags/files/Manuals/4.x, 2

\bibitem[{Plummer(2019)}]{plummer2019}
---. 2019, rjags: Bayesian Graphical Models using MCMC.
\newblock \url{https://CRAN.R-project.org/package=rjags}

\bibitem[{Poncet(2019)}]{modeest2019}
Poncet, P. 2019, modeest: Mode Estimation.
\newblock \url{https://CRAN.R-project.org/package=modeest}

\bibitem[{{PostgreSQL Global Development Group}(2020)}]{PostgreSQL}
{PostgreSQL Global Development Group}. 2020, PostgreSQL 12.4.
\newblock \url{https://www.postgresql.org/}

\bibitem[{{Povich} {et~al.}(2016){Povich}, {Townsley}, {Robitaille}, {Broos},
  {Orbin}, {King}, {Naylor}, \& {Whitney}}]{Povich16}
{Povich}, M.~S., {Townsley}, L.~K., {Robitaille}, T.~P., {et~al.} 2016, \apj,
  825, 125, \dodoi{10.3847/0004-637X/825/2/125}

\bibitem[{{Povich} {et~al.}(2009){Povich}, {Churchwell}, {Bieging}, {Kang},
  {Whitney}, {Brogan}, {Kulesa}, {Cohen}, {Babler}, {Indebetouw}, {Meade}, \&
  {Robitaille}}]{Povich09}
{Povich}, M.~S., {Churchwell}, E., {Bieging}, J.~H., {et~al.} 2009, \apj, 696,
  1278, \dodoi{10.1088/0004-637X/696/2/1278}

\bibitem[{{Povich} {et~al.}(2011){Povich}, {Smith}, {Majewski}, {Getman},
  {Townsley}, {Babler}, {Broos}, {Indebetouw}, {Meade}, {Robitaille},
  {Stassun}, {Whitney}, {Yonekura}, \& {Fukui}}]{Povich11}
{Povich}, M.~S., {Smith}, N., {Majewski}, S.~R., {et~al.} 2011, \apjs, 194, 14,
  \dodoi{10.1088/0067-0049/194/1/14}

\bibitem[{{Povich} {et~al.}(2013){Povich}, {Kuhn}, {Getman}, {Busk},
  {Feigelson}, {Broos}, {Townsley}, {King}, \& {Naylor}}]{Povich13}
{Povich}, M.~S., {Kuhn}, M.~A., {Getman}, K.~V., {et~al.} 2013, The
  Astrophysical Journal Supplement Series, 209, 31,
  \dodoi{10.1088/0067-0049/209/2/31}

\bibitem[{{R Core Team}(2019)}]{rcore19}
{R Core Team}. 2019, R: A Language and Environment for Statistical Computing, R
  Foundation for Statistical Computing, Vienna, Austria.
\newblock \url{https://www.R-project.org/}

\bibitem[{{Rebull} {et~al.}(2011){Rebull}, {Guieu}, {Stauffer}, {Hillenbrand },
  {Noriega-Crespo}, {Stapelfeldt}, {Carey}, {Carpenter}, {Cole}, {Padgett},
  {Strom}, \& {Wolff}}]{Rebull2011}
{Rebull}, L.~M., {Guieu}, S., {Stauffer}, J.~R., {et~al.} 2011, \apjs, 193, 25,
  \dodoi{10.1088/0067-0049/193/2/25}

\bibitem[{{Rebull} {et~al.}(2014){Rebull}, {Cody}, {Covey}, {G{\"u}nther},
  {Hillenbrand}, {Plavchan}, {Poppenhaeger}, {Stauffer}, {Wolk}, {Gutermuth},
  {Morales-Calder{\'o}n}, {Song}, {Barrado}, {Bayo}, {James}, {Hora}, {Vrba},
  {Alves de Oliveira}, {Bouvier}, {Carey}, {Carpenter}, {Favata}, {Flaherty},
  {Forbrich}, {Hernandez}, {McCaughrean}, {Megeath}, {Micela}, {Smith},
  {Terebey}, {Turner}, {Allen}, {Ardila}, {Bouy}, \& {Guieu}}]{Rebull2014}
{Rebull}, L.~M., {Cody}, A.~M., {Covey}, K.~R., {et~al.} 2014, \aj, 148, 92,
  \dodoi{10.1088/0004-6256/148/5/92}

\bibitem[{{Reid} {et~al.}(2019){Reid}, {Menten}, {Brunthaler}, {Zheng}, {Dame},
  {Xu}, {Li}, {Sakai}, {Wu}, {Immer}, {Zhang}, {Sanna}, {Moscadelli}, {Rygl},
  {Bartkiewicz}, {Hu}, {Quiroga-Nu{\~n}ez}, \& {van Langevelde}}]{Reid2019}
{Reid}, M.~J., {Menten}, K.~M., {Brunthaler}, A., {et~al.} 2019, \apj, 885,
  131, \dodoi{10.3847/1538-4357/ab4a11}

\bibitem[{{Reiter} {et~al.}(2015){Reiter}, {Marengo}, {Hora}, \&
  {Fazio}}]{Reiter2015}
{Reiter}, M., {Marengo}, M., {Hora}, J.~L., \& {Fazio}, G.~G. 2015, \mnras,
  447, 3909, \dodoi{10.1093/mnras/stu2725}

\bibitem[{{Rieke} \& {Lebofsky}(1985)}]{Rieke1985}
{Rieke}, G.~H., \& {Lebofsky}, M.~J. 1985, \apj, 288, 618,
  \dodoi{10.1086/162827}

\bibitem[{{Ripley}(1976)}]{Ripley1976}
{Ripley}, B.~D. 1976, Journal of Applied Probability, 13, 255–266,
  \dodoi{10.2307/3212829}

\bibitem[{{Robitaille}(2017)}]{Robitaille17}
{Robitaille}, T.~P. 2017, Astronomy and Astrophysics, 600, A11,
  \dodoi{10.1051/0004-6361/201425486}

\bibitem[{{Robitaille} {et~al.}(2007){Robitaille}, {Whitney}, {Indebetouw}, \&
  {Wood}}]{Robitaille2007}
{Robitaille}, T.~P., {Whitney}, B.~A., {Indebetouw}, R., \& {Wood}, K. 2007,
  \apjs, 169, 328, \dodoi{10.1086/512039}

\bibitem[{{Robitaille} {et~al.}(2006){Robitaille}, {Whitney}, {Indebetouw},
  {Wood}, \& {Denzmore}}]{Robitaille2006}
{Robitaille}, T.~P., {Whitney}, B.~A., {Indebetouw}, R., {Wood}, K., \&
  {Denzmore}, P. 2006, \apjs, 167, 256, \dodoi{10.1086/508424}

\bibitem[{{Robitaille} {et~al.}(2008){Robitaille}, {Meade}, {Babler},
  {Whitney}, {Johnston}, {Indebetouw}, {Cohen}, {Povich}, {Sewilo}, {Benjamin},
  \& {Churchwell}}]{Robitaille08}
{Robitaille}, T.~P., {Meade}, M.~R., {Babler}, B.~L., {et~al.} 2008, The
  Astronomical Journal, 136, 2413, \dodoi{10.1088/0004-6256/136/6/2413}

\bibitem[{{Roche} \& {Aitken}(1984)}]{Roche1984}
{Roche}, P.~F., \& {Aitken}, D.~K. 1984, \mnras, 208, 481,
  \dodoi{10.1093/mnras/208.3.481}

\bibitem[{Rokach \& Maimon(2014)}]{rokach2014}
Rokach, L., \& Maimon, O. 2014, Data Mining With Decision Trees: Theory and
  Applications, 2nd edn. (USA: World Scientific Publishing Co., Inc.)

\bibitem[{{Bonito} {et~al.}(2018){Bonito}, {Hartigan}, {Venuti},
  {Guarcello}, {Prisinzano}, {Argiroffi}, {Messina}, {Johns-Krull},
  {Feigelson}, {Stauffer}, {Giannini}, {Antoniucci}, {Sciortino}, {Micela},
  {Pillitteri}, {Fedele}, {Podio}, {Damiani}, {McGehee}, {Street}, {Gizis},
  {Sacco}, {Magrini}, {Flaccomio}, {Orlando}, {Miceli}, {Stelzer}, {Fuchs},
  {Chen}, {Pikuz}, {Frasca}, {Biazzo}, {Codella}, {Pastorello}, {Alcala'},
  {Covino}, {Bianchi}, \& {Nisini}}]{Rosaria2018}
{Bonito}, R., {Hartigan}, P., {Venuti}, L., {et~al.} 2018, arXiv e-prints,
  arXiv:1812.03135.
\newblock \doarXiv{1812.03135}

\bibitem[{Sagi \& Rokach(2018)}]{sagi2018}
Sagi, O., \& Rokach, L. 2018, WIREs Data Mining and Knowledge Discovery, 8,
  e1249, \dodoi{10.1002/widm.1249}

\bibitem[{{Samal} {et~al.}(2010){Samal}, {Pandey}, {Ojha}, {Ghosh}, {Kulkarni},
  {Kusakabe}, {Tamura}, {Bhatt}, {Thompson}, \& {Sagar}}]{Samal2010}
{Samal}, M.~R., {Pandey}, A.~K., {Ojha}, D.~K., {et~al.} 2010, \apj, 714, 1015,
  \dodoi{10.1088/0004-637X/714/2/1015}

\bibitem[{{Samal} {et~al.}(2014){Samal}, {Zavagno}, {Deharveng}, {Molinari},
  {Ojha}, {Paradis}, {Tig{\'e}}, {Pandey}, \& {Russeil}}]{Samal2014}
{Samal}, M.~R., {Zavagno}, A., {Deharveng}, L., {et~al.} 2014, \aap, 566, A122,
  \dodoi{10.1051/0004-6361/201321794}

\bibitem[{{Sasdelli} {et~al.}(2016){Sasdelli}, {Ishida}, {Vilalta}, {Aguena},
  {Busti}, {Camacho}, {Trindade}, {Gieseke}, {de Souza}, {Fantaye}, \&
  {Mazzali}}]{Sasdelli2016}
{Sasdelli}, M., {Ishida}, E.~E.~O., {Vilalta}, R., {et~al.} 2016, \mnras, 461,
  2044, \dodoi{10.1093/mnras/stw1228}

\bibitem[{{Sato} {et~al.}(2011){Sato}, {Ichiki}, \& {Takeuchi}}]{Sato2011}
{Sato}, M., {Ichiki}, K., \& {Takeuchi}, T.~T. 2011, \prd, 83, 023501,
  \dodoi{10.1103/PhysRevD.83.023501}

\bibitem[{Schwarz(1978)}]{schwarz1978}
Schwarz, G. 1978, Ann. Statist., 6, 461, \dodoi{10.1214/aos/1176344136}

\bibitem[{Scrucca {et~al.}(2016)Scrucca, Fop, Murphy, \& Raftery}]{mclust}
Scrucca, L., Fop, M., Murphy, T.~B., \& Raftery, A.~E. 2016, The {R} Journal,
  8, 289.
\newblock \url{https://doi.org/10.32614/RJ-2016-021}

\bibitem[{{Shao} {et~al.}(2018){Shao}, {Jiang}, {Li}, {Gao}, {Lv}, \&
  {Yao}}]{Shao2018}
{Shao}, Z., {Jiang}, B.~W., {Li}, A., {et~al.} 2018, \mnras, 478, 3467,
  \dodoi{10.1093/mnras/sty1267}

\bibitem[{{Shu} {et~al.}(1987){Shu}, {Adams}, \& {Lizano}}]{Shu1987}
{Shu}, F.~H., {Adams}, F.~C., \& {Lizano}, S. 1987, \araa, 25, 23,
  \dodoi{10.1146/annurev.aa.25.090187.000323}

\bibitem[{{Simon} {et~al.}(2007){Simon}, {Bolatto}, {Whitney}, {Robitaille},
  {Shah}, {Makovoz}, {Stanimirovi{\'c}}, {Barb{\'a}}, \& {Rubio}}]{Simon2007}
{Simon}, J.~D., {Bolatto}, A.~D., {Whitney}, B.~A., {et~al.} 2007, \apj, 669,
  327, \dodoi{10.1086/521544}

\bibitem[{{Simpson} {et~al.}(2012){Simpson}, {Povich}, {Kendrew}, {Lintott},
  {Bressert}, {Arvidsson}, {Cyganowski}, {Maddison}, {Schawinski}, {Sherman},
  {Smith}, \& {Wolf-Chase}}]{Simpson2012}
{Simpson}, R.~J., {Povich}, M.~S., {Kendrew}, S., {et~al.} 2012, \mnras, 424,
  2442, \dodoi{10.1111/j.1365-2966.2012.20770.x}

\bibitem[{{Skrutskie} {et~al.}(2006){Skrutskie}, {Cutri}, {Stiening},
  {Weinberg}, {Schneider}, {Carpenter}, {Beichman}, {Capps}, {Chester},
  {Elias}, {Huchra}, {Liebert}, {Lonsdale}, {Monet}, {Price}, {Seitzer},
  {Jarrett}, {Kirkpatrick}, {Gizis}, {Howard}, {Evans}, {Fowler}, {Fullmer},
  {Hurt}, {Light}, {Kopan}, {Marsh}, {McCallon}, {Tam}, {Van Dyk}, \&
  {Wheelock}}]{Skrutskie06}
{Skrutskie}, M.~F., {Cutri}, R.~M., {Stiening}, R., {et~al.} 2006, \aj, 131,
  1163, \dodoi{10.1086/498708}

\bibitem[{{Smith} {et~al.}(2018){Smith}, {Lucas}, {Kurtev}, {Smart}, {Minniti},
  {Borissova}, {Jones}, {Zhang}, {Marocco}, {Contreras Pe{\~n}a}, {Gromadzki},
  {Kuhn}, {Drew}, {Pinfield}, \& {Bedin}}]{Smith18}
{Smith}, L.~C., {Lucas}, P.~W., {Kurtev}, R., {et~al.} 2018, \mnras, 474, 1826,
  \dodoi{10.1093/mnras/stx2789}

\bibitem[{{Soto} {et~al.}(2013){Soto}, {Barb{\'a}}, {Gunthardt}, {Minniti},
  {Lucas}, {Majaess}, {Irwin}, {Emerson}, {Gonzalez-Solares}, {Hempel},
  {Saito}, {Gurovich}, {Roman-Lopes}, {Moni-Bidin}, {Santucho}, {Borissova},
  {Kurtev}, {Toledo}, {Geisler}, {Dominguez}, \& {Beamin}}]{Soto2013}
{Soto}, M., {Barb{\'a}}, R., {Gunthardt}, G., {et~al.} 2013, \aap, 552, A101,
  \dodoi{10.1051/0004-6361/201220046}

\bibitem[{{Starck} {et~al.}(2003){Starck}, {Donoho}, \&
  {Cand{\`e}s}}]{2003A&A...398..785S}
{Starck}, J.~L., {Donoho}, D.~L., \& {Cand{\`e}s}, E.~J. 2003, \aap, 398, 785,
  \dodoi{10.1051/0004-6361:20021571}

\bibitem[{{Stauffer} {et~al.}(2017){Stauffer}, {Collier Cameron}, {Jardine},
  {David}, {Rebull}, {Cody}, {Hillenbrand}, {Barrado}, {Wolk}, {Davenport}, \&
  {Pinsonneault}}]{Stauffer2017}
{Stauffer}, J., {Collier Cameron}, A., {Jardine}, M., {et~al.} 2017, \aj, 153,
  152, \dodoi{10.3847/1538-3881/aa5eb9}

\bibitem[{{Stern} {et~al.}(2005){Stern}, {Eisenhardt}, {Gorjian}, {Kochanek},
  {Caldwell}, {Eisenstein}, {Brodwin}, {Brown}, {Cool}, {Dey}, {Green},
  {Jannuzi}, {Murray}, {Pahre}, \& {Willner}}]{Stern2005}
{Stern}, D., {Eisenhardt}, P., {Gorjian}, V., {et~al.} 2005, \apj, 631, 163,
  \dodoi{10.1086/432523}

\bibitem[{{Stolovy} {et~al.}(2006){Stolovy}, {Ramirez}, {Arendt}, {Cotera},
  {Yusef-Zadeh}, {Law}, {Gezari}, {Sellgren}, {Karr}, {Moseley}, \&
  {Smith}}]{Stolovy2006}
{Stolovy}, S., {Ramirez}, S., {Arendt}, R.~G., {et~al.} 2006, in Journal of
  Physics Conference Series, Vol.~54, Journal of Physics Conference Series,
  176--182, \dodoi{10.1088/1742-6596/54/1/030}

\bibitem[{{Suh}(2020)}]{Suh2020}
{Suh}, K.-W. 2020, \apj, 891, 43, \dodoi{10.3847/1538-4357/ab6609}

\bibitem[{{Sung} {et~al.}(2009){Sung}, {Stauffer}, \& {Bessell}}]{Sung2009}
{Sung}, H., {Stauffer}, J.~R., \& {Bessell}, M.~S. 2009, \aj, 138, 1116,
  \dodoi{10.1088/0004-6256/138/4/1116}

\bibitem[{{Taylor}(2005)}]{Taylor2005}
{Taylor}, M.~B. 2005, in Astronomical Society of the Pacific Conference Series,
  Vol. 347, Astronomical Data Analysis Software and Systems XIV, ed.
  P.~{Shopbell}, M.~{Britton}, \& R.~{Ebert}, 29

\bibitem[{{Townsley} {et~al.}(2011){Townsley}, {Broos}, {Corcoran},
  {Feigelson}, {Gagn{\'e}}, {Montmerle}, {Oey}, {Smith}, {Garmire}, {Getman},
  {Povich}, {Remage Evans}, {Naz{\'e}}, {Parkin}, {Preibisch}, {Wang}, {Wolk},
  {Chu}, {Cohen}, {Gruendl}, {Hamaguchi}, {King}, {Mac Low}, {McCaughrean},
  {Moffat}, {Oskinova}, {Pittard}, {Stassun}, {ud-Doula}, {Walborn}, {Waldron},
  {Churchwell}, {Nichols}, {Owocki}, \& {Schulz}}]{Townsley2011}
{Townsley}, L.~K., {Broos}, P.~S., {Corcoran}, M.~F., {et~al.} 2011, \apjs,
  194, 1, \dodoi{10.1088/0067-0049/194/1/1}

\bibitem[{{van Breemen} {et~al.}(2011){van Breemen}, {Min}, {Chiar}, {Waters},
  {Kemper}, {Boogert}, {Cami}, {Decin}, {Knez}, {Sloan}, \&
  {Tielens}}]{vanBreemen2011}
{van Breemen}, J.~M., {Min}, M., {Chiar}, J.~E., {et~al.} 2011, \aap, 526,
  A152, \dodoi{10.1051/0004-6361/200811142}

\bibitem[{van Buuren \& Groothuis-Oudshoorn(2011)}]{Stef2011}
van Buuren, S., \& Groothuis-Oudshoorn, K. 2011, Journal of Statistical
  Software, Articles, 45, 1, \dodoi{10.18637/jss.v045.i03}

\bibitem[{{van den Bergh}(1964)}]{vanDenBergh1964}
{van den Bergh}, S. 1964, \apjs, 9, 65, \dodoi{10.1086/190097}

\bibitem[{{van der Schaaf} \& {van Hateren}(1996)}]{VANDERSCHAAF1996}
{van der Schaaf}, A., \& {van Hateren}, J. 1996, Vision Research, 36, 2759 ,
  \dodoi{https://doi.org/10.1016/0042-6989(96)00002-8}

\bibitem[{Venter(1967)}]{venter1967}
Venter, J.~H. 1967, Ann. Math. Statist., 38, 1446,
  \dodoi{10.1214/aoms/1177698699}

\bibitem[{{Vioque} {et~al.}(2020){Vioque}, {Oudmaijer}, {Schreiner},
  {Mendigut{\'\i}a}, {Baines}, {Mowlavi}, \&
  {P{\'e}rez-Mart{\'\i}nez}}]{Vioque20}
{Vioque}, M., {Oudmaijer}, R.~D., {Schreiner}, M., {et~al.} 2020, arXiv
  e-prints, arXiv:2005.01727.
\newblock \doarXiv{2005.01727}

\bibitem[{{Watson} {et~al.}(2008){Watson}, {Povich}, {Churchwell}, {Babler},
  {Chunev}, {Hoare}, {Indebetouw}, {Meade}, {Robitaille}, \&
  {Whitney}}]{Watson2008}
{Watson}, C., {Povich}, M.~S., {Churchwell}, E.~B., {et~al.} 2008, \apj, 681,
  1341, \dodoi{10.1086/588005}

\bibitem[{{Werner} {et~al.}(2004){Werner}, {Roellig}, {Low}, {Rieke}, {Rieke},
  {Hoffmann}, {Young}, {Houck}, {Brandl}, {Fazio}, {Hora}, {Gehrz}, {Helou},
  {Soifer}, {Stauffer}, {Keene}, {Eisenhardt}, {Gallagher}, {Gautier}, {Irace},
  {Lawrence}, {Simmons}, {Van Cleve}, {Jura}, {Wright}, \&
  {Cruikshank}}]{Werner2004}
{Werner}, M.~W., {Roellig}, T.~L., {Low}, F.~J., {et~al.} 2004, \apjs, 154, 1,
  \dodoi{10.1086/422992}

\bibitem[{{Whitney} {et~al.}(2013){Whitney}, {Robitaille}, {Bjorkman}, {Dong},
  {Wolff}, {Wood}, \& {Honor}}]{Whitney2013}
{Whitney}, B.~A., {Robitaille}, T.~P., {Bjorkman}, J.~E., {et~al.} 2013, \apjs,
  207, 30, \dodoi{10.1088/0067-0049/207/2/30}

\bibitem[{{Williams} \& {Cieza}(2011)}]{Williams2011}
{Williams}, J.~P., \& {Cieza}, L.~A. 2011, \araa, 49, 67,
  \dodoi{10.1146/annurev-astro-081710-102548}

\bibitem[{{Winston} {et~al.}(2019){Winston}, {Hora}, {Gutermuth}, \&
  {Tolls}}]{Winston19}
{Winston}, E., {Hora}, J., {Gutermuth}, R., \& {Tolls}, V. 2019, \apj, 880, 9,
  \dodoi{10.3847/1538-4357/ab27c8}

\bibitem[{{Winston} {et~al.}(2020){Winston}, {Hora}, \& {Tolls}}]{Winston2020}
{Winston}, E., {Hora}, J.~L., \& {Tolls}, V. 2020, \aj, 160, 68,
  \dodoi{10.3847/1538-3881/ab99c8}

\bibitem[{{Wright} {et~al.}(2010){Wright}, {Eisenhardt}, {Mainzer}, {Ressler},
  {Cutri}, {Jarrett}, {Kirkpatrick}, {Padgett}, {McMillan}, {Skrutskie},
  {Stanford}, {Cohen}, {Walker}, {Mather}, {Leisawitz}, {Gautier}, {McLean},
  {Benford}, {Lonsdale}, {Blain}, {Mendez}, {Irace}, {Duval}, {Liu}, {Royer},
  {Heinrichsen}, {Howard}, {Shannon}, {Kendall}, {Walsh}, {Larsen}, {Cardon},
  {Schick}, {Schwalm}, {Abid}, {Fabinsky}, {Naes}, \& {Tsai}}]{Wright2010}
{Wright}, E.~L., {Eisenhardt}, P. R.~M., {Mainzer}, A.~K., {et~al.} 2010, \aj,
  140, 1868, \dodoi{10.1088/0004-6256/140/6/1868}

\bibitem[{{Xu} {et~al.}(2016){Xu}, {Reid}, {Dame}, {Menten}, {Sakai}, {Li},
  {Brunthaler}, {Moscadelli}, {Zhang}, \& {Zheng}}]{Xu2016}
{Xu}, Y., {Reid}, M., {Dame}, T., {et~al.} 2016, Science Advances, 2, e1600878,
  \dodoi{10.1126/sciadv.1600878}

\bibitem[{{Xue} {et~al.}(2016){Xue}, {Jiang}, {Gao}, {Liu}, {Wang}, \&
  {Li}}]{Xue2016}
{Xue}, M., {Jiang}, B.~W., {Gao}, J., {et~al.} 2016, \apjs, 224, 23,
  \dodoi{10.3847/0067-0049/224/2/23}

\bibitem[{{Yang} {et~al.}(2012){Yang}, {Nie}, {Xu}, {Luo}, {Zhuang}, \&
  {Pan}}]{Yang2012}
{Yang}, Y., {Nie}, F., {Xu}, D., {et~al.} 2012, IEEE Transactions on Pattern
  Analysis and Machine Intelligence, 34, 723

\bibitem[{{Zari} {et~al.}(2018){Zari}, {Hashemi}, {Brown}, {Jardine}, \& {de
  Zeeuw}}]{Zari2018}
{Zari}, E., {Hashemi}, H., {Brown}, A.~G.~A., {Jardine}, K., \& {de Zeeuw},
  P.~T. 2018, \aap, 620, A172, \dodoi{10.1051/0004-6361/201834150}

\bibitem[{{Zasowski} {et~al.}(2009){Zasowski}, {Majewski}, {Indebetouw},
  {Meade}, {Nidever}, {Patterson}, {Babler}, {Skrutskie}, {Watson}, {Whitney},
  \& {Churchwell}}]{Zasowski2009}
{Zasowski}, G., {Majewski}, S.~R., {Indebetouw}, R., {et~al.} 2009, \apj, 707,
  510, \dodoi{10.1088/0004-637X/707/1/510}

\bibitem[{{Zavagno} {et~al.}(2006){Zavagno}, {Deharveng}, {Comer{\'o}n},
  {Brand}, {Massi}, {Caplan}, \& {Russeil}}]{Zavagno2006}
{Zavagno}, A., {Deharveng}, L., {Comer{\'o}n}, F., {et~al.} 2006, \aap, 446,
  171, \dodoi{10.1051/0004-6361:20053952}

\bibitem[{{Zucker} {et~al.}(2020){Zucker}, {Speagle}, {Schlafly}, {Green},
  {Finkbeiner}, {Goodman}, \& {Alves}}]{Zucker2020}
{Zucker}, C., {Speagle}, J.~S., {Schlafly}, E.~F., {et~al.} 2020, \aap, 633,
  A51, \dodoi{10.1051/0004-6361/201936145}

\end{thebibliography}

\clearpage\clearpage

\begin{deluxetable}{rcl}[ht]
\tablecaption{Candidate YSOs\label{tab:spicy}}
\tabletypesize{\tiny}\tablewidth{0pt}
\tablehead{
  \colhead{Column} &  \colhead{Column ID}   &  \colhead{Description} 
}
\startdata
1 & SPICY & Candidate YSO designation\\
2 & RAdeg & ICRS R.A.\ coordinate in decimal degrees\\
3 & DEdeg & ICRS decl.\ coordinate in decimal degrees\\
4 & GLON & Galactic longitude\\
5 & GLAT & Galactic latitude\\
6 & p1 & YSO random forest score$^a$ from IRAC+2MASS photometry\\
7 & p2 & YSO random forest score$^a$ from IRAC+UKIDSS photometry\\
8 & p3 & YSO random forest score$^a$ from IRAC+VVV photometry\\
9 & class & YSO class$^b$\\
10 & silicate & Flag for a possible strong silicate feature\\
11 & pah & Flag for a possible strong PAH feature\\
12 & alpha & Spectral index used for YSO class$^c$\\
13 & alpha8 & Spectral index derived from the $[4.5]-[8.0]$ color\\
14 & alpha24 & Spectral index derived from the $[4.5]-[24]$ color\\
15 & alphaW4 & Spectral index derived from the $[4.5]-W4$ color\\
16  & env & Classification of the YSO environment$^d$ from the $3^\prime\times3^\prime$ IRAC cutout\\ 
17 & group & HDBSCAN group to which the star is assigned$^e$\\
18 & var & ZTF light curve variability flag$^f$ \\
19 & n\_ZTFrmag & Number of good ZTF $r$-band observations used\\
20 & ZTFrmag & ZTF mean magnitude in the $r$-band\\
21 & sigma & ZTF light curve $\sigma_\mathrm{var}$ standard deviation in the $r$-band\\
22 & skewness & ZTF light curve skew in the $r$-band\\
\hline
\multicolumn{3}{l}{GLIMPSE (and Extensions) Catalog Columns}\\
23 & Spitzer & Spitzer source designation\\
24 & 3.6mag & Spitzer/IRAC channel 1 magnitude\\
25 & e\_3.6mag & Error on Spitzer/IRAC channel 1 magnitude\\
26 & 4.5mag & Spitzer/IRAC channel 2 magnitude\\
27 & e\_4.5mag & Error on Spitzer/IRAC channel 2 magnitude\\
28 & 5.8mag & Spitzer/IRAC channel 3 magnitude\\
29 & e\_5.8mag & Error on Spitzer/IRAC channel 3 magnitude\\
30 & 8.0mag & Spitzer/IRAC channel 4 magnitude\\
31 & e\_8.0mag & Error on Spitzer/IRAC channel 4 magnitude\\
32 & csf &  Close source flag \\
33 & n\_3.6mag & number of detections in the 3.6~$\mu$m band  \\
34 & n\_4.5mag & number of detections in the 4.5~$\mu$m band   \\
35 & n\_5.8mag & number of detections in the 5.8~$\mu$m band   \\
36 & n\_8.0mag & number of detections in the 8.0~$\mu$m band   \\
\hline
\multicolumn{3}{l}{Cross-Matched Catalogs}\\
37 & 2MASS & 2MASS source designation\\
38 & UKIDSS & UKIDSS source designation\\
39 & VIRAC & VIRAC DR1 source designation\\
40 & GaiaDR2 & Gaia DR2 source designation\\
41 & MIPS & Spitzer/MIPS source designation\\
42 & AllWISE & AllWISE source designation\\
43 & ZTFDR3 & ZTF DR3 source designation\\
\enddata
\tablecomments{ 
In addition to the quantities derived in this paper, for the convenience of the user, this table also provides select columns from the GLIMPSE (and extensions) catalogs. \\
(This table is available in its entirety in a machine-readable form in the online journal. The list of columns is shown here for guidance regarding its form and content.)
}
\vspace*{-0.05in}
\tablenotetext{$a$}{Random forest scores range from 0 to 1, where higher scores indicate a greater chance that an object is a YSO. We include sources with scores $>$0.5 from one of the classifiers.}
\vspace*{-0.05in}
\tablenotetext{$b$}{YSO classes are ``Class I,'' ``FS'' (flat SED), ``Class II,'' and ``Class III.''}
\vspace*{-0.05in}
\tablenotetext{$c$}{This $\alpha$ combines the results from Columns~13--15 as described in \autoref{sec:yso_class}.}
\vspace*{-0.05in}
\tablenotetext{$d$}{The environment classes are ``EnvI'' (no or minimal nebulosity), ``EnvII'' (mixed category), ``EnvIII'' (cloud-like environment).}
\vspace*{-0.05in}
\tablenotetext{$e$}{The list of groups is provided in Table~\ref{tab:groups}.}
\vspace*{-0.05in}
\tablenotetext{$f$}{The variability classes are 1 (weak or statistically insignificant variability), 2 (moderate variability), 3 (high variability).}
\end{deluxetable}

\begin{deluxetable}{rcl}[ht]
\tablecaption{YSO Groups from HDBSCAN\label{tab:groups}}
\tabletypesize{\small}\tablewidth{0pt}
\tablehead{
  \colhead{Column} &  \colhead{Column ID}   &  \colhead{Description} 
}
\startdata
1 & group & Group designation\\
2 & l0 & Central Galactic longitude $\ell_0$ [deg]\\
3 & b0 & Central Galactic latitude $b_0$ [deg]\\
4 & plx & Mean parallax [mas]\\
5 & e\_plx & Error on mean parallax [mas]\\
6 & pml & Mean proper motion in $\ell$ [mas yr$^{-1}$]\\
7 & e\_pml & Error on mean proper motion in $\ell$ [mas yr$^{-1}$]\\
8 & pmb & Mean proper motion in $b$ [mas yr$^{-1}$]\\
9 & e\_pmb & Error on mean proper motion in $b$ [mas yr$^{-1}$]\\
10 & n & Total number of constituents\\
11 & nG & Number of constituents with\\
&&5-parameter Gaia astrometric solutions\\
12 & flag & Flag for potential model problems
\enddata
\tablecomments{Properties of YSO groups identified from the HDBSCAN algorithm. Median astrometric properties, including group parallax and proper motion, are inferred from the hierarchical Bayesian modeling of the Gaia DR2 astrometry. 
The group parallaxes and proper motions in this table are in the Gaia DR2 system, with no correction for zero-point offsets. We report formal (MAD) uncertainties from our model added in quadrature to the $\pm$0.04~mas and $\pm$0.07~mas~yr$^{-1}$ spatially correlated systematic errors on DR2 zero points \citep{Lindegren2018}. Groups are flagged if potential problems could affect interpretation of the Bayesian model as described in \autoref{sec:xy}.
}
\end{deluxetable}

\begin{deluxetable}{RRRR}[ht]
\tablecaption{Regions Near the Galactic Midplane without Significant YSO Populations \label{tab:rectangles}}
\tabletypesize{\small}\tablewidth{0pt}
\tablehead{
  \colhead{$\ell_\mathrm{min}$} &  \colhead{$\ell_\mathrm{max}$}   &  \colhead{$b_\mathrm{min}$}   &  \colhead{$b_\mathrm{max}$} \\
  \colhead{(deg)} &  \colhead{(deg)}   &  \colhead{(deg)} &  \colhead{(deg)}
}
\startdata
275.63 & 277.85 & -1.52 & 0.51\\
328.95 & 331.37 & -2.97 & -1.17\\
349.02 & 349.40 & -2.22 & -0.09\\
358.32 & 2.08 & 0.81 & 3.87\\
1.25 & 1.44 & -0.33 & -0.08\\
0.12 & 2.28 & -4.43 & -3.11\\
0.12 & 6.43 & -3.12 & -1.88\\
23.89 & 26.47 & -3.01 & -1.22\\
28.84 & 31.46 & -2.97 & -1.30\\
32.06 & 32.73 & -1.14 & -0.23\\
43.22 & 44.37 & 0.28 & 1.13\\
47.79 & 48.46 & 0.20 & 0.86\\
\enddata
\tablecomments{This table provides the lower left and upper right 
boundaries of the rectangular regions from which we obtained our labeled list 
of ``field'' objects. Rectangles 
are drawn with lines of constant Galactic 
$\ell$ and $b$.
}
\end{deluxetable}
\clearpage\clearpage

\end{document}